\documentclass[aps,amsmath,amssymb,showpacs,showkeys,longbibliography]{revtex4-2}
\usepackage[dvips]{graphicx}
\usepackage{times}
\usepackage{braket}
\usepackage{xcolor}
\usepackage{orcidlink}
\usepackage{hyperref}
\hypersetup{
  colorlinks=true,
  urlcolor=magenta,
  linkcolor=red,
  citecolor=blue
}
\usepackage{float}
\usepackage{multirow}
\usepackage{mathptmx}
\usepackage[sort&compress]{natbib}
\usepackage[shortcuts]{extdash}

\begin{document}
\title{Quasinormal Modes and Shadows of Black Holes in Infinite Derivative 
Theory of Gravity}

\author{Rupam Jyoti Borah\orcidlink{0009-0005-5134-0421}}
\email[Email: ]{rupamjyotiborah856@gmail.com}

\author{Umananda Dev Goswami\orcidlink{0000-0003-0012-7549}}
\email[Email: ]{umananda@dibru.ac.in}

\affiliation{Department of Physics, Dibrugarh University, Dibrugarh 786004, 
Assam, India}

\begin{abstract}
In this work, we studied the quasinormal modes (QNM) and shadow of a Schwarzschild Black hole (BH) with higher-order metric corrections, in the framework of Infinite Derivative theory of Gravity (IDG). We studied the effects on QNMs and shadow of the BH, which arises from the IDG's corrections to the BH's metric. We used 6th order Padé averaged WKB approximation method to study the QNMs of the BH perturbed by a scalar field. We analyze the dependence of the amplitude and damping of QNMs with respect to the free parameters. Next, we studied the time evolution of a scalar field around the BH spacetime. The QNMs have been calculated from the time profile of the evolution. Then we studied the shadow of the BH. The variation of the shadow radius due to the inclusion of higher-order corrections was studied and the variation of the shadow radius with respect to the free parameters also studied. Furthermore, the dependence of the shadow radius on the mass of the BH is also studied. Finally, we constrained the free parameters associated with the correction terms using the data from the  Keck and VLTI observation, and we obtained some bounds on the parameters.
\end{abstract}


\maketitle                                                                      

\section{Introduction}
General Relativity (GR) has been the most successful theory of gravity from 
its beginning. It has been extensively tested through various experiments 
and astronomical observations, which confirmed its predictions with high 
precision. The recent detection of gravitational waves (GWs) by the LIGO 
collaboration \cite{L1,L2,L3,L4,L5} and the groundbreaking images of black 
holes (BHs) captured lately by the Event Horizon Telescope (EHT) group 
\cite{E1,E2,E3,E4,E5,E6} stand as two remarkable milestones in this direction. 
However, despite its successes GR has been facing several fundamental 
challenges. As an example, we can say that GR struggles to explain why the 
Universe is expanding at an accelerated rate and why a significant portion 
of its mass seems to be missing \cite{F1,F2,F3,F4}. 
Similarly, when gravity is viewed as a fundamental interaction, at a very high 
energy scale where the quantum effects of gravity become dominant, GR fails to 
incorporate the Quantum Gravity (QG) effects into its solutions. The Standard 
Model (SM) of particle physics successfully describes the three fundamental 
interactions: strong, electromagnetic and weak interactions, in the framework 
of Quantum Field Theory (QFT), but it is unable to incorporate the 
gravitational interaction within it because GR is not a renormalizable theory. 
The quest to merge GR with SM continues to be an area of active research, and 
new theoretical constructs show increasing promise. Thus, QG theories 
\cite{QG1,QG2,QG3,QG4,QG5,QG6} are being formulated to bridge the gap between 
GR and QFT. Loop Quantum Gravity (LQG) \cite{RLQ1,RLQ2,RLQ3,RLQ4} and String 
Theory \cite{RST1,RST2,RST3,RST4,RST5} are two widely studied theories in this 
direction. String Theory tries to unify the four fundamental interactions of 
nature in a single framework. String theory predicts the existence of a 
particle called graviton, which is the mediator of gravitational interaction. 
On the other hand, LQG only deals with quantizing gravity. In LQG, spacetime 
is quantized into discrete loops or quanta. 

Although LQG and String Theory have been widely studied to investigate the QG
approaches, their predictions have not been tested yet. The fact is that to 
test the predictions of String Theory and LQG, an enormous amount of energy, 
far beyond current experimental facilities is needed, which limits the 
experimental verification of these theories to an unachievable task. Because 
of this difficult challenge, alternative approaches have been looked into to 
study QG effects. One such approach is the framework of the Effective Field 
Theory (EFT). The EFT provides a way to explore QG effects within the energy 
scale accessible to current experiments without requiring extremely high 
energies. By treating GR as an effective theory of gravity valid up to a 
certain cut-off, this approach allows us to incorporate quantum corrections 
while maintaining consistency with GR at low energy. Several studies have been 
performed by considering GR as an EFT 
\cite{GREFT1,GREFT2,GREFT3,GREFT4,GREFT5,GREFT6}. It is to be noted that 
although EFT can predict only the low energy quantum corrections to a system 
and these low energy effects are very small, these effects become 
significant in extreme gravity regions such as near the event horizon of a BH 
or in the early Universe. The most promising testing ground for these theories 
with our current understanding lies in the extreme environments surrounding 
BHs. Accordingly, a number of studies have been done on the BHs 
in the framework of EFT \cite{M1,M2,M3,M4,M5,M6,M7,M8,M9}.

Considering GR as an EFT, we can incorporate the low energy QG effects into 
the solutions of Einstein field equations, and hence we can test our 
calculations using the observational data obtained from BHs and Cosmic 
Microwave Background (CMB) \cite{cmb1,cmb2}. Since traditional approaches 
of EFT ignore the contributions from the high energy scale, we require 
some modifications to traditional EFTs to cover the UV incompleteness. 
Recently, the Infinite Derivative theory of Gravity (IDG) 
\cite{Tirthabir1,Tirthabir3} has gained significant attention due to its 
remarkable properties. This framework offers an asymptotically free and 
ghost-free extension of GR, ensuring that the only propagating degree of 
freedom in spacetime is the massless graviton. Moreover, IDG is considered as 
a promising candidate for UV complete modifications of GR, as its nonlocal 
structure helps to suppress divergences at high energy scales, potentially 
resolving singularities within the theory. The theory incorporates all 
possible higher derivative terms that are quadratic in curvature into the 
Einstein-Hilbert action, leading to modifications of Einstein's gravity at 
short distances. Furthermore, such a higher derivative formulation naturally 
arises in String Field Theory (SFT), where it emerges as a consequence of 
higher-order $\alpha'$ corrections \cite{alpha1}, refining the gravitational 
dynamics at small scales.

Studying BH's observables in such a theory can give us valuable information 
about how gravity behaves at the quantum scale. By analyzing these 
observables, one can test whether our current theories hold or not. BH 
thermodynamics provides a key framework where gravity, quantum physics and 
thermodynamics intersect, which is significant in unveiling the quantum aspect 
of gravity. Observation of BH phase shifts and changes in entropy can provide 
additional information on QG. These measurable quantities not only help to 
test the validity of existing theories but also guide the development of new 
models that aim to unify GR with quantum physics. BH thermodynamics have been 
extensively studied in Refs.~\cite{C1,C8,Thermo0,Thermo1,Thermo4,Thermo5,
Thermo6,Thermo7,Thermo8} from the perspectives of different theories of 
gravity.

The quasinormal modes (QNMs) of BHs allow the direct extraction of the 
frequency associated with the characteristic oscillation of the BHs. These 
oscillation frequencies are dependent on the mass, spin, and charge of the 
BHs and can be altered due to the impact of the QG effects acting on the BH 
spacetime. In many approaches to QG, such as LQG and String Theory, 
modifications to the classical BH metric lead to shifts in QNM frequencies. 
Additionally, the AdS/CFT correspondence suggests a deep connection between 
QNMs and the relaxation timescales of strongly coupled QFTs \cite{ac1,ac2}. 
This duality implies that understanding the QNMs in asymptotically AdS 
spacetime can provide insights into the quantum properties of gravity. Because 
of these reasons, the study of QNMs of BHs can provide QG signatures around 
the BH spacetime. The QNMs of BHs have been studied extensively in different
theories of gravity in recent times 
\cite{QN1,QN3,QN4,QN5,QN6,QN7,QN8,QN9,QN10,QN11}. 

Similarly, the study of the BH shadow offers valuable information on the
quantum nature of gravity. The nature of BH shadows is based on the spacetime 
geometry, which is affected by quantum corrections. In various QG models, 
modifications to the BH metric lead to deviations in the shadow's size and 
shape from GR predictions. In LQG, corrections to the Schwarzschild or Kerr 
metric can result in a slightly altered photon sphere, leading to measurable 
changes in the shadow’s angular diameter \cite{shad1,shad2,shad3}. The study 
of BH shadow is also linked to the holographic principle, which suggests that 
gravity in higher-dimensional spacetime can be described by a 
lower-dimensional QFT \cite{ads1,ads2}. The EHT provides us the opportunity 
to observe the BH shadows and analyze the differences between predictions 
made with classical GR and the inclusion of possible effects of QG. The various
studies on shadows of BH based on different forms of theories of gravity are
currently available in the literature \cite{S1,S2,S3,S4,S5,S6,S7}.

Inspired by the above mentioned factors, our work aims to study the QNMs and 
shadows of BHs in the framework of IDG. Specifically, a recent development in
relation to BH solutions within this theoretical framework \cite{Yong3377} has 
inspired us to investigate the QNMs and shadows of the BHs in this gravity 
framework. To calculate the QNMs we implement the Padé averaged 6th-order WKB 
approximation method and for the study of shadows, we adopt the techniques 
used in Refs.~\cite{C8,J1,J2}. 

Our work is organized as follows: In Section~\ref{2}, we briefly discuss 
IDG and the solution of Schwarzschild BHs in this framework of gravity. In 
Section~\ref{3}, we compute the QNMs of the BHs. In Section~\ref{4}, we study 
the evolution of a scalar field perturbation around our considered BHs 
spacetime and calculate the QNMs of the BHs using time domain analysis of 
the evolution. In Section~\ref{5}, we study the shadows of the BHs. In 
Section~\ref{6}, we summarize and conclude our work. 

\section{Infinite derivative Gravity and Schwarzschild Black Holes} \label{2}
IDG is a modification of GR that incorporates an infinite number of curvature 
derivatives into the Einstein-Hilbert action, distinguishing it from 
conventional higher-derivative gravity theories \cite{J3,J4}. As mentioned 
earlier the motivation behind the development of IDG was to construct a theory 
free from ghosts and singularities. Moreover, IDG exhibits better UV behavior 
than other higher-derivative gravity theories because its specific nonlocal 
modifications lead to an exponential suppression of high-energy modes while 
avoiding ghosts. The details of the derivations of the field equations and BH 
solutions in IDG are available in the literature \cite{Yong3377,Tirthabir3}. 
In this section, we highlight some important steps of those derivations. 

The most general form of the action with second order in curvature is given by 
\citep{Yong3377,Tirthabir1,Tirthabir3}\\ \begin{equation}
S = \int{d^{4}x}\sqrt{-g}\left[\frac{R}{16\pi}+RF_{1}(\Box)R+R_{\mu\nu}F_{2}(\Box)R^{\mu\nu}+C_{\mu\nu\sigma\rho}F_{3}(\Box)C^{\mu\nu\sigma\rho}\right], \label{eq1}
\end{equation}
where $g_{\mu\nu}$ is the spacetime metric, $\Box$ is the d'Alemberian 
operator, $R$ is the Ricci scalar, $R_{\mu\nu}$ is the Ricci tensor and 
$C_{\mu\nu\sigma\rho}$ is the Weyl tensor. $F_1$, $F_2$ and $F_3$ are 
analytic functions and are given by 
\begin{equation}
F_i(\Box) = \sum_{n\,=\,0}^{\infty} f_{i,n}{\Box}^n.
\end{equation} 
Here, $f_{i,n}$ are Taylor series coefficients with $i=1,\,2,\,3$ for three 
$Fs$ and $f_{i,n}$ can be considered as the free parameters of the theory. The 
operator $\Box$ always comes with an energy scale $M_{i,n}$, which can be 
expressed as ${\Box}/M_{i,n}$ with $M_{i,n}$ as a high-energy scale at which 
the influence of the additional term becomes significant. For simplicity here 
we consider $M_{i,n} = 1$ and we choose the unit system with 
$G = c = \hbar = k_B = 1$. The variation of the action \eqref{eq1} with 
respect to metric gives the gravitational field equations as \cite{Tirthabir3}
\begin{equation}
G^{\alpha\beta}=8{\pi}T^{\alpha\beta}, \label{eq3}
\end{equation} 
where $G^{\alpha\beta}=R^{\alpha\beta}-\frac{1}{2}g^{\alpha\beta}R$ is the 
Einstein tensor and $T^{\alpha\beta}$ is an effective energy-momentum tensor, 
which is express as \cite{Yong3377,Tirthabir3} 
\begin{align}
T^{\alpha\beta} = &\; 4G^{\alpha\beta}F_{1}(\Box)R 
+ g^{\alpha\beta}RF_{1}(\Box)R 
- 4\big(\nabla^{\alpha}\nabla^{\beta} - g^{\alpha\beta}\Box\big)F_{1}R 
\notag \\[3pt]
& - 2\Omega_{1}^{\alpha\beta} 
+ g^{\alpha\beta}\big(\Omega_{1\sigma}^{\sigma} + \bar{\Omega}_{1}\big) 
+ 4R^{\alpha}_{\mu}F_{2}(\Box)R^{\mu\beta} 
- g^{\alpha\beta}R_{\mu\nu}F_2(\Box)R^{\mu\nu} \notag \\[3pt]
& - 4\nabla_{\mu}\nabla^{\beta}F_{2}(\Box)R^{\mu\alpha} 
+ 2\Box\big(F_2(\Box)R^{\alpha\beta}\big) 
+ 2g^{\alpha\beta}\nabla_\mu\nabla_\nu\big(F_{2}(\Box)R^{\mu\nu}\big) 
\notag\\[3pt]
& - 2\Omega^{\alpha\beta}_{2} 
+ g^{\alpha\beta}\big(\Omega^{\sigma}_{2\sigma} + \bar{\Omega}_{2}\big) 
- 4\Delta^{\alpha\beta}_{2} 
- g^{\alpha\beta}C_{\mu\nu\sigma\rho}F_{3}(\Box)C^{\mu\nu\sigma\rho} 
\notag \\[3pt]
& + 4C^{\alpha\mu\nu\sigma}F_{3}(\Box)C^{\beta}_{\mu\nu\sigma} 
- 4\big(2\nabla_{\mu}\nabla_{\nu} + R_{\mu\nu}\big)\big(F_{3}(\Box)C^{\beta\mu\nu\alpha}\big) \notag \\[3pt]
& - 2\Omega^{\alpha\beta}_{3} 
+ g^{\alpha\beta}\big(\Omega^{\sigma}_{3\sigma} + \bar{\Omega}_{3}\big) 
- 8\Delta^{\alpha\beta}_{3}.
\end{align}
Where,
\begin{equation}
\Omega_{1}^{\alpha\beta} = \sum_{n = 1}^{\infty}f_{1,n}\sum_{l = 0}^{\infty}\nabla^{\alpha}R^{(l)}\nabla^{\beta}R^{(n-l-1)},  \quad \bar{\Omega_{1}} = \sum_{n = 1}^{\infty}f_{1,n}\sum_{l = 0}^{n-1}R^{(l)}R^{(n-l)},
\end{equation}
\begin{equation}
\Omega_{2}^{\alpha\beta} = \sum_{n = 1}^{\infty}f_{2,n}\sum_{l = 0}^{n-1}\nabla^{\alpha}R^{{\mu\nu}(l)}\nabla^{\beta}R_{\mu\nu}^{(n-l-1)}, \quad \bar{\Omega_{2}} = \sum_{n = 1}^{\infty}f_{2,n}\sum_{l = 0}^{n-1}R_{\mu\nu}^{(l)}R^{{\mu}{\nu}(n-l)},
\end{equation}
\begin{equation}
\Omega_{3}^{\alpha\beta} = \sum_{n-1}^{\infty}f_{3,n}\sum_{l = 0}^{n-1}\nabla^{\alpha}C_{{\mu\nu\sigma\rho}(l)}\nabla^{\beta}C_{\mu\nu\sigma\rho}^{(n-l-1)}, \quad \bar{\Omega_3} = \sum_{n = 1}^{\infty}f_{3,n}\sum_{l = 0}^{n-1}C_{\mu\nu\sigma\rho}^{(l)}C^{{\mu\nu\sigma\rho}(n-l)},
\end{equation}
\begin{equation}
\Omega_{1\sigma}^{\sigma} = \sum_{n = 1}^{\infty} f_{1,n} \sum_{l = 0}^{n-1} \nabla^\sigma R^{(l)} \nabla_\sigma R^{(n-l-1)},
\end{equation}
\begin{equation}
\Omega_{2\sigma}^{\sigma} = \sum_{n = 1}^{\infty} f_{2,n} \sum_{l = 0}^{n-1} \nabla^\sigma R_{\mu}^{\ \nu (l)} \nabla_\sigma R^{\mu}_{\ \nu (n-l-1)},
\end{equation}
\begin{equation}
\Omega_{3\sigma}^{\sigma} = \sum_{n = 1}^{\infty} f_{3,n} \sum_{l = 0}^{n-1} \nabla^\sigma C^{\mu\nu\rho\sigma (l)} \nabla_\sigma C_{\mu\nu\rho\sigma}^{(n-l-1)},
\end{equation}
\begin{equation}
\Delta_{2}^{\alpha\beta} = \sum_{n = 1}^{\infty}f_{2,n}\sum_{l = 0}^{n-1}[R_{\mu}^{{\nu}(l)}\nabla^{\alpha}R^{{\beta\mu}(n-l-1)}-\nabla^{\alpha}R_{\mu}^{{\nu}(l)}R^{{\beta\mu}(n-l-1)}];\nu,
\end{equation}
\begin{equation}
\Delta_{3}^{\alpha\beta} = \sum_{n = 1}^{\infty}f_{3,n}\sum_{l = 0}^{n-1}[C_{\sigma\mu}^{{\rho\nu}(l)}\nabla^{\alpha}C_{\rho}^{{\beta\sigma\mu}(n-l-1)}-\nabla^{\alpha}C_{\sigma\mu}^{{\rho\nu}(l)}C_{\rho}^{{\beta\sigma\mu}(n-l-1)}];\nu
\end{equation}
with $R^{(l)}$ is stands for ${\Box}^{l}R$, and the same notation applies to 
the other terms. The solution of Eq.~\eqref{eq3} i.e.~the Schwarzschild BH 
solution in IDG was derived in Ref.~\citep{Yong3377}, which can be expressed as 
\begin{equation}
ds^2 = -f(r)dt^2+\frac{1}{g(r)}dr^2+r^2d\Omega, \label{eq5}
\end{equation}
where $d\Omega = d\theta^2 + \sin^2\theta d\phi^2$ and 
\begin{align}
f(r) & = 1-\frac{2M}{r} + \sum_{i=1}^{3} \sum_{n=0}^{\infty} f_{i,n}\, a_{i,n}(r), \label{eq6}\\[5pt]
g(r) & = 1-\frac{2M}{r} + \sum_{i=1}^{3} \sum_{n=0}^{\infty} f_{i,n}\, b_{i,n}(r).\label{eq7}
\end{align}
Here, $a_{i,n}$ and $b_{i,n}$ are infinite number of unknown coefficients, 
which are to be found from Eq.~\eqref{eq3}. In Ref.~\cite{Yong3377} it 
is shown that $T^{\alpha\beta}$ is reduced to only $i=3$ dependent terms and 
hence $a_{3,n}$ and $b_{3,n}$ are obtained as follows: 

For the metric correction obtained from the term $f_{3,1}\Box$, we can write 
\begin{align}
a_{3,1}(r) & = \frac{384M^{2}\pi(7M-4r)}{r^7},\\[5pt]
b_{3,1}(r) & = \frac{384M^2\pi(5M-3r)}{r^7}.
\end{align}
For the metric correction obtained from the term $f_{3,2}{\Box}^2$, we can
write 
\begin{align}
a_{3,2}(r) & = \frac{7680M^2\pi(13M^2-13Mr+3r^2)}{r^{10}},\\[5pt]
b_{3,2}(r) & = -\frac{1536M^2\pi(313M^2-279Mr+60r^2)}{r^{10}}.
\end{align}
For the metric correction obtained from the term $f_{3,3}{\Box}^3$, we can 
wirte 
\begin{align}
a_{3,3}(r) & = \frac{4608M^2\pi(-36531M^3+51405M^2r-22592Mr^2+3080r^3)}{11r^{13}},\\[5pt]
b_{3,3}(r) & = \frac{4608M^2\pi(276309M^3-330270M^2r+126720Mr^2-15400r^3)}{11r^{13}}.
\end{align}
For the metric correction obtained from the term $f_{3,4}{\Box}^4$, we can 
write 
\begin{align}
a_{3,4}(r) & = \frac{184320M^2\pi(2183454M^4-3824925M^3r+238405M^2r^2-624106Mr^3+57330r^4)}{91r^{16}},\\[5pt]
b_{3,4}(r) & = -\frac{36864M^2\pi(107657550M^4-162607731M^3r+89066033M^2r^2-20781670Mr^3+1719900r^4)}{91r^{16}}.
\end{align}
Similarly, other corrections can be obtained. With each higher-order 
correction, the form of the solution becomes increasingly complicated. In our 
work, we will stick only up to the metric correction obtained from the term 
$f_{3,3}{\Box}^3$, i.e.~up to the 3rd-order correction term. This also can 
capture the low energy tiny QG effects into the solution. Moreover, 
throughout the calculations we use the value of $M=1$ unless stated 
otherwise.

\begin{figure}[!h]
        \centerline{
        \includegraphics[scale = 0.6]{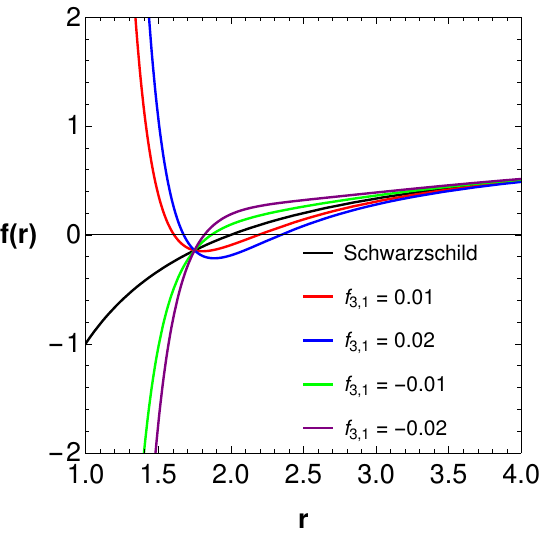}\hspace{0.3cm}
        \includegraphics[scale = 0.6]{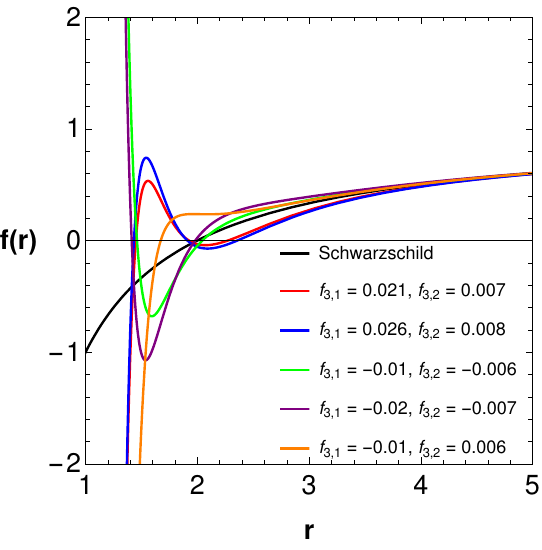}\hspace{0.3cm}
        \includegraphics[scale = 0.6]{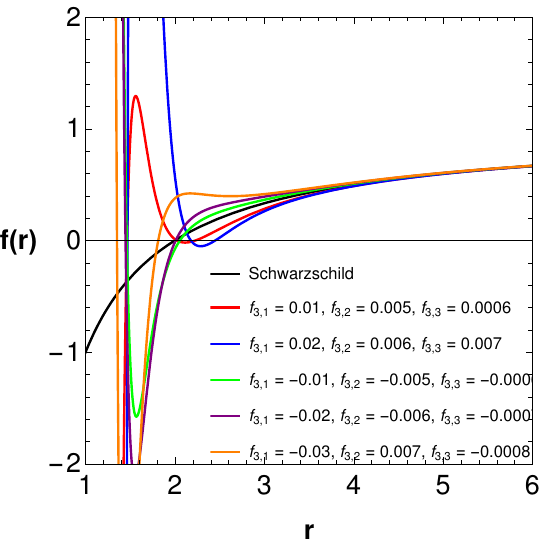}}\vspace{0.5cm}
      \centerline{
     \includegraphics[scale=0.6]{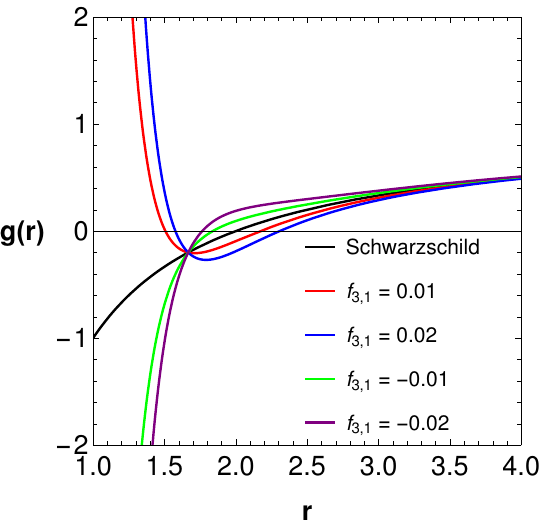}\hspace{0.3cm}
     \includegraphics[scale=0.6]{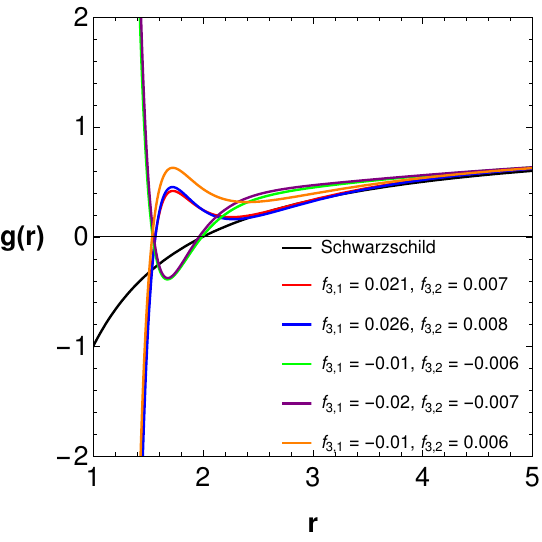}\hspace{0.3cm}
     \includegraphics[scale=0.6]{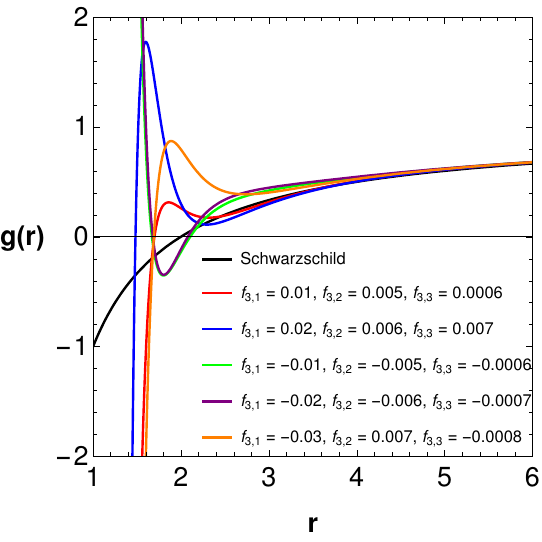}}
        \vspace{-0.2cm}
\caption{ The horizon structures of our chosen BH for higher-order 
corrections with different values of $f_{3,n}$ ($n = 1, 2, 3$). In this
figure and subsequent ones, $M = 1$ is used unless stated otherwise.}
\label{figA}
\end{figure}

Before proceed further, to account for the effects of higher-order 
corrections in the metric, we now examine the behavior of the functions $f(r)$ 
and $g(r)$, which can admit multiple real positive roots depending on the 
values of the parameters $f_{3,n}$ ($n=1,\,2,\,3$). These roots correspond to 
locations of possible horizons. Fig.~\ref{figA} illustrates $f(r)$ and $g(r)$ 
as a function of $r$, highlighting the horizon structure for different cases. 
The upper row shows the horizon structure for $f(r)$ and the lower row depicts 
the horizon structure for $g(r)$. In the upper row, the left panel shows the 
horizon structure for the 1st-order correction with different values of 
$f_{3,1}$, the middle panel shows the horizon structure for the 2nd-order 
correction, considering different combinations of $f_{3,1}$ and $f_{3,2}$, 
and the right panel shows the horizon structure for the 3rd-order correction, 
corresponding to different values of $f_{3,1}$, $f_{3,2}$, and $f_{3,3}$. It 
can be observed from the figure that for the 1st-order correction, there is 
only one horizon when $f_{3,1} < 0$, while two horizons appear when 
$f_{3,1} > 0$. In the case of the 2nd-order correction, three horizons are 
present when both $f_{3,1} > 0$ and $f_{3,2} > 0$, and two horizons are 
present when both are negative i.e.~$f_{3,1} < 0$ and $f_{3,2} < 0$. 
Interestingly, when $f_{3,1} < 0$ and $f_{3,2} > 0$, only one horizon is 
observed. For the 3rd-order correction, three horizons appear when all 
parameters are positive i.e.~$f_{3,1} > 0$, $f_{3,2} > 0$, and $f_{3,3} > 0$. 
When all three parameters are negative, $f_{3,1} < 0$, $f_{3,2} < 0$, and 
$f_{3,3} < 0$, two horizons are found. Additionally, in the case where 
$f_{3,1} < 0$ and $f_{3,3} < 0$ but $f_{3,2} > 0$, two horizons are also 
observed. In all subsequent analyses, the largest real positive root of 
$f(r) = 0$ is identified as the event horizon, which serves as the starting 
point for computing the effective potential and QNMs. Similarly, in the lower 
row, we have shown the horizon structure for $g(r)$ with the same sign 
combinations of $f_{3,n}$. It is seen that for the 1st-order correction $g(r)$ 
has the same horizon structure as that of $f(r)$, whereas for the 
2nd-order and 3rd-order corrections it has some different horizon structures 
than that of $f(r)$ in such cases. However, this behavior of the horizon 
structure of $g(r)$ does not affect our calculation, as it is clear from the 
figure that the largest real positive root of $f(r) = 0$ will remain outside 
the horizons of $g(r)$ for those two cases.

\section{Quasinormal modes of black holes} \label{3}
In this section, we study the QNMs of the BH solution \eqref{eq5} using 
the 6th-order Pad\'e averaged WKB approximation method for a massless scalar 
field perturbation around the BH spacetime. For this, we study the effective 
BH potential up to 3rd-order metric corrections, whereas for the QNMs we 
consider the metric corrections up to 2nd-order term only because of the
complexity involved in the calculations.

For a massless scalar field $\Phi$ the equation of motion is given by 
\cite{Konoplya1}
\begin{equation}
\frac{1}{\sqrt{-g}}\partial_{\mu}(\sqrt{-g}g^{\mu\nu}\partial_{\nu}\Phi). \label{eq14}
\end{equation} 
Considering a spherical symmetry with $R_{{\omega},l}(r)$ as the radial part
and $Y_{l}({\theta},{\phi})$ as the spherical harmonic, the scalar field 
$\Phi$ can be represented as 
\begin{equation}
\Phi = e^{{\pm}\,i{\omega}t}\frac{{\Psi_{\omega,l}(r)}}{r}\,Y_{l}({\theta},{\phi}), 
\label{eq15}
\end{equation} 
where $l$ is the multipole number, $\omega$ is the frequency of oscillation 
and $\Psi_{\omega,l}(r)$ is define as 
\begin{equation}
\Psi_{\omega,l}(r) = r\,{R_{{\omega},l}(r)}.
\end{equation} 
Substituting Eq.~\eqref{eq15} in Eq.~\eqref{eq14} we arrive at the following 
Schrödinger-like equation:
\begin{equation}
\frac{{d^2}\Psi}{d{r_{\star}}^2} + ({\omega}^2-V(r))\Psi=0, 
\label{eq17}
\end{equation} 
where $r_{\star}$ is the tortoise coordinate, and it can be expressed as 
\begin{equation}
r_{\star} = \int{\frac{dr}{\sqrt{f(r)g(r)}}}.
\end{equation} 
In Eq.~\eqref{eq17}, $V(r)$ is the effective potential, the expression for 
which is given by \cite{Konoplya1} 
\begin{equation}
V(r) = \frac{l(l+1)f(r)}{r^2} + \frac{g(r)\frac{\partial{f(r)}}{\partial{r}} + f(r)\frac{\partial{g(r)}}{\partial{r}}}{2r} \label{eq19}
\end{equation} 
For an asymptotically flat spacetime Eq.~\eqref{eq17} has to satisfy the 
following criteria:
\begin{equation}
\Psi(r) \to 
\begin{cases} 
A e^{+i\omega r} & \text{if } r \to -\infty, \\[3pt]
B e^{-i\omega r} & \text{if } r \to +\infty.
\end{cases}
\end{equation} 
Here, the coefficients $A$ and $B$ are the amplitudes of the wave. These 
incoming and outgoing waves satisfy the physical principle that nothing can 
escape from the BH event horizon and no radiation can be received from 
infinity. Additionally, these guarantee the existence of an infinite 
collection of discrete complex numbers, commonly referred to as QNMs.

\begin{figure}[!h]
        \centerline{
        \includegraphics[scale = 0.27]{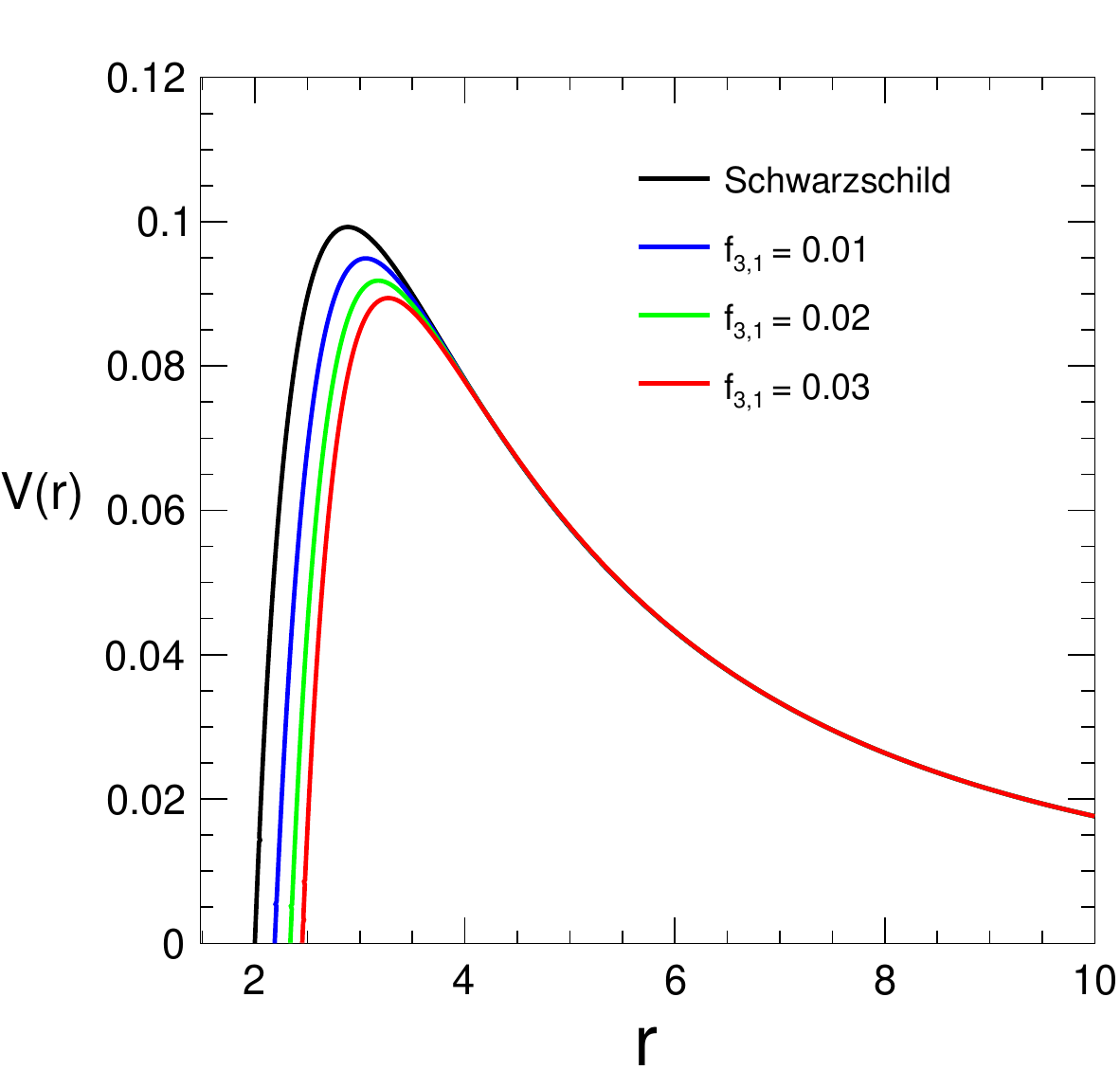}\hspace{0.3cm}
        \includegraphics[scale = 0.27]{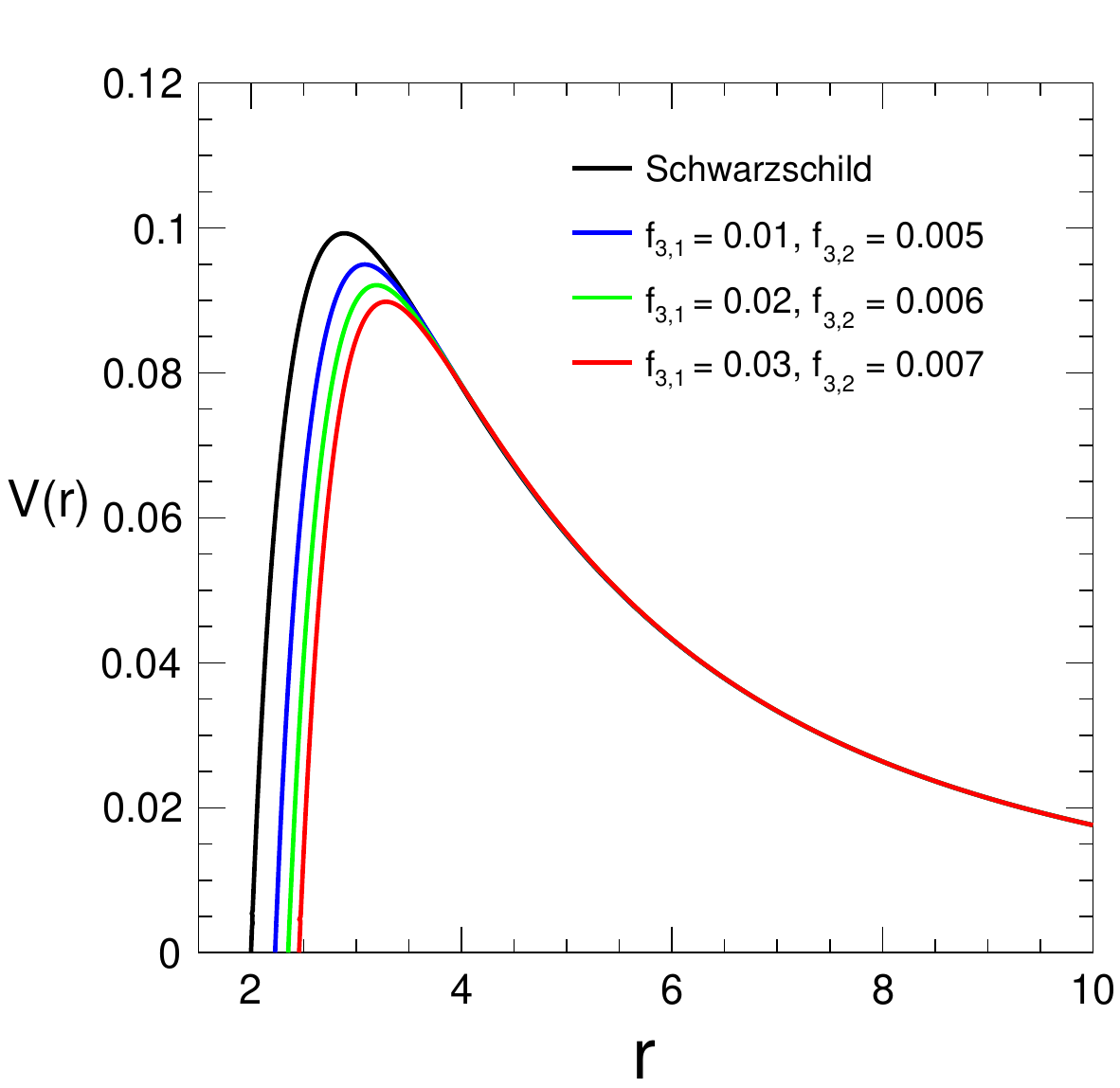}\hspace{0.3cm}
        \includegraphics[scale = 0.27]{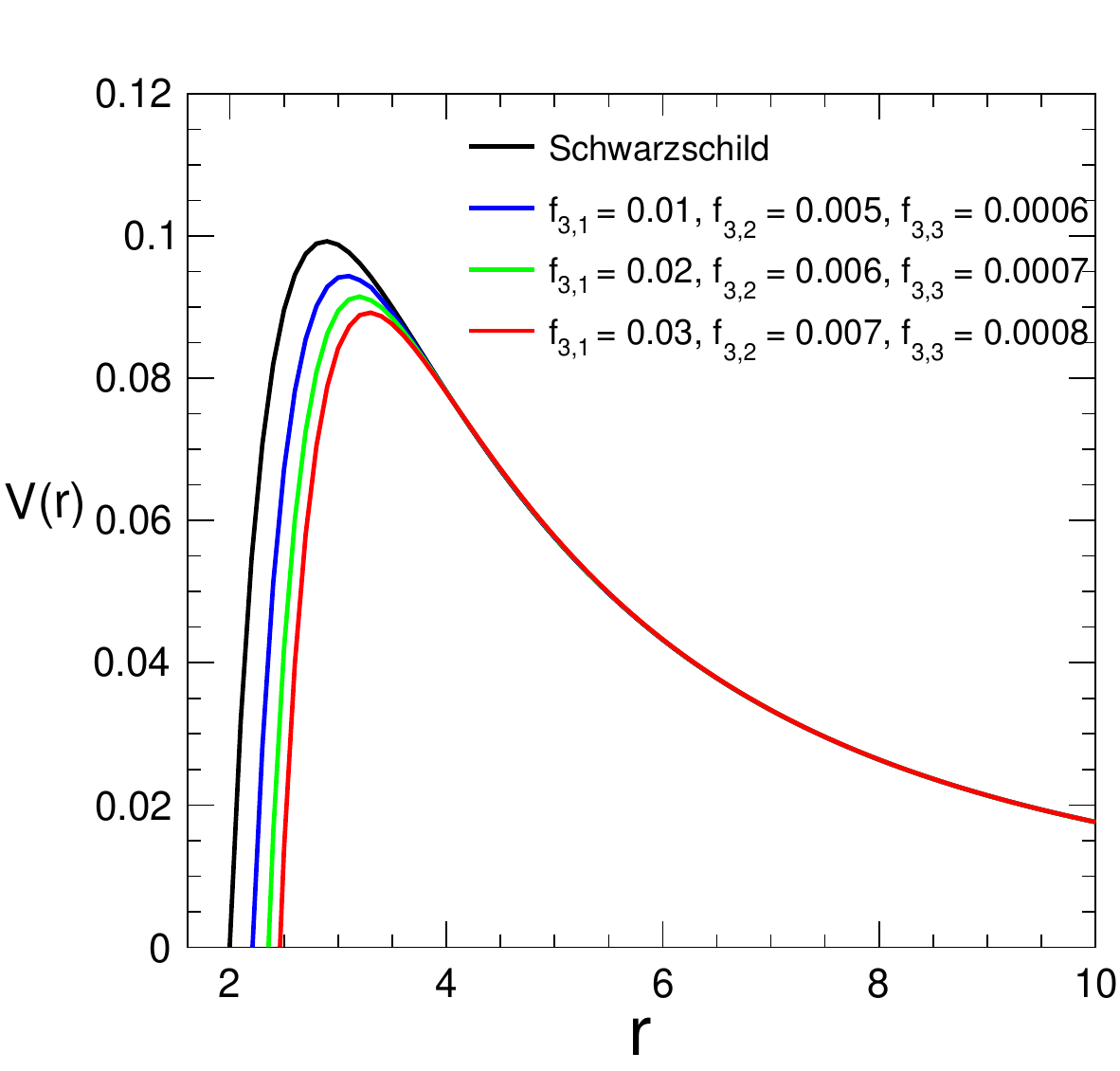}}
        \vspace{-0.2cm}
\caption{Behavior of black hole potential \eqref{eq19} with radial distance 
$r$ for different metric corrections obtained by considering the positive 
values of the coefficients $f_{3,n}$. In this figure and subsequent 
ones, $l = 1$ is used if it is not mentioned otherwise.}
\label{fig1}
\end{figure}

\begin{figure}[!h]
 	\centerline{
 	\includegraphics[scale = 0.27]{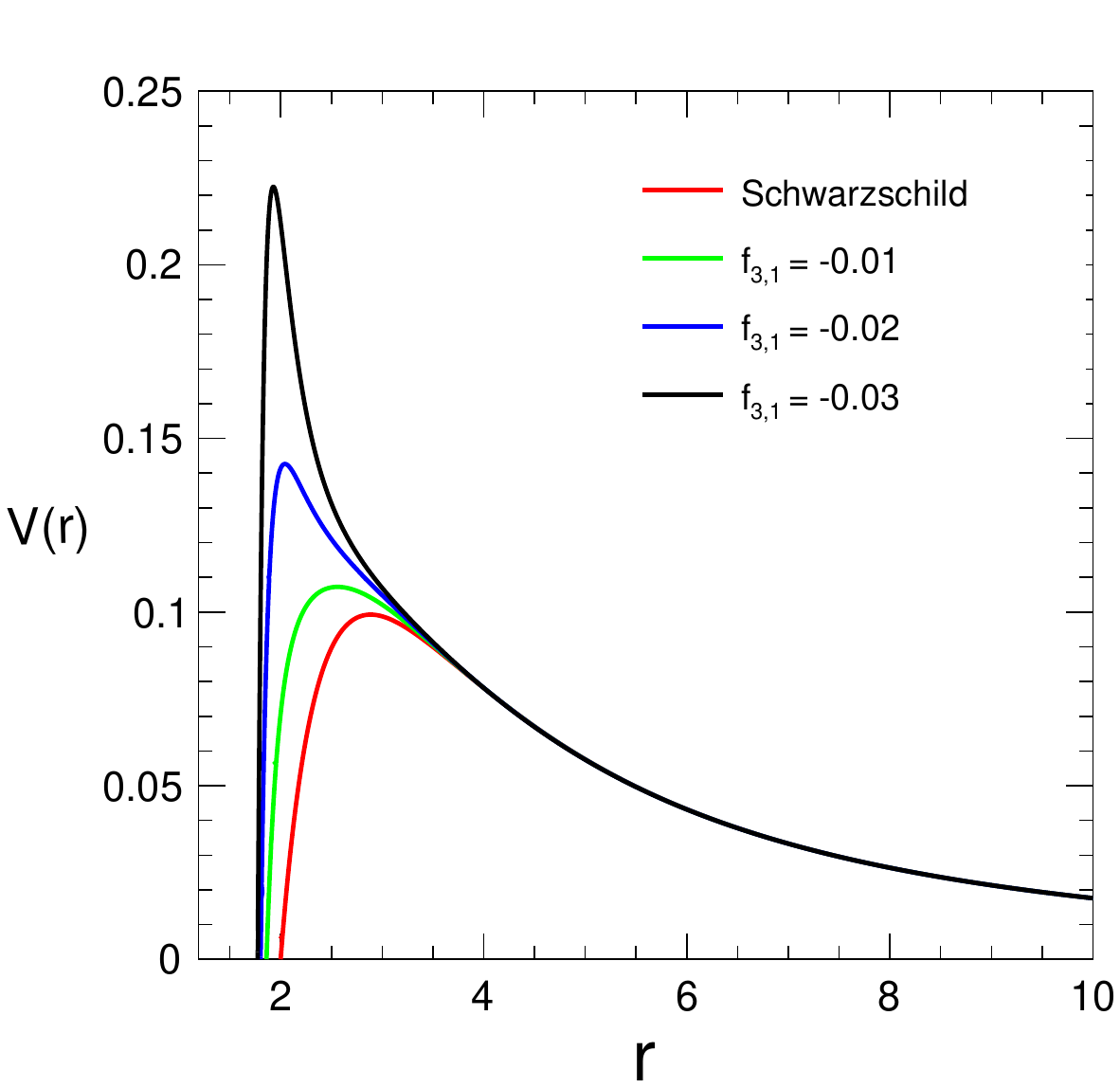}\hspace{.3cm}
 	\includegraphics[scale = 0.27]{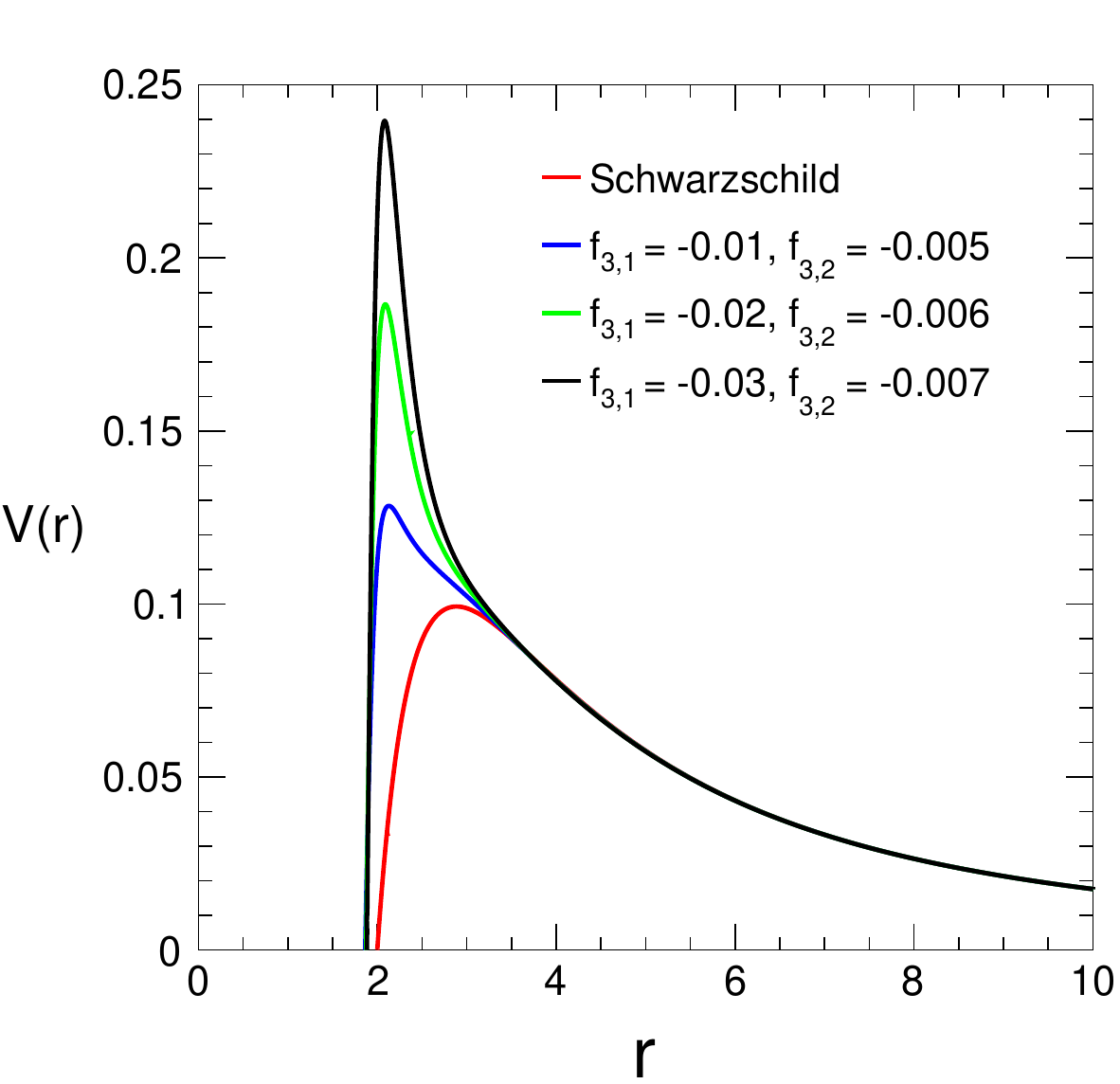}\hspace{.3cm}
 	\includegraphics[scale = 0.27]{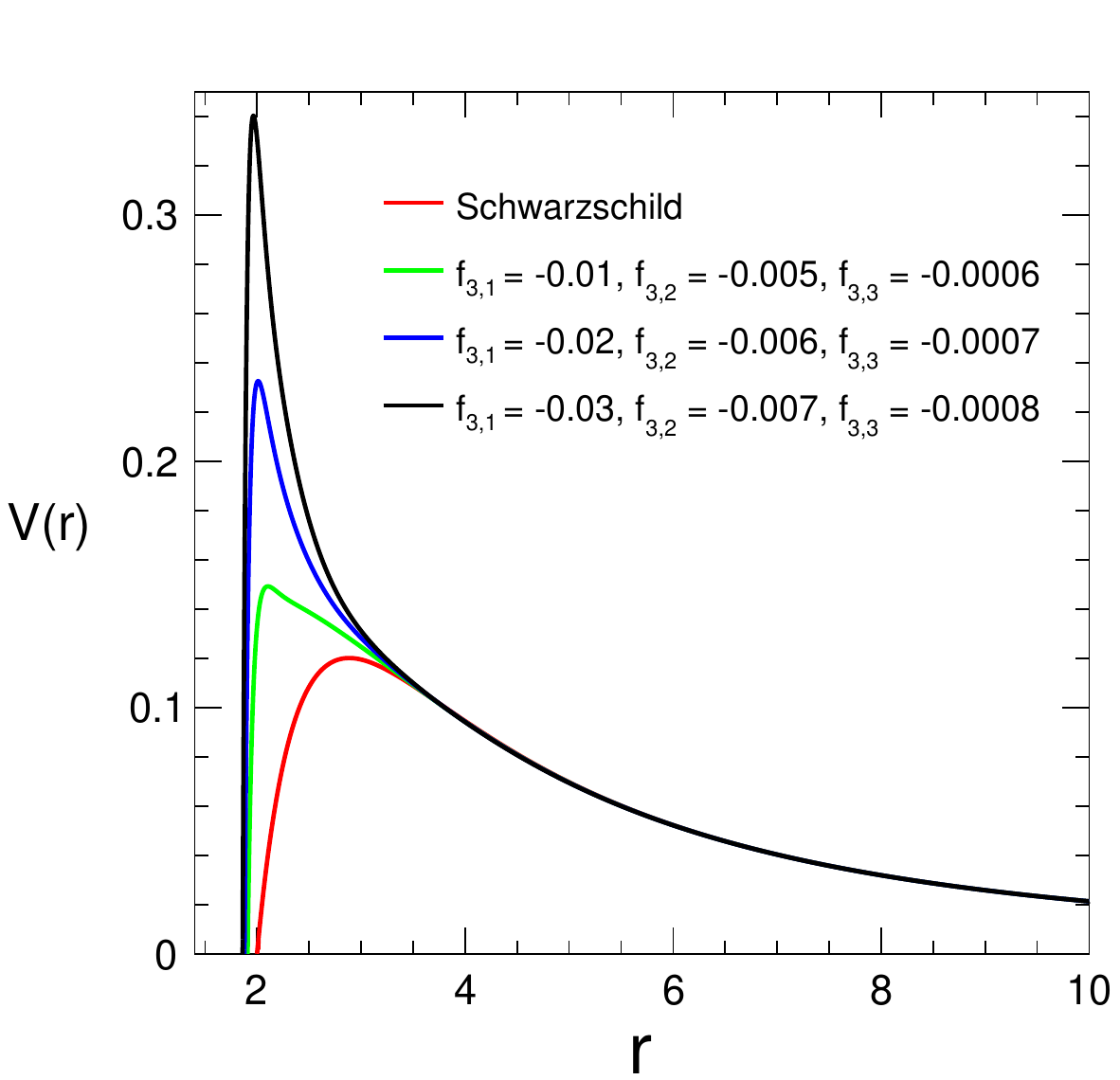}}	
        \vspace{-0.2cm}
 	\caption{Behavior of black hole potential \eqref{eq19} with radial 
distance $r$ for different metric corrections obtained by considering the 
negative values of the coefficients $f_{3,n}$.} 
\label{fig2}
\end{figure}

\begin{figure}[!h]
    \centerline{
    \includegraphics[scale = 0.28]{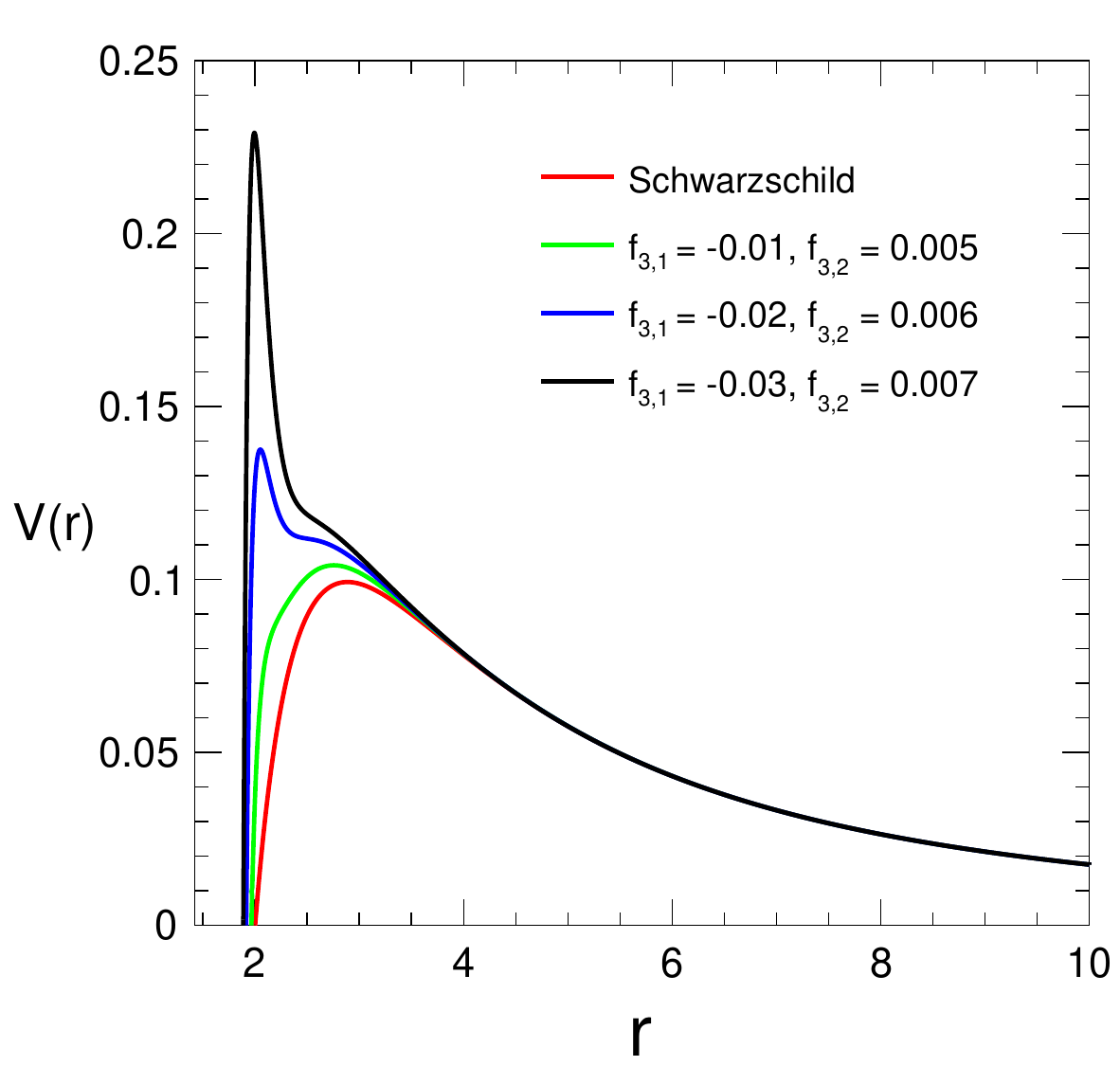}\hspace{0.3cm}
    \includegraphics[scale = 0.28]{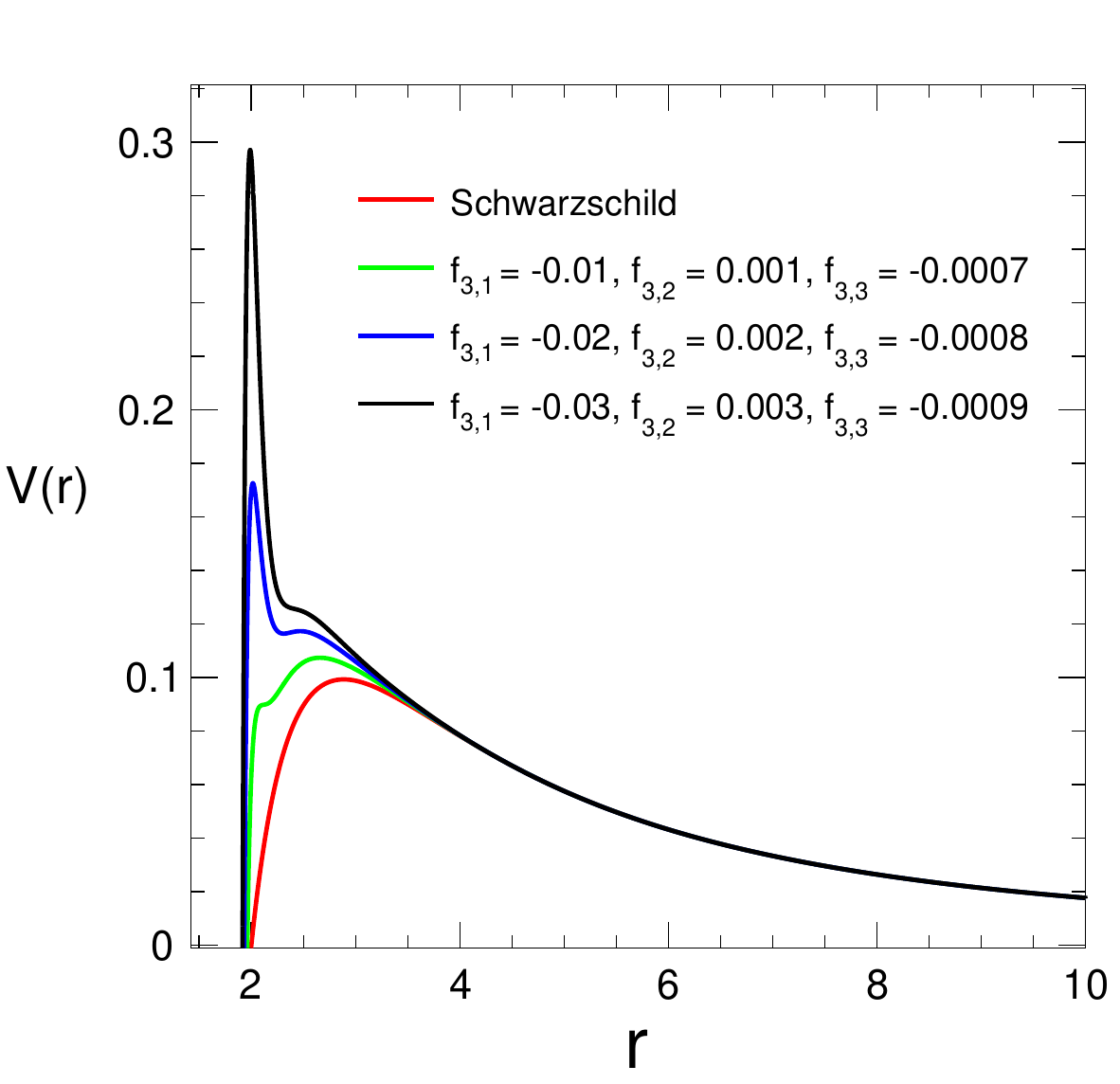}\hspace{0.3cm}
    \includegraphics[scale=0.28]{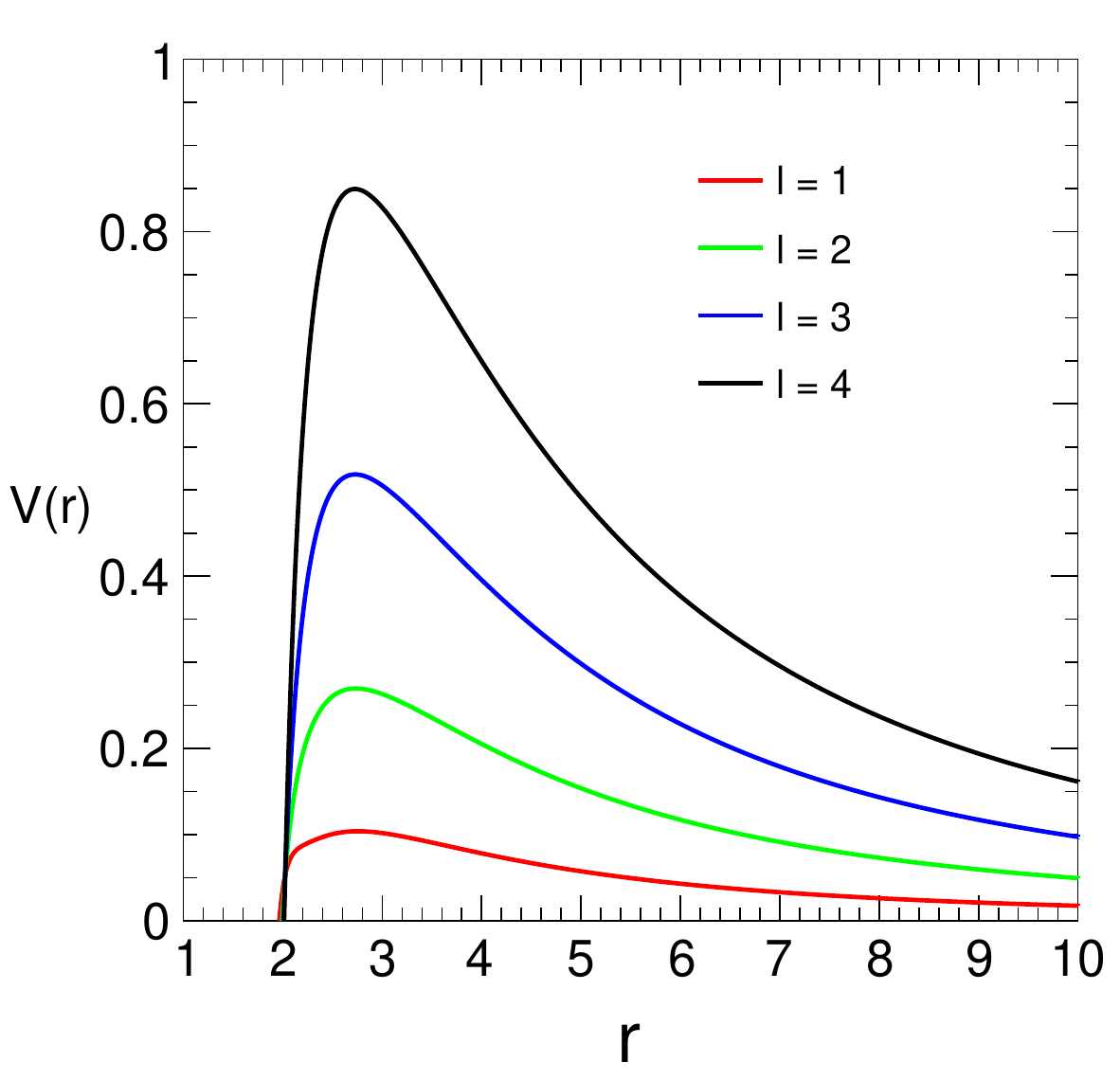}}
    \vspace{-0.2cm}
    \caption{ Variation of potential \eqref{eq19} with respect to $r$. The 
left plot is for up to 2nd-order metric corrections with negative values of 
$f_{3,1}$ and positive values of $f_{3,2}$, and the middle plot is for up to 
3rd-order metric corrections with negative values of $f_{3,1}$, positive 
values of $f_{3,2}$ and negative values of $f_{3,3}$. The right 
plot is for different values of $l$ with a fixed set of $f_{3,n}$ as 
$f_{3,1} = -0.01$, $f_{3,2} = 0.001$ and $f_{3,3} = -0.00007$.}
    \label{fig3}
\end{figure}

We study the behavior of the effective potential $V(r)$ of 
Eq.~\eqref{eq19} with respect to the radial distance $r$ considering both 
positive and negative values of the parameters $f_{3,n}$. For this purpose, 
we consider up to the 3rd-order correction to the Schwarzschild metric as 
mentioned already. Fig.~\ref{fig1} shows the behavior of $V(r)$ with respect 
to $r$ for the positive values of the parameter $f_{3,n}$ taking 
$l = 1$. At this point, it needs to be mentioned that in the 
following analysis and figures, we used $l = 1$ unless stated 
otherwise. Moreover, all calculations that are related to and with $V(r)$ are 
made considering the outer horizon as the starting point of $r$ as already 
mentioned. In the left plot of the figure, we show $V(r)$ versus $r$ behavior 
for the 1st-order correction to the metric, in the middle plot we show the 
same for up to the 2nd-order corrections to the metric and in the right plot 
we show it for up to the 3rd-order correction to the metric. It can be 
observed from Fig.~\ref{fig1} 
that the peak of the potential decreases with increasing positive values of the 
parameters $f_{3,n}$, and the height of the peak remains at the same position 
and no change of values as well as behavior of the potential even after 
the inclusion of higher order corrections. Although the metric corrected peak
value of the potential is lower than the Schwarzschild BH potential, after a 
certain value of the radial distance $r$ all potential values with corrections
and without corrections are merged together. 
In Fig.~\ref{fig2} we plot $V(r)$ with respect to $r$ for negative values of 
the coefficients $f_{3,n}$. Here the left plot is for the 1st-order metric 
correction, the middle one is for up to 2nd-order metric corrections and the 
right one is for up to 3rd-order metric corrections. From Fig.~\ref{fig2} it 
is clear that the height of the peak of the potential increases with more 
negative values of the coefficients $f_{3,n}$. Moreover, it can also be 
observed that the height of the peak increases with the inclusion of 
higher order corrections to the metric. Further, in this case, the peak values 
of metric corrected potential are higher than the uncorrected Schwarzschild 
case. However, in this case also all potentials are merged together after a 
certain value of $r$ as in the previous case.
In Fig.~\ref{fig3} we plot $V(r)$ as a function of $r$ considering 
both positive and negative values of the coefficients $f_{3,n}$, and for both 
2nd-order and 3rd-order metric corrections. Here the left plot is for up to 
2nd-order metric corrections and the middle one is for up to 3rd-order 
metric corrections. In this case also the peak of the potential increases 
with the magnitude of the parameters $f_{3,n}$ and the height of the 
potential increases with the inclusion of higher order metric corrections. 
The rest of the behaviors are similar to the previous case. For 
completeness, in the right plot we show the variation of $V(r)$ with respect 
to $r$ for different values of $l$ considering up to 3rd-order corrections 
with $f_{3,1} = -0.01$, $f_{3,2} = 0.001$ and $f_{3,3} = -0.00007$. It is seen 
that the peak as well as other values of the effective potential increase 
with the multiple number $l$, a general pattern applicable to all cases.
 
\begin{figure}[!h]
    \centerline{
    \includegraphics[height=6.2cm,width=6.2cm]{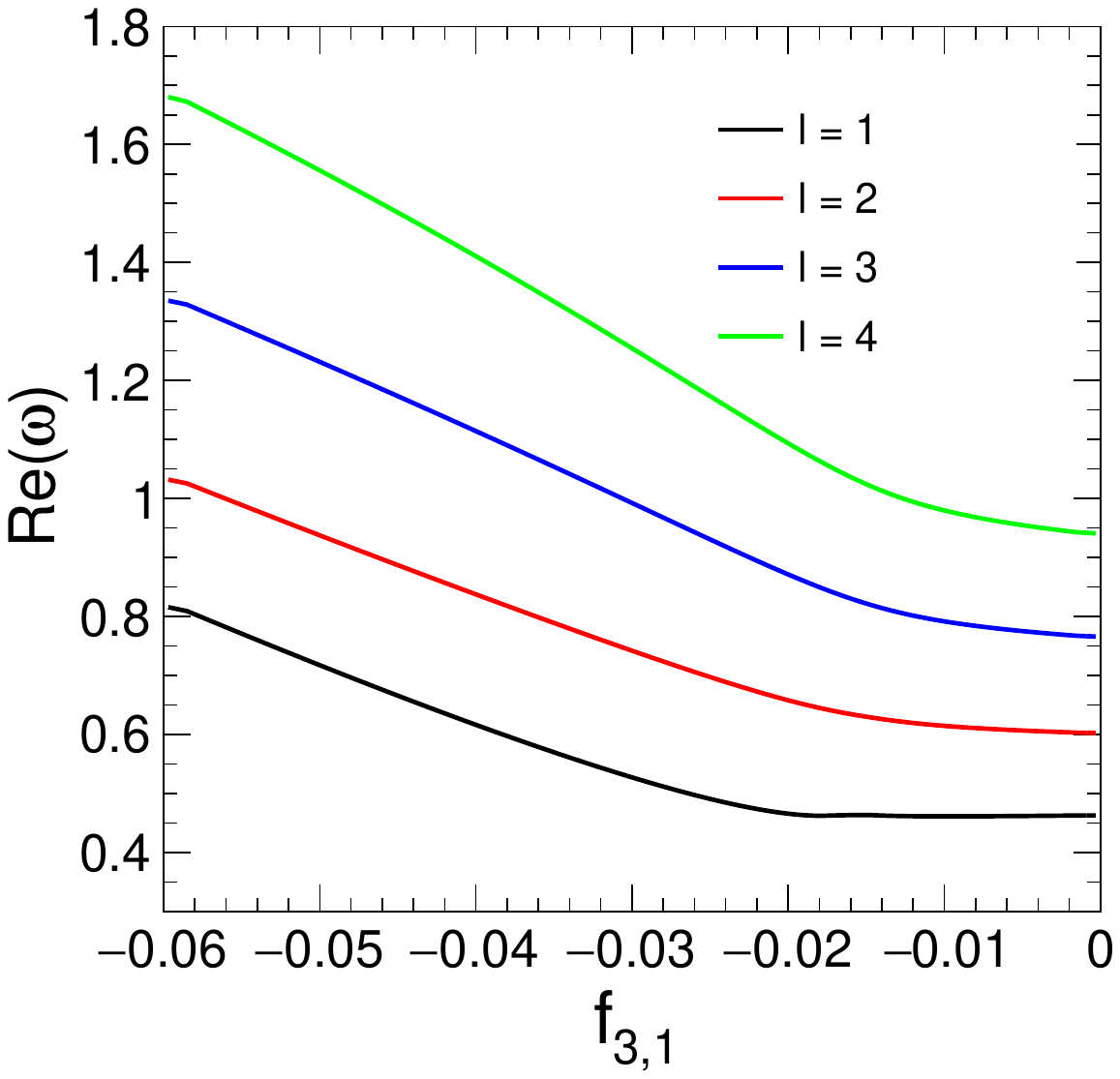}\hspace{1cm}
    \includegraphics[height=6cm,width=6.2cm]{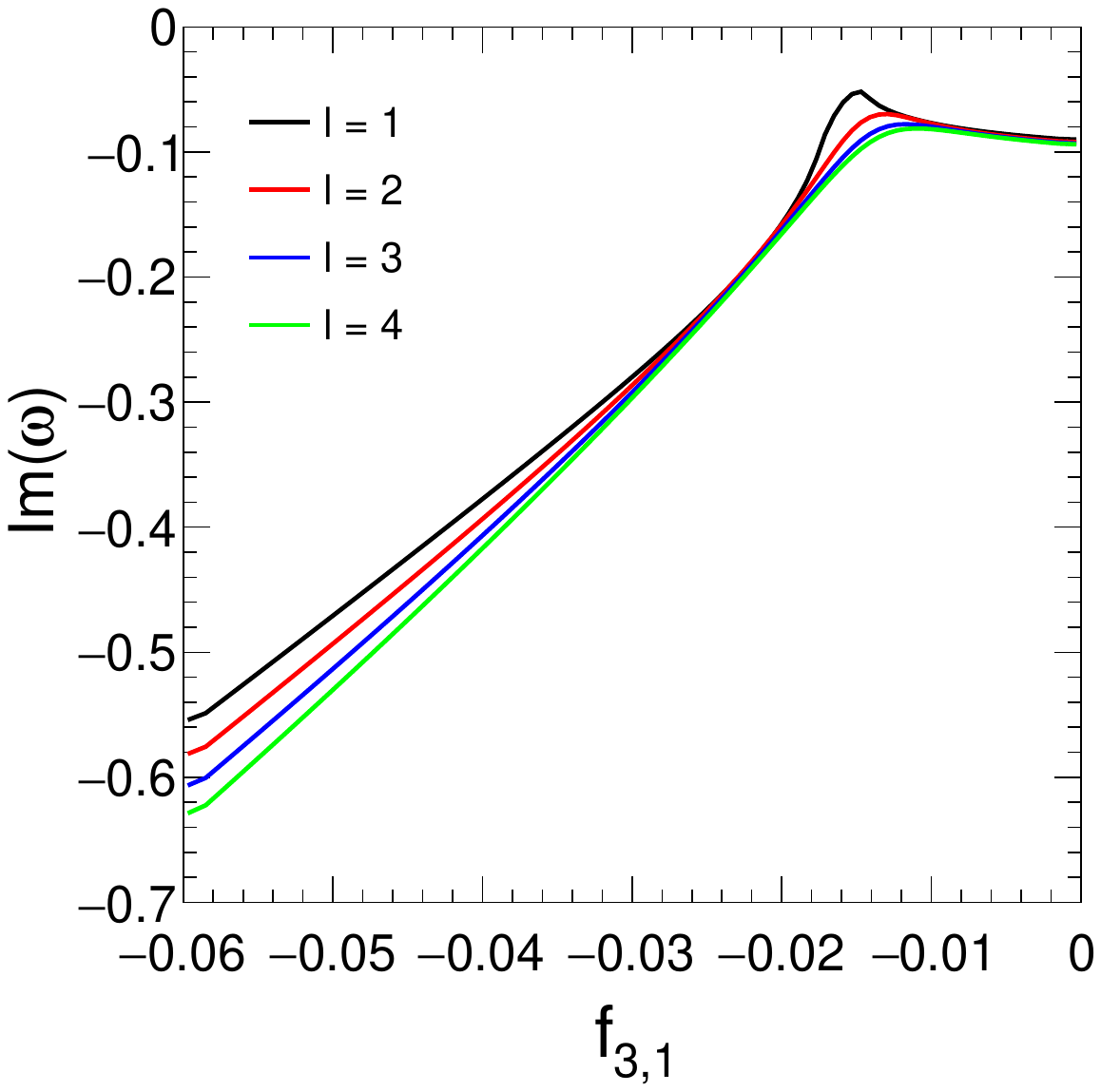}}
    \vspace{-0.2cm}
    \caption{Variation of real and imaginary parts of QNMs with respect 
to negative values of $f_{3,1}$ for the 1st-order correction and for different 
values of $l$.}
    \label{fig4}
\end{figure}

\begin{figure}[!h]
    \centerline{
    \includegraphics[scale = 0.32]{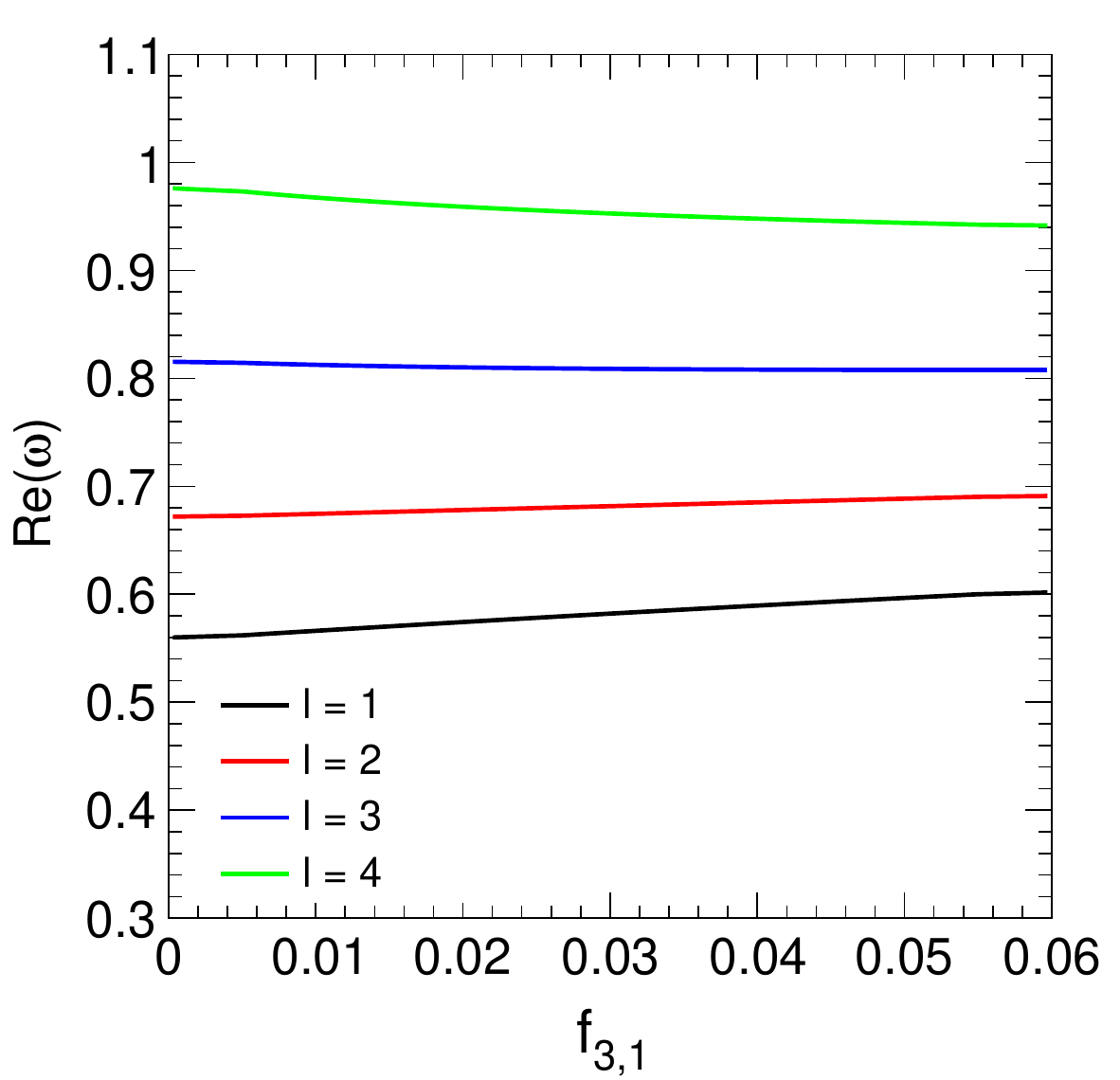}\hspace{1cm}
    \includegraphics[scale = 0.32]{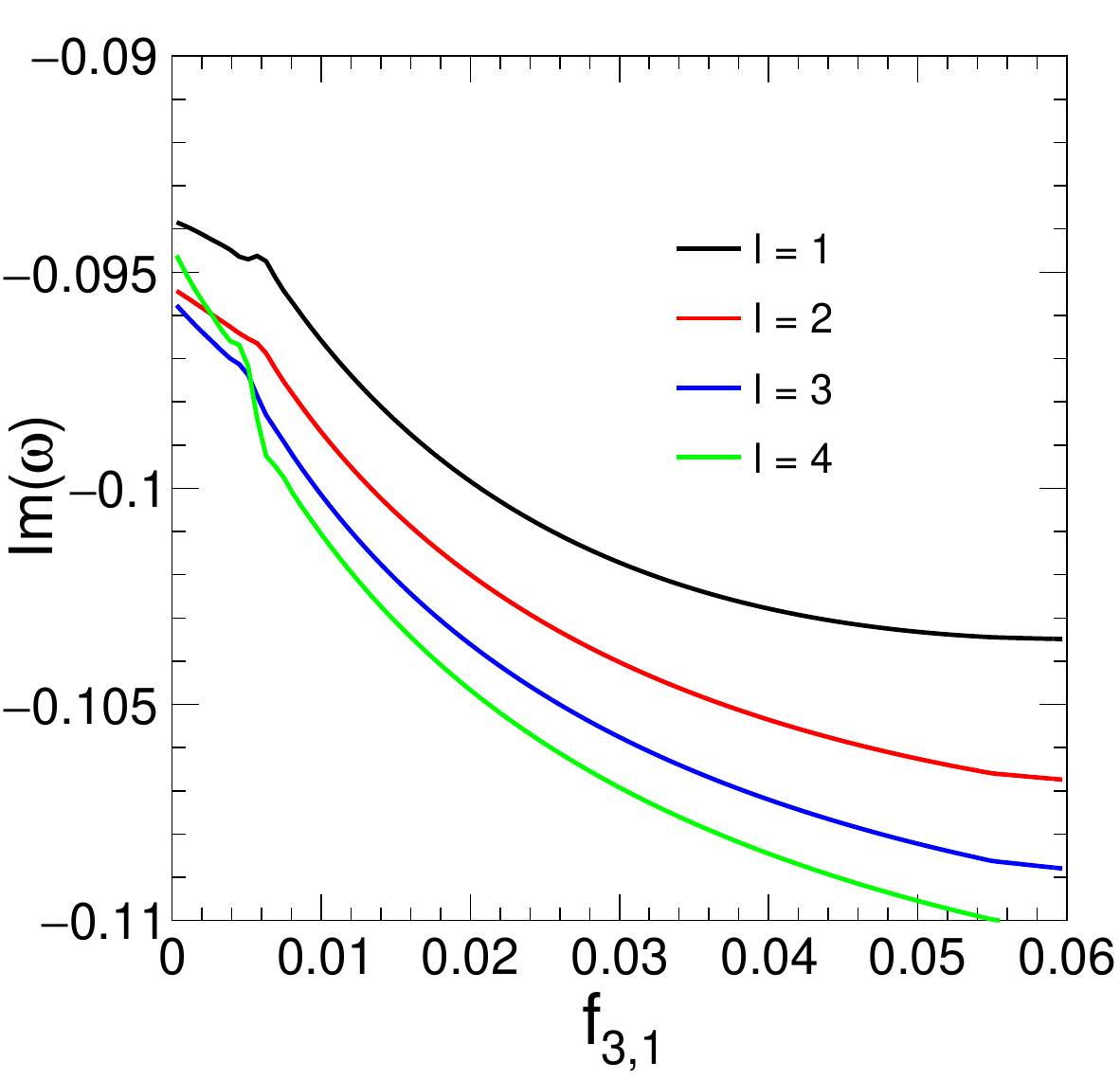}}
    \vspace{-0.2cm}
    \caption{Variation of real and imaginary parts of QNMs with respect 
to positive values of $f_{3,1}$ for the 1st-order correction and for different 
values of $l$.}
    \label{fig4A}
\end{figure}

\begin{figure}[!h]
    \centerline{
    \includegraphics[scale = 0.32]{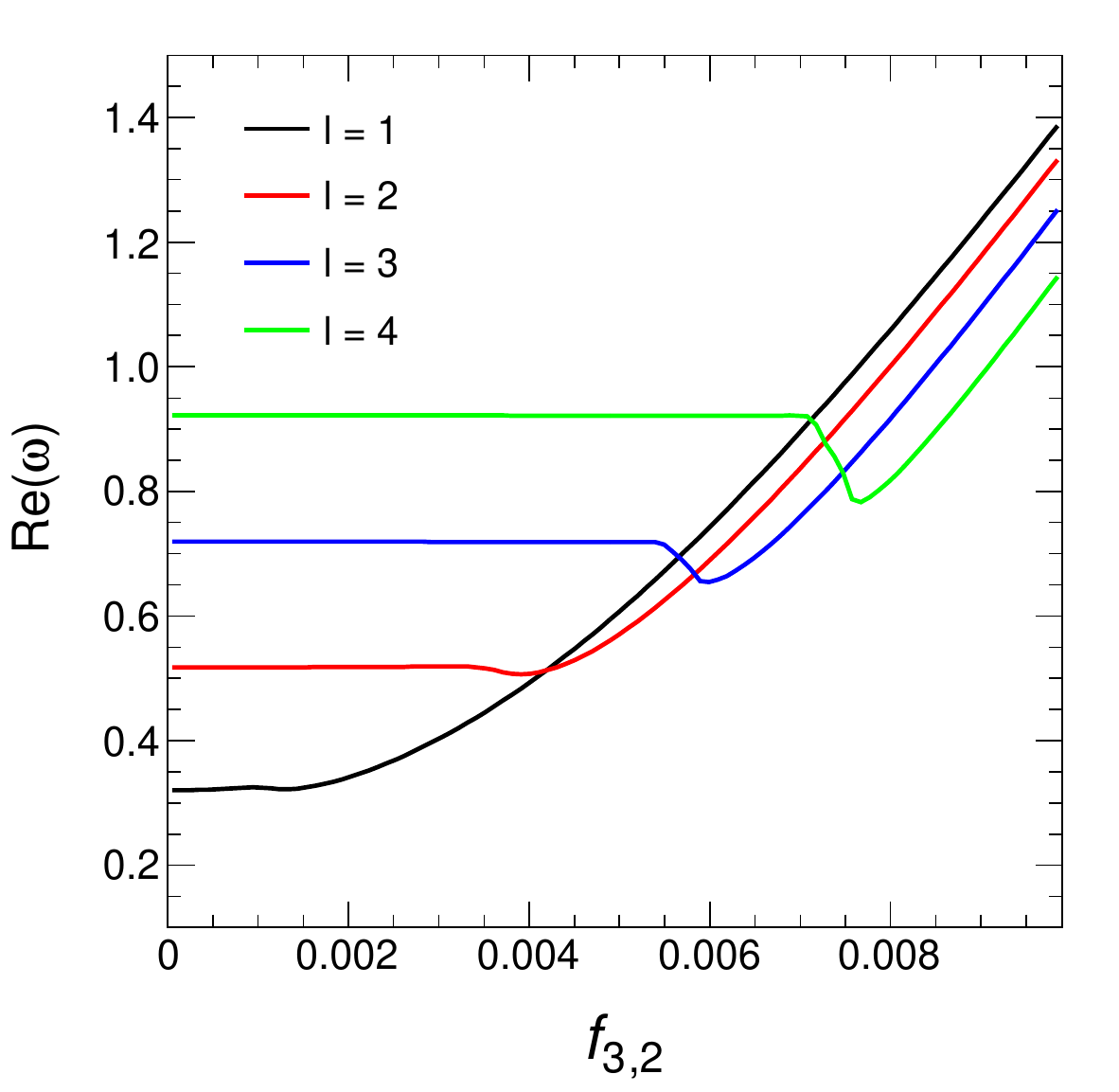}\hspace{1cm}
    \includegraphics[scale = 0.32]{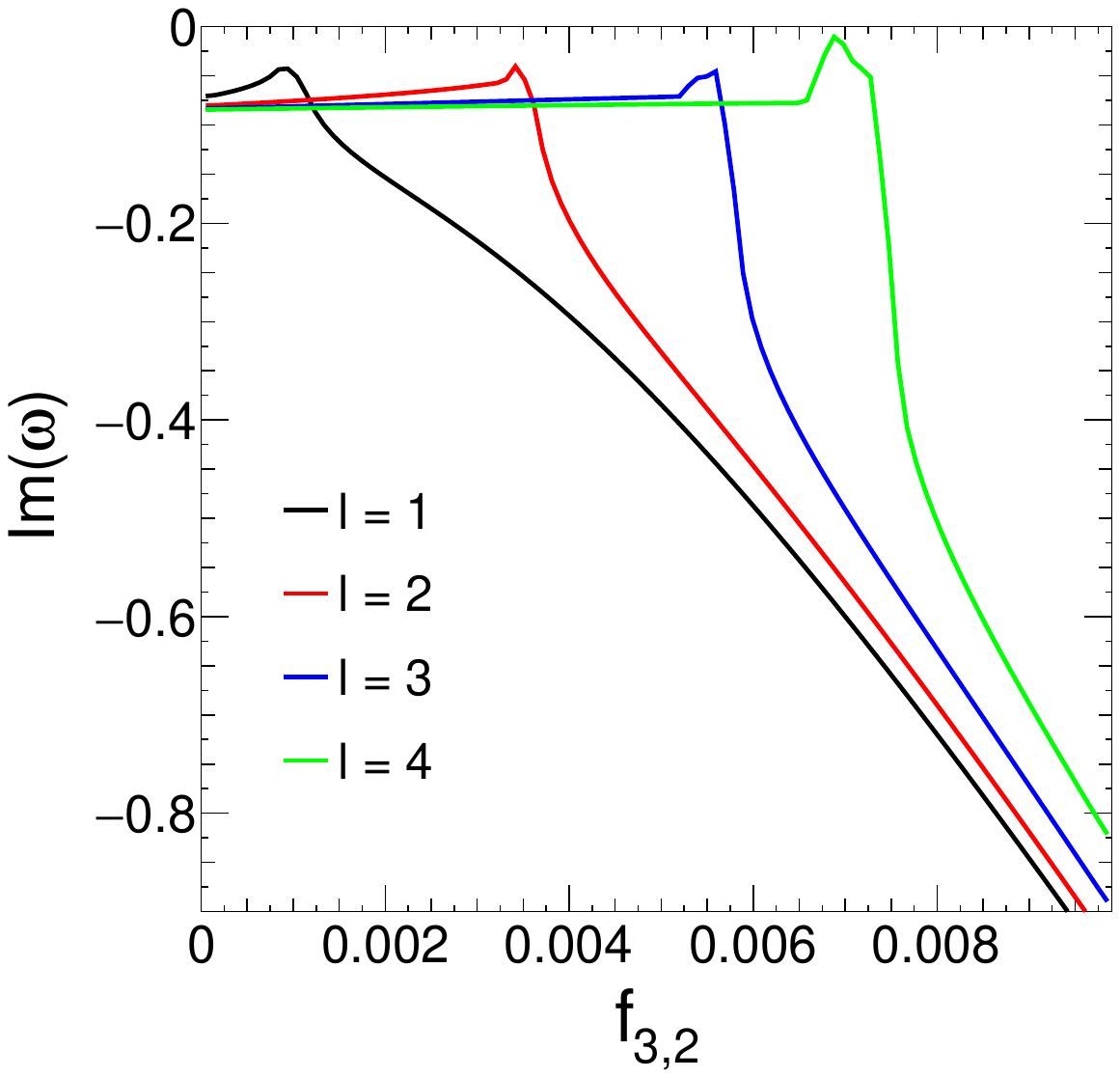}}
    \vspace{-0.2cm}
    \caption{Variation of real and imaginary parts of QNMs with respect 
to positive values of $f_{3,2}$ for the 2nd-order correction and for different 
values of $l$. Here we consider $f_{3,1} = -\,0.01$.}
    \label{fig5}
\end{figure}

\begin{figure}[!h]
    \centerline{
    \includegraphics[scale = 0.32]{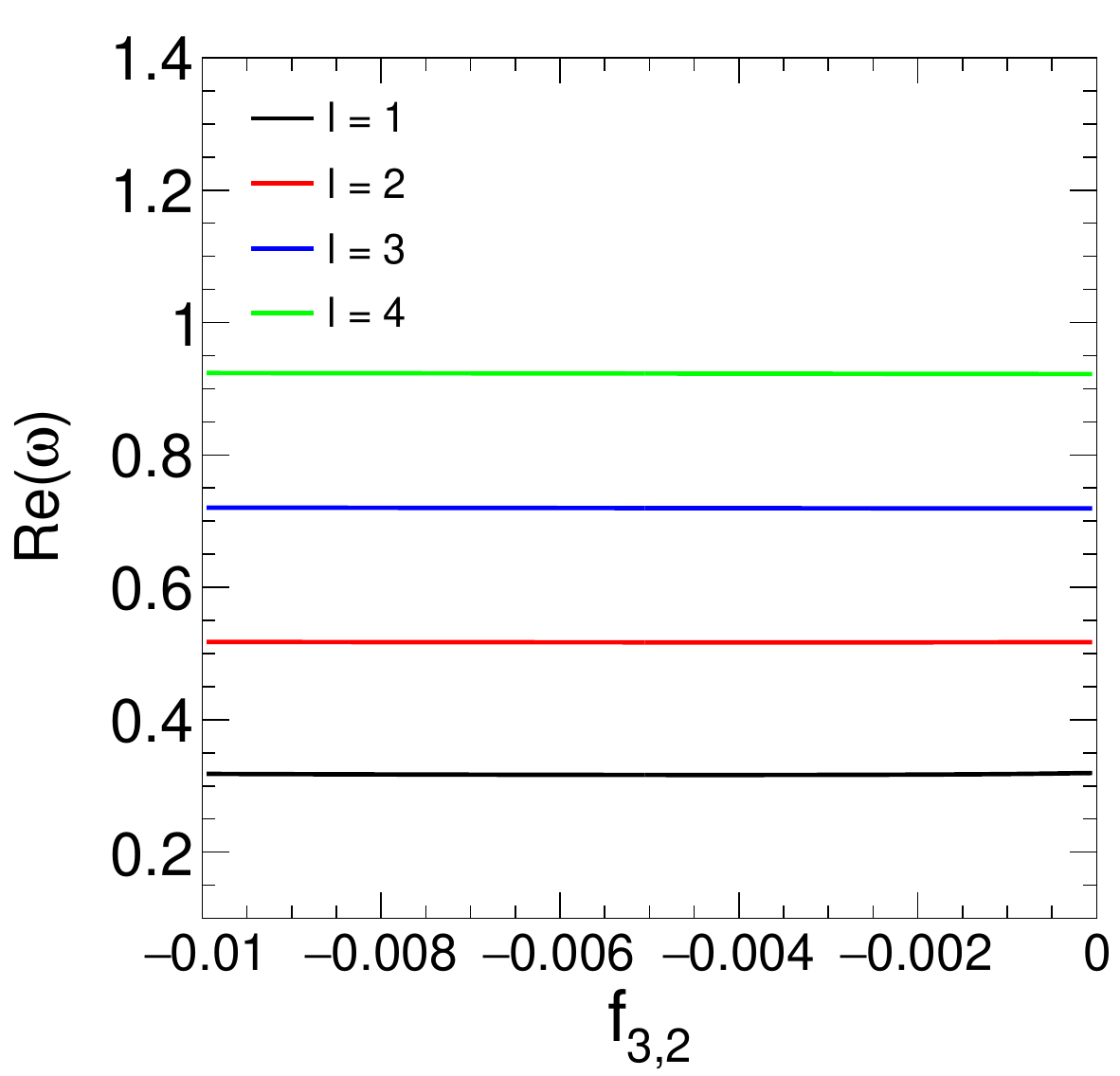}\hspace{1cm}
    \includegraphics[scale = 0.32]{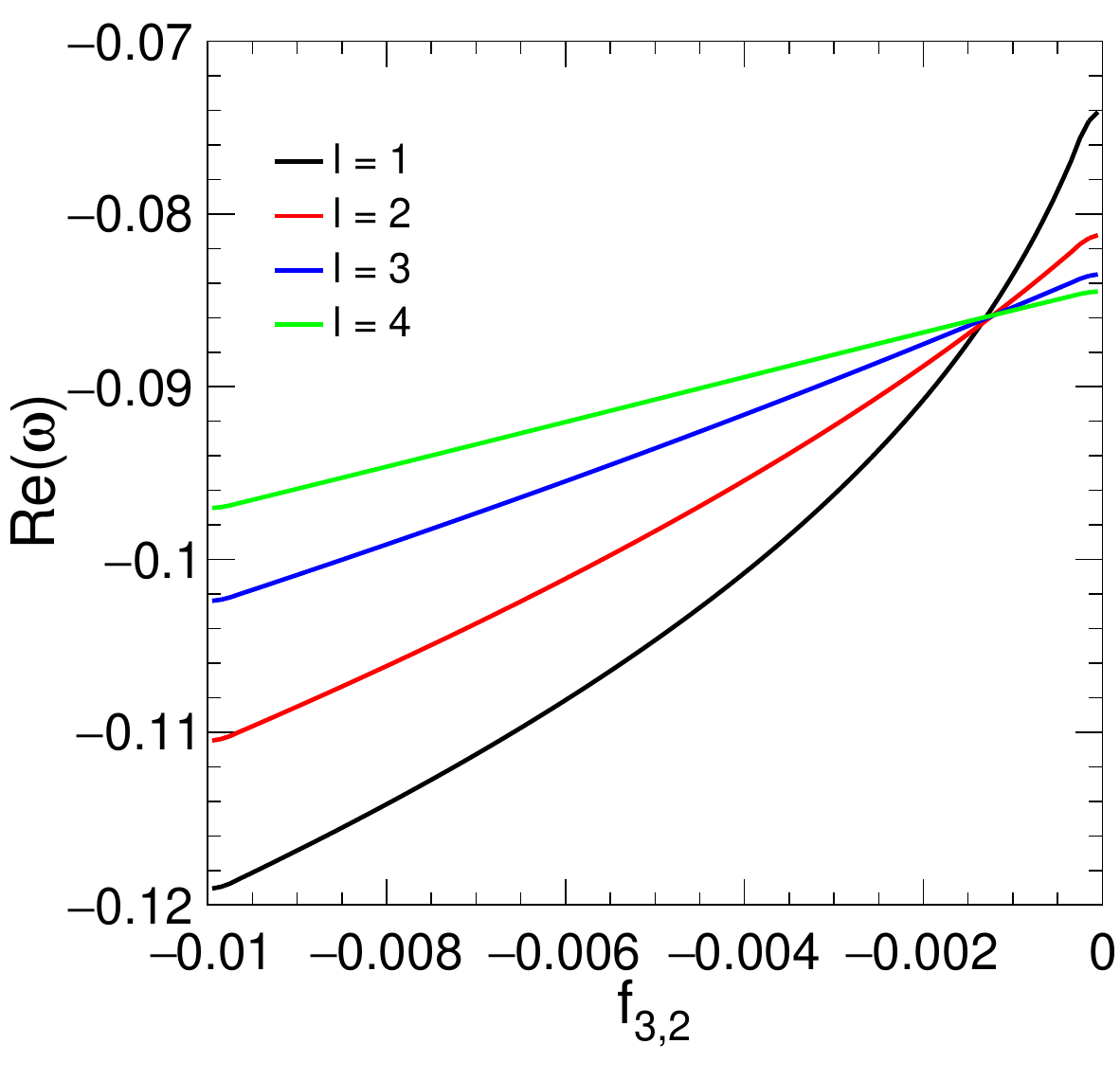}}
    \vspace{-0.2cm}
    \caption{Variation of real and imaginary parts of QNMs with respect 
to negative values of $f_{3,2}$ for the 2nd-order correction and for different 
values of $l$. Here we consider $f_{3,1} = -\,0.01$.}
    \label{fig5A}
\end{figure}

\begin{figure}[!h]
    \centerline{
    \includegraphics[scale = 0.32]{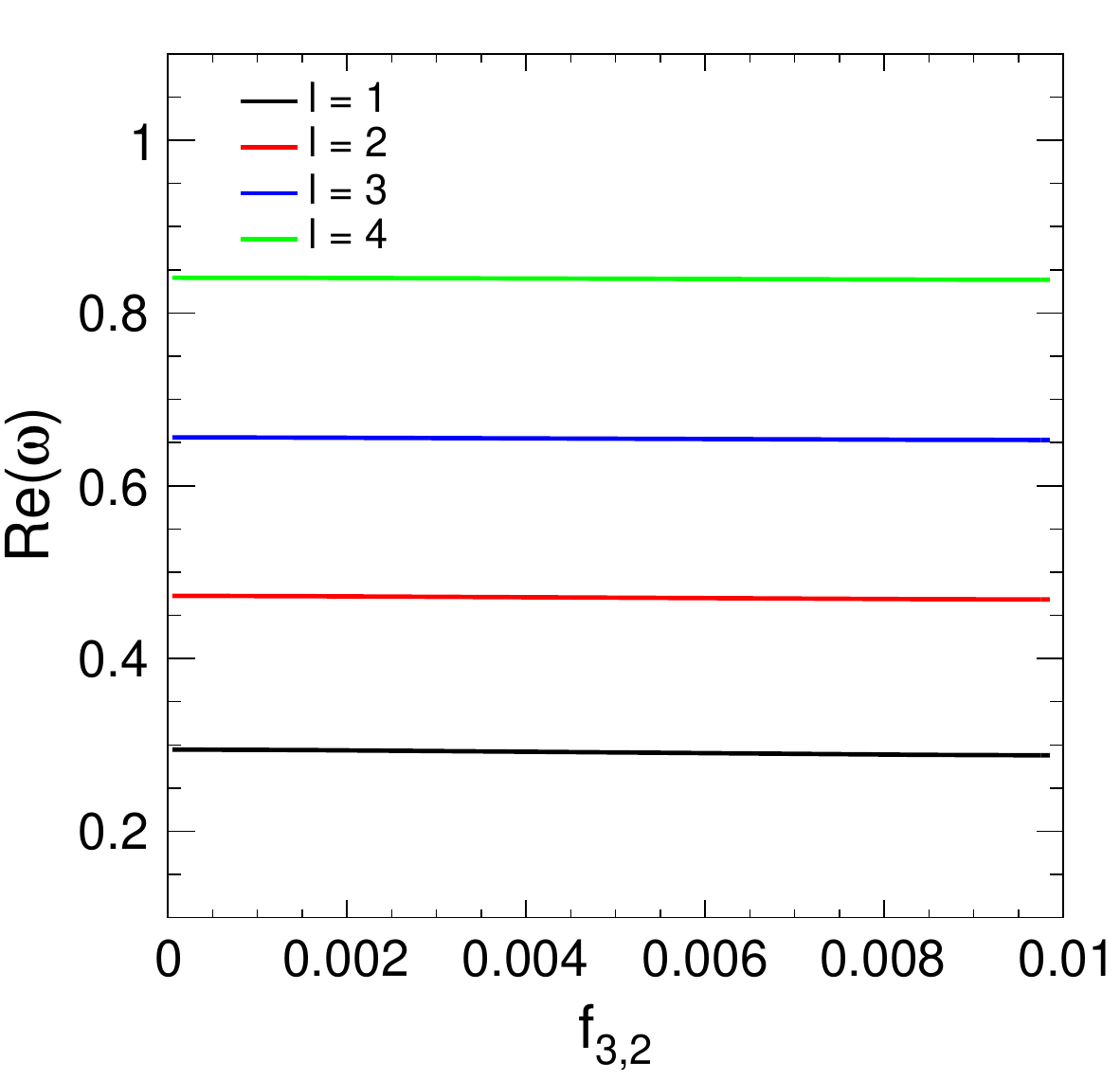}\hspace{1cm}
    \includegraphics[scale = 0.32]{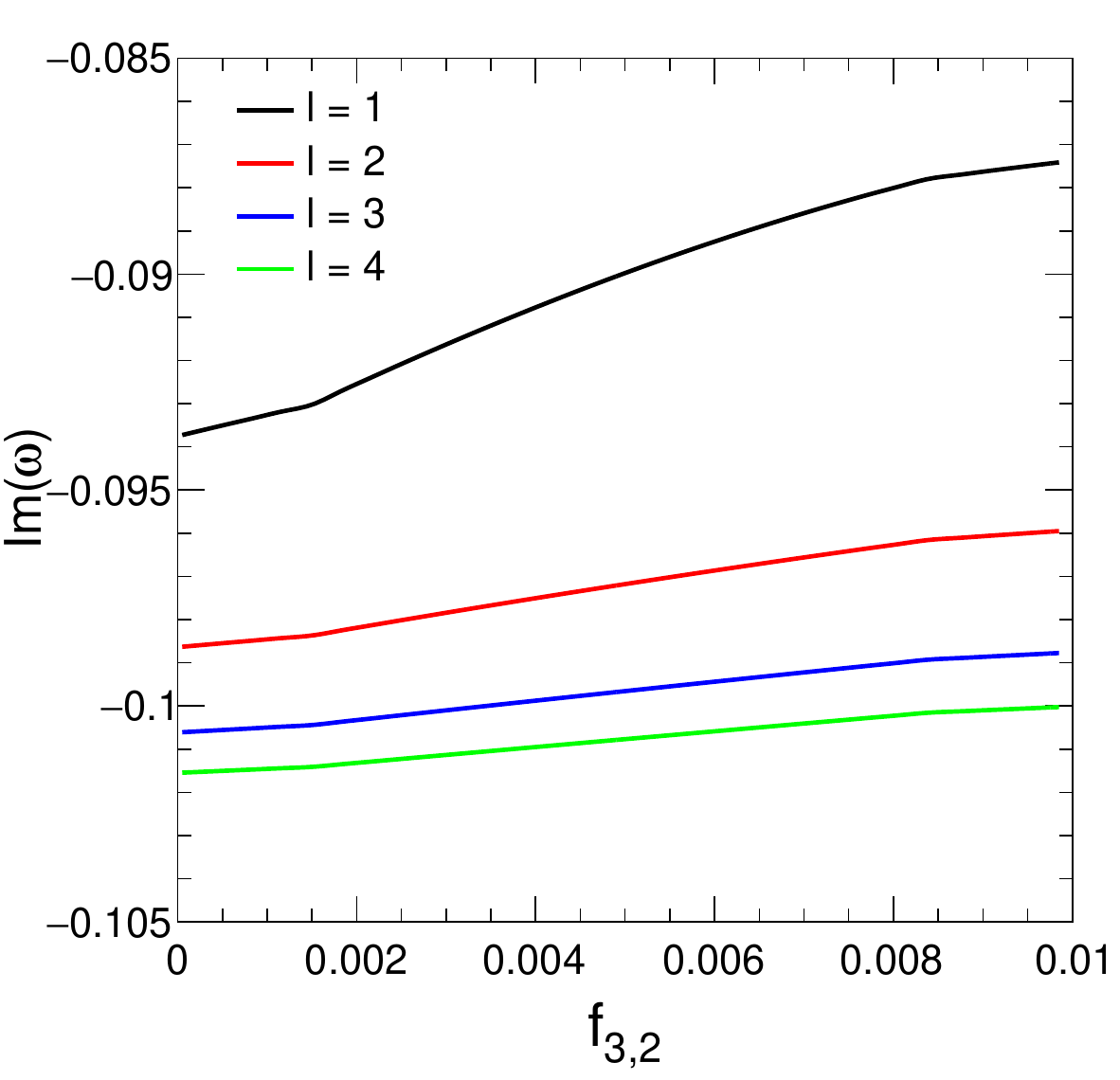}}
    \vspace{-0.2cm}
    \caption{Variation of real and imaginary parts of QNMs with respect 
to positive values of $f_{3,2}$ for the 2nd-order correction and for different 
values of $l$. Here we consider $f_{3,1} = 0.01$.}
    \label{fig5B}
\end{figure}

As the horizon structure and effective potential depend on the sign 
combinations of the coefficients $f_{3,1}$ and $f_{3,2}$, it is 
important to study the behavior of QNMs with respect to different sign choices 
of these two coefficients. We calculate the QNMs of the BH using the 6th-order 
Pad\'e-averaged WKB approximation method in terms of their amplitude and 
damping, and analyze their variation with respect to different sign 
combinations of the coefficients $f_{3,1}$ and $f_{3,2}$.  Fig.~\ref{fig4} 
shows the variation of the real and imaginary parts of the QNMs with respect 
to negative values of $f_{3,1}$ for the 1st-order correction, while 
Fig.~\ref{fig4A} presents the corresponding variation for positive values of 
$f_{3,1}$. It is observed from Fig.~\ref{fig4} that both amplitude and damping 
of the QNMs increase with both the multiple number $l$ and magnitude of the 
parameters $f_{3,1}$. Fig.~\ref{fig4A} also reveals that both the amplitude 
and damping of the QNMs increase with $l$. However, the oscillation frequency 
increases with $f_{3,1}$ up to $l = 2$, after which it begins to decrease 
slowly with further increases in $f_{3,1}$. In contrast, the damping 
consistently increases with both $l$ and $f_{3,1}$ (See Table~\ref{table3} 
for a clearer illustration). For 2nd-order correction, the variations of real 
and imaginary parts of QNMs are shown in Fig.~\ref{fig5} -- \ref{fig5B}. 
Fig.~\ref{fig5} depicts the variation of the real and imaginary parts of the 
QNMs with respect to positive values of $f_{3,2}$ for a fixed negative value 
of $f_{3,1}$. It can be observed from Fig.~\ref{fig5} that both the 
oscillation frequency and damping increase with the magnitude of $f_{3,2}$. 
Whereas the frequency increases, the damping decreases with $l$. 
Fig.~\ref{fig5A} displays the variation of the real 
and imaginary parts of the QNMs with respect to negative values of $f_{3,2}$ 
for a fixed negative value of $f_{3,1}$, while Fig.~\ref{fig5B} shows the 
variation with respect to positive values of $f_{3,2}$ for a fixed positive 
value of $f_{3,1}$. It can be observed from Fig.~\ref{fig5A} that the 
oscillation decreases with the increasing magnitude of $f_{3,2}$. However, 
the variation is too small to be clearly visible in the figure. 
Table~\ref{table4} more clearly illustrates this trend. Additionally, the 
oscillation amplitude increases with the multipole number $l$. In this case, 
the damping increases with the magnitude of $f_{3,2}$ but decreases with 
increasing $l$. Fig.~\ref{fig5B} shows that the oscillation frequency and 
damping decrease with the magnitude of $f_{3,2}$, whereas they increase with 
higher values of $l$. In this case also the variation of the real part of QNMs
with the magnitude of $f_{3,2}$ is subtle and not easily visible in the 
figure, but is clearly reflected in Table~\ref{table5}.
The error associated with the WKB method of calculation of QNMs has been 
estimated using a well-established formula, commonly found in the literature, 
which is as follows \cite{DJ1,Konoplya2,C8}:
\begin{equation}
\Delta_6 = \frac{|\text{WKB}_7 - \text{WKB}_5|}{2},
\end{equation} 
where $\text{WKB}_7$ and $\text{WKB}_5$ are the 7th and 5th-order QNMs 
obtained from the Pad\'e averaged WKB method respectively. 

For more clarity in the behaviors QNMs of the BH for the scalar field 
perturbations with different orders of metric corrections as discussed in 
relation to Figs.~\ref{fig4}--\ref{fig5B}, we list the corresponding values 
of QNMs together with the associated errors in Tables 
\ref{table1}--\ref{table5}.  
Table~\ref{table1} shows the QNMs of the BH with the corresponding 
calculated errors for the 1st-order metric correction and for negative values 
of the coefficients $f_{3,1}$ with three different values of multipole moment 
$l$. From this table, it can be observed that as seen in Fig.~\ref{fig4} both 
the oscillation and damping of the QNMs increase with the multipole moment and 
magnitude of $f_{3,1}$. It is also seen that the errors of the calculations of 
the QNMs are very small for smaller values of $l$, which increase with the 
increasing value of $l$ and are in general smaller than the uncorrected 
Schwarzschild case. Moreover, in general, the error decreases with the 
increasing magnitude of the coefficient $f_{3,1}$.  
\begin{table}[!h]
\centering
\caption{6th-order Pad\'e averaged WKB QNMs of the BH defined by the metric 
\eqref{eq5} for the 1st-order correction, with multipoles $l = 1,\,2,\,3$ and 
$n=0$. We use different negative values of the parameter $f_{3,1}$.}
\vspace{5pt}
\begin{tabular}{c@{\hskip 10pt}c@{\hskip 10pt}c@{\hskip 10pt}c@{\hskip 10pt}c@{\hskip 10pt}c}
\hline\hline
& Multipole & $f_{3,1}$ & Pad\'e averaged 6th-order WKB & $\Delta_{6}$ \\
\hline
& Schwarzschild ($l=1$) & 0 & $0.292936 - 0.097660i$ & $0.3140 \times 10^{-5}$ & \\ \hline
&\multirow{4}{4em}{$l=1$} 
  & $-0.012$ & $0.461274 - 0.071067i$ & $0.3187 \times 10^{-6}$ &\\
&  & $-0.015$ & $0.463277 - 0.050489i$ & $0.7825 \times 10^{-6}$ &\\
&  & $-0.018$ & $0.461195 - 0.118660i$ & $0.6511 \times 10^{-5}$ &\\
&  & $-0.021$ & $0.468746 - 0.174250i$ & $0.3463 \times 10^{-6}$ &\\
\hline
& Schwarzschild ($l=2$) & 0 & $0.483647 - 0.096718i$ & $0.5298 \times 10^{-4}$ &\\ \hline
&\multirow{4}{4em}{$l=2$} 
  & $-0.012$ & $0.619113 - 0.071169i$  & $0.2159 \times 10^{-3}$ &\\
&  & $-0.015$ & $0.629781 - 0.076175i$  & $0.0710 \times 10^{-5}$ &\\
&  & $-0.018$ & $0.644586 - 0.126197i$  & $0.0721 \times 10^{-6}$ &\\
&  & $-0.021$ & $0.664840 - 0.173427i$  & $0.5194 \times 10^{-6}$ &\\
\hline
& Schwarzschild ($l=3$) & 0 & $0.599440 - 0.09270i$ & $0.4719 \times 10^{-3}$ &\\ \hline
&\multirow{4}{4em}{$l=3$} 
  & $-0.012$ & $0.800980 - 0.076887i$  & $0.2513 \times 10^{-3}$ &\\
&  & $-0.015$ & $0.821367 - 0.092311i$  & $0.3175 \times 10^{-3}$ &\\
&  & $-0.018$ & $0.849275 - 0.133391i$  & $0.6540 \times 10^{-3}$ &\\
&  & $-0.021$ & $0.882445 - 0.176460i$  & $0.1763 \times 10^{-4}$ &\\
\hline\hline
\end{tabular}
\label{table1}
\end{table}
Table~\ref{table2} presents the QNMs for the 1st-order metric 
correction with positive values of the coefficient $f_{3,1}$. It more clearly 
illustrates the variation pattern of QNMs with respect to both $f_{3,1}$ and 
$l$, as shown in Fig.~\ref{fig4A}. Furthermore, the errors in the calculation 
of QNMs are small and remain nearly constant for all values of $f_{3,1}$ and 
$l$.
\begin{table}[!h]
\centering
\caption{6th-order Pad\'e averaged WKB QNMs of the BH defined by the 
metric \eqref{eq5} for the 1st-order correction, multipoles $l = 1,\,2,\,3$
and $n=0$, with positive values of the parameter $f_{3,1}$.}
\vspace{5pt}
\begin{tabular}{c@{\hskip 10pt}c@{\hskip 10pt}c@{\hskip 10pt}c}
\hline\hline
Multipole & $f_{3,1}$ & Pad\'e averaged 6th-order WKB & $\Delta_{6}$ \\
\hline
\multirow{4}{*}{$l=1$} 
   & $0.01$ & 0.566109 - 0.072774i  & $0.017 \times 10^{-6}$ \\
   & $0.03$ & 0.582811 - 0.081964i  & $0.370 \times 10^{-6}$ \\
   & $0.05$ & 0.596585 - 0.088189i & $0.011 \times 10^{-6}$ \\
\hline
\multirow{4}{*}{$l=2$} 
   & $0.01$ & 0.674438 - 0.082673i  & $0.410 \times 10^{-6}$ \\
   & $0.03$ & 0.681567 - 0.089697i  & $0.243 \times 10^{-5}$ \\
   & $0.05$ & 0.688514 - 0.093858i  & $0.240 \times 10^{-6}$ \\
\hline
\multirow{4}{*}{$l=3$} 
   & $0.01$ & 0.812411 - 0.089384i  & $0.012 \times 10^{-6}$ \\
   & $0.03$ & 0.808747 - 0.095519i  & $1.243 \times 10^{-6}$ \\
   & $0.05$ & 0.807713 - 0.098776i  & $0.221 \times 10^{-4}$ \\
\hline\hline
\end{tabular}
\label{table2}
\end{table}

Table~\ref{table3} shows the QNMs with the associated errors of calculations 
for the 2nd-order metric correction with positive values of $f_{3,2}$ while 
keeping $f_{3,1} = -\,0.01$ with the same values of $l$ as in the previous 
case. The results shown in this table are similar to that shown in 
Fig.~\ref{fig5}. From this table, it can be clearly observed that in general, 
with an increase in $l$ and $f_{3,2}$ the oscillation and damping of the QNMs 
also increase. However, for higher values of $l$ difference of the values of 
the QNMs for increasing values of $f_{3,2}$ is very small up to a certain 
higher value of $f_{3,2}$, and such a small difference could not be observed 
in Fig.~\ref{fig5} due to the size of scale used to plot the QNMs. Although, 
in general, the errors of calculations of QNMs are samll, the error increases 
randomly without following a particular pattern due to the peculiar nature of 
the QNMs in this case as mentioned.   
\begin{table}[h!]
\centering
\caption{6th-order Pad\'e averaged WKB QNMs of the BH defined by the 
metric \eqref{eq5} for the 2nd-order correction, with multipoles 
$l = 1,\,2,\,3$ and $n=0$. Here, we use different positive values of the 
parameter $f_{3,2}$ keeping $f_{3,1} = -\,0.01$.}
\vspace{5pt}
\begin{tabular}{c@{\hskip 10pt}c@{\hskip 10pt}c@{\hskip 10pt}c}
\hline\hline
Multipole & $f_{3,2}$ & Pad\'e averaged 6th-order WKB & $\Delta_{6}$ \\
\hline
\multirow{4}{*}{$l=1$} 
   & $0.001$ & 0.326565 - 0.030865i  & $0.1732 \times 10^{-3}$ \\
   & $0.003$ & 0.402747 - 0.217869i  & $0.6154 \times 10^{-6}$\\
   & $0.005$ & 0.608456 - 0.385280i  & $0.5710 \times 10^{-6}$ \\
   & $0.007$ & 0.896927 - 0.598825i  & $0.0571 \times 10^{-4}$ \\
   & $0.008$ & 1.060840 - 0.721304i  & $0.2162 \times 10^{-6}$\\
\hline
\multirow{4}{*}{$l=2$} 
   & $0.001$ & 0.517537 - 0.075501i  & $0.0071 \times 10^{-6}$ \\
   & $0.003$ & 0.518963 - 0.060009i  & $1.7260 \times 10^{-4}$ \\
   & $0.005$ & 0.570409 - 0.331340i  & $1.7260 \times 10^{-4}$ \\
   & $0.007$ & 0.835102 - 0.564314i  & $3.5760 \times 10^{-2}$ \\
   & $0.008$ & 1.002850 - 0.691294i  & $0.6430 \times 10^{-6}$\\
\hline
\multirow{4}{*}{$l=3$} 
   & $0.001$ & 0.719024 - 0.080950i  & $1.7220 \times 10^{-6}$ \\
   & $0.003$ & 0.718888 - 0.076272i  & $0.4220 \times 10^{-3}$ \\
   & $0.005$ & 0.718777 - 0.071456i  & $0.3210 \times 10^{-6}$ \\
   & $0.007$ & 0.757422 - 0.489539i  & $4.1200 \times 10^{-3}$ \\
   & $0.008$ & 0.918746 - 0.634811i  & $0.3724 \times 10^{-6}$\\
\hline\hline
\end{tabular}
\label{table3}
\end{table}
Tables~\ref{table4} and \ref{table5} also present the QNMs of the BH 
for the 2nd-order metric correction with different sign combinations of 
$f_{3,1}$ and $f_{3,2}$. In Table \ref{table4} we keep $f_{3,1} = -\,0.01$ and 
consider the negative values of $f_{3,2}$ and in Table \ref{table5} we keep 
$f_{3,1} = 0.01$ and consider the positive values of $f_{3,2}$. Both tables 
clearly show the variation pattern of QNMs, as illustrated by Fig.~\ref{fig5A} 
and Fig.~\ref{fig5B} respectively.

\begin{table}[h!]
\centering
\caption{QNMs computed using 6th-order Pad\'e-averaged WKB approximation 
for the BH described by the metric~\eqref{eq5} with 2nd-order corrections are 
shown for $l = 1,\,2,\,3$ and $n = 0$. Various negative values of the 
parameter $f_{3,2}$ are considered, while fixing $f_{3,1} = -0.01$.
}
\vspace{5pt}
\begin{tabular}{c@{\hskip 10pt}c@{\hskip 10pt}c@{\hskip 10pt}c}
\hline\hline
Multipole & $f_{3,2}$ & Pad\'e averaged 6th-order WKB & $\Delta_{6}$ \\
\hline
\multirow{4}{*}{$l=1$} 
   & $-0.001$ & 0.317939 - 0.085563i & $1.732 \times 10^{-4}$ \\
   & $-0.002$ & 0.317086 - 0.090776i & $0.072 \times 10^{-4}$ \\
   & $-0.003$ & 0.316683 - 0.096261i & $0.081 \times 10^{-2}$ \\
\hline
\multirow{4}{*}{$l=2$} 
   & $-0.001$ & 0.516991 - 0.084976i & $0.001 \times 10^{-4}$ \\
   & $-0.002$ & 0.516872 - 0.088808i & $1.310 \times 10^{-3}$ \\
   & $-0.003$ & 0.516820 - 0.092262i & $0.112 \times 10^{-3}$ \\
\hline
\multirow{4}{*}{$l=3$} 
   & $-0.001$ & 0.719196 - 0.083400i & $0.001 \times 10^{-4}$ \\
   & $-0.002$ & 0.719298 - 0.087531i & $1.310 \times 10^{-3}$ \\
   & $-0.003$ & 0.719409 - 0.089601i & $0.112 \times 10^{-3}$ \\
\hline\hline
\end{tabular}
\label{table4}
\end{table}
\begin{table}[!h]
\centering
\caption{6th-order Pad\'e averaged WKB QNMs of the BH defined by the 
metric \eqref{eq5} for 2nd-order correction with multipoles $l = 1,\,2,\,3$ 
and $n=0$. We use different positive values of the parameter $f_{3,2}$ 
keeping $f_{3,1} = 0.01$.}
\vspace{5pt}
\begin{tabular}{c@{\hskip 10pt}c@{\hskip 10pt}c@{\hskip 10pt}c}
\hline\hline
Multipole & $f_{3,2}$ & Pad\'e averaged 6th-order WKB & $\Delta_{6}$ \\
\hline
\multirow{4}{*}{$l=1$} 
   & $0.001$ & 0.295069 - 0.092857i  & $0.173 \times 10^{-6}$ \\
   & $0.002$ & 0.294984 - 0.090672i  & $0.015 \times 10^{-6}$ \\
   & $0.003$ & 0.294957 - 0.088497i  & $0.065 \times 10^{-6}$ \\
\hline
\multirow{4}{*}{$l=2$} 
   & $0.001$ & 0.472728 - 0.098018i  & $0.112 \times 10^{-5}$ \\
   & $0.002$ & 0.472474 - 0.097105i  & $0.865 \times 10^{-5}$ \\
   & $0.003$ & 0.472223 - 0.096079i  & $4.214 \times 10^{-6}$ \\
\hline
\multirow{4}{*}{$l=3$} 
   & $0.001$ & 0.656097 - 0.100244i  & $0.112 \times 10^{-5}$ \\
   & $0.002$ & 0.655861 - 0.099691i  & $0.865 \times 10^{-5}$ \\
   & $0.003$ & 0.655601 - 0.099147i  & $4.214 \times 10^{-6}$ \\
\hline\hline
\end{tabular}
\label{table5}
\end{table}

\section{Evolution of a Scalar Perturbation Around the Black Hole} \label{4}
In this section, we study the evolution of a scalar perturbation around the BH 
spacetime for the 1st-order and 2nd-order corrections to the 
metric. To study the evolution of the scalar field perturbation we use the 
time domain integration method described in Refs.~\citep{DJ2,Gundlach}. Thus, 
defining $\Psi(r_{\star},t) = \Psi(i{\Delta}r_{\star},j{\Delta}t) = 
\Psi_{i,j}$ and $V(r(r_\star)) = V(r_{\star},t) = V_{i,j}$, Eq.~\eqref{eq14} 
can be expressed in the following form:
\begin{equation}
\frac{\Psi_{i+1,j}-2\Psi_{i,j}+\Psi_{i-1,j}}{{\Delta}r_{\star}^2}-\frac{\Psi_{i,j+1}-2\Psi_{i,j}+\Psi_{i,j-1}}{{\Delta}t^2}-V_{i}\Psi_{i,j}=0. 
\label{eq22}
\end{equation} 
Now using the initial conditions: $\Psi(r_{\star},t) = \exp\left[-(r_{\star}-k)^2/2\lambda^2\right]$ and $\left|\Psi(r_{\star},t)\right|_{t<0} = 0$, where $k$ 
is the median and $\lambda$ is the width of the initial wave packet, the final 
expression for the time evolution of the scalar field can be written as 
\begin{equation}
\Psi_{i,j+1} = -\,\Psi_{i,j-1}+\left(\frac{{\Delta}t}{{\Delta}r_{\star}}\right)^{\!2}\left(\Psi_{i+1,j}+\Psi_{i-1,j}\right)+\left(2-2\left(\frac{{\Delta}t}{{\Delta}r_{\star}}\right)^{\!2}-V_{i\,}\Delta{t}^2\right)\Psi_{i,j}. 
\label{eq23}
\end{equation} 
Moreover, we apply the Von Neumann stability condition: 
$\frac{{\Delta}t}{{\Delta}r_{\star}}<1$ during the numerical procedure to 
ensure stable results and compute the time profiles.
\begin{figure}[!h]
    \centerline{
    \includegraphics[scale = 0.29]{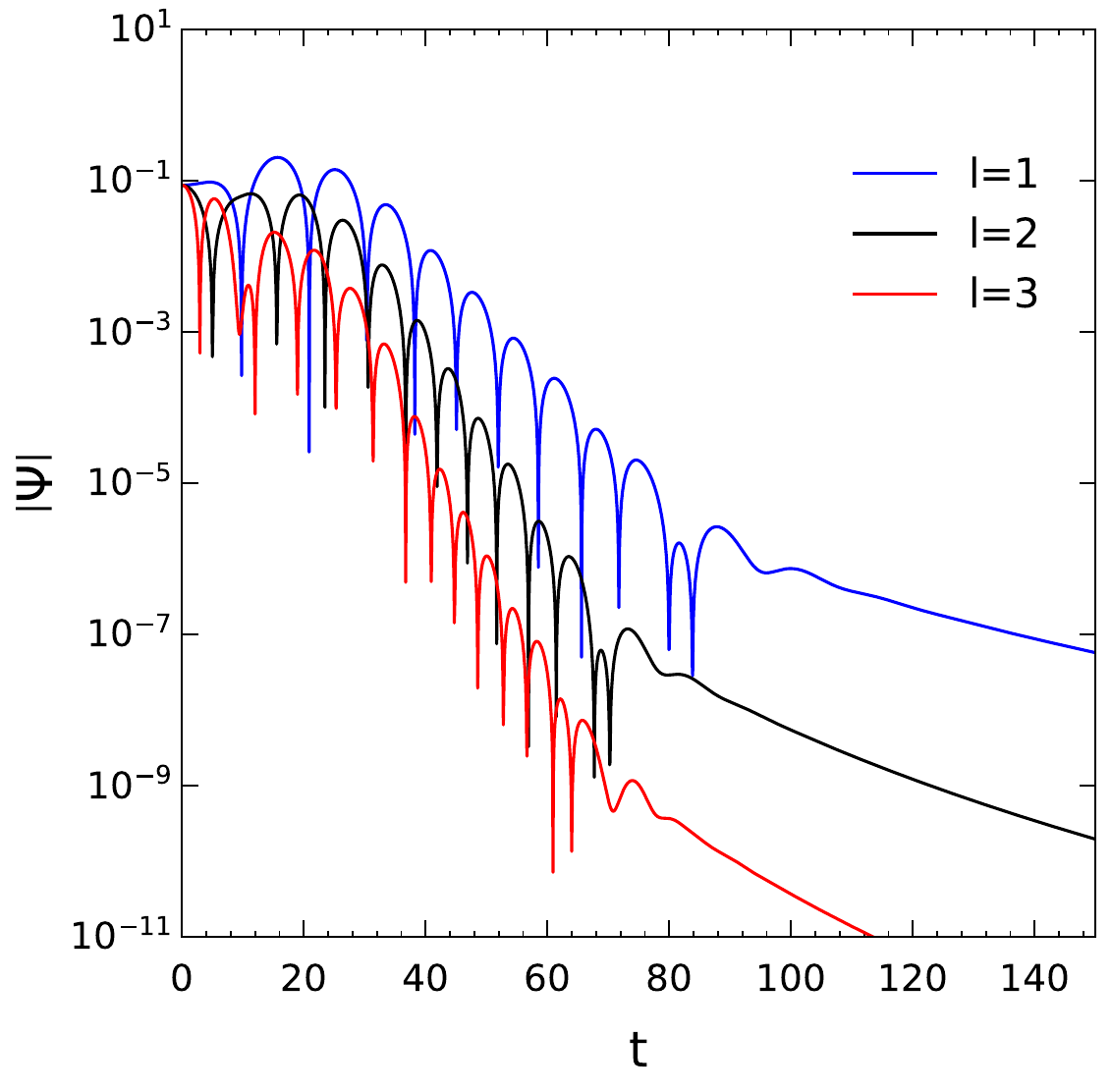}\hspace{0.3cm}
    \includegraphics[scale = 0.29]{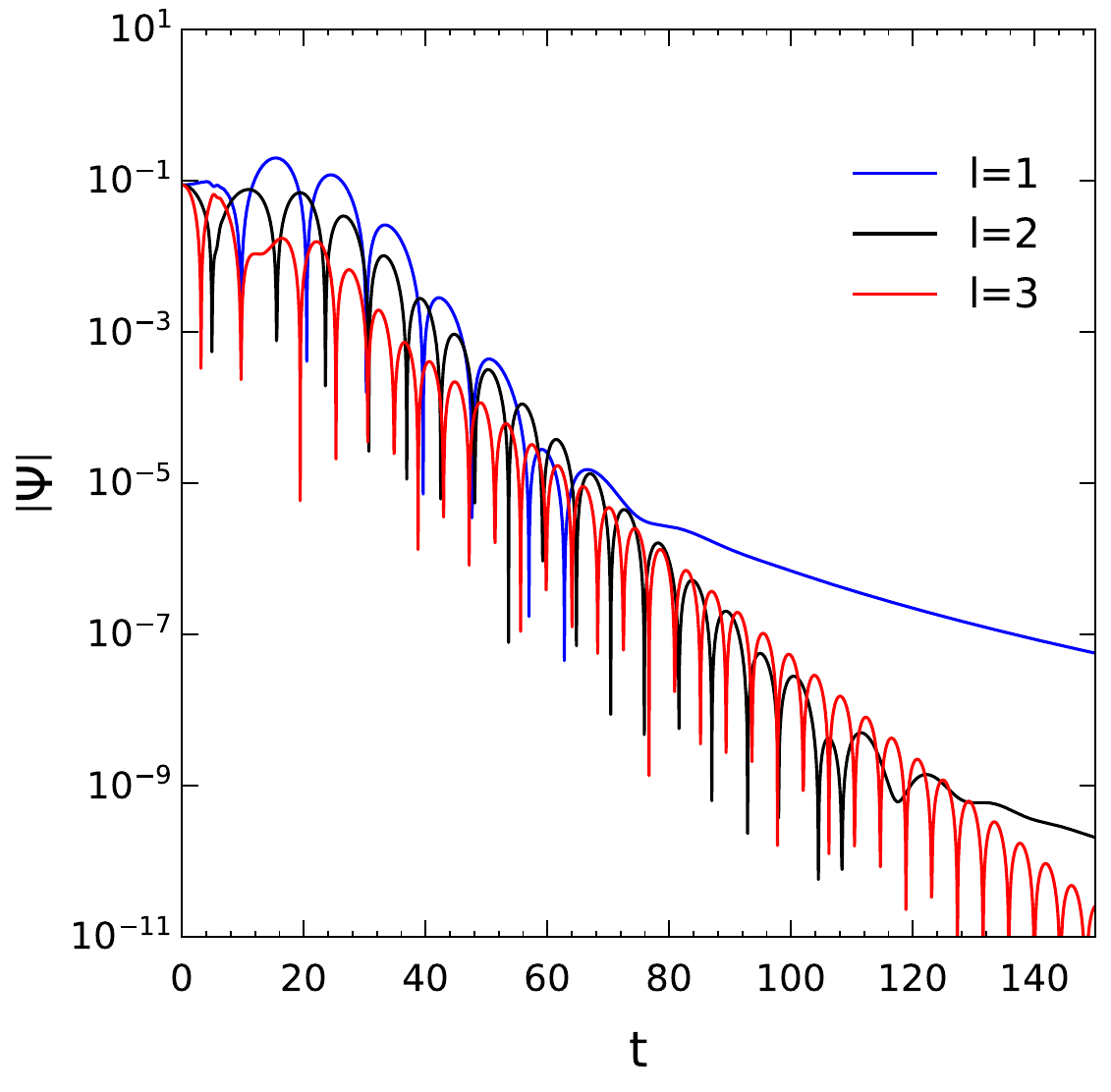}\hspace{0.3cm}
    \includegraphics[scale = 0.29]{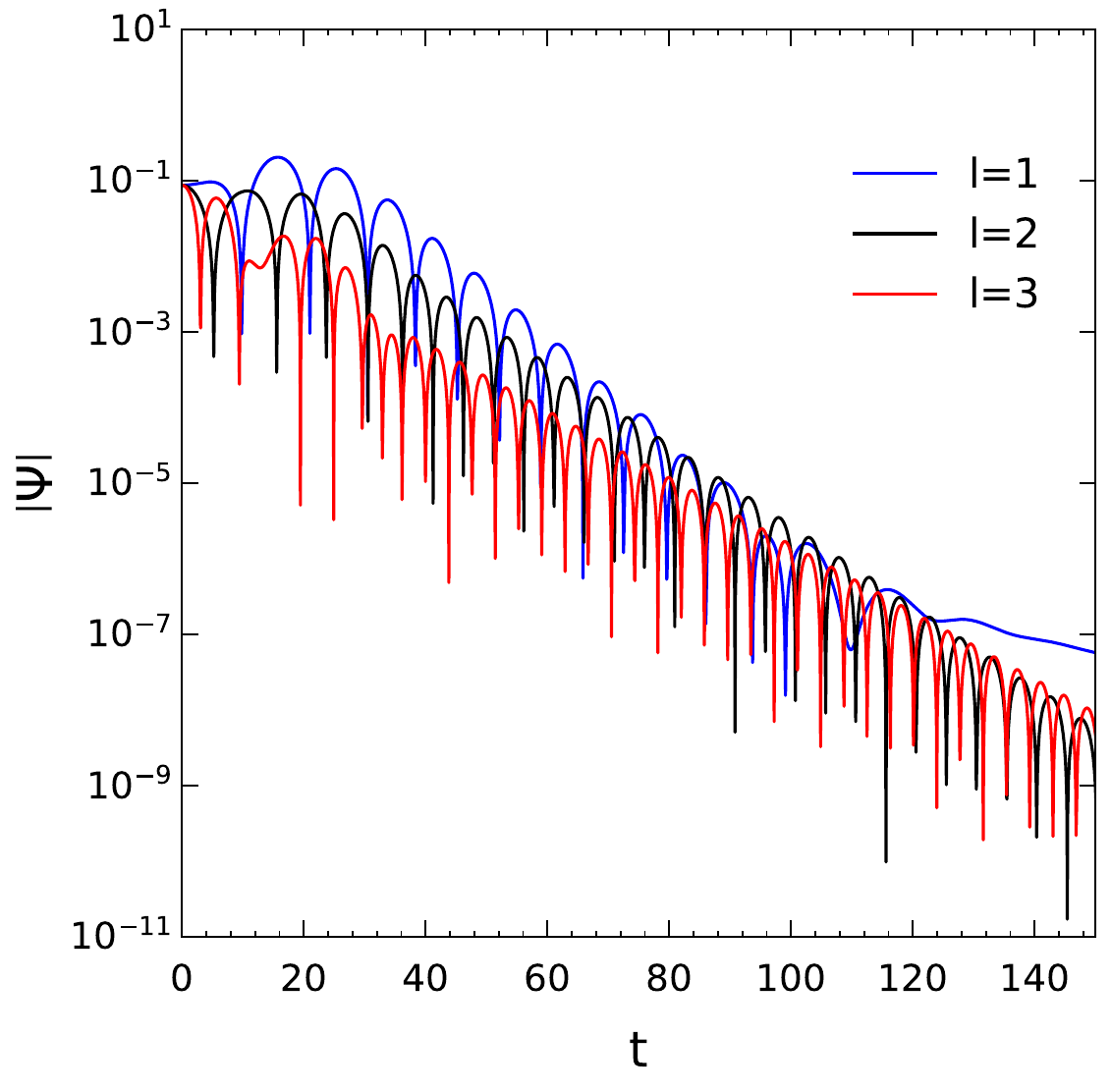}}\vspace{0.5cm}
      \centerline{
     \includegraphics[scale=0.29]{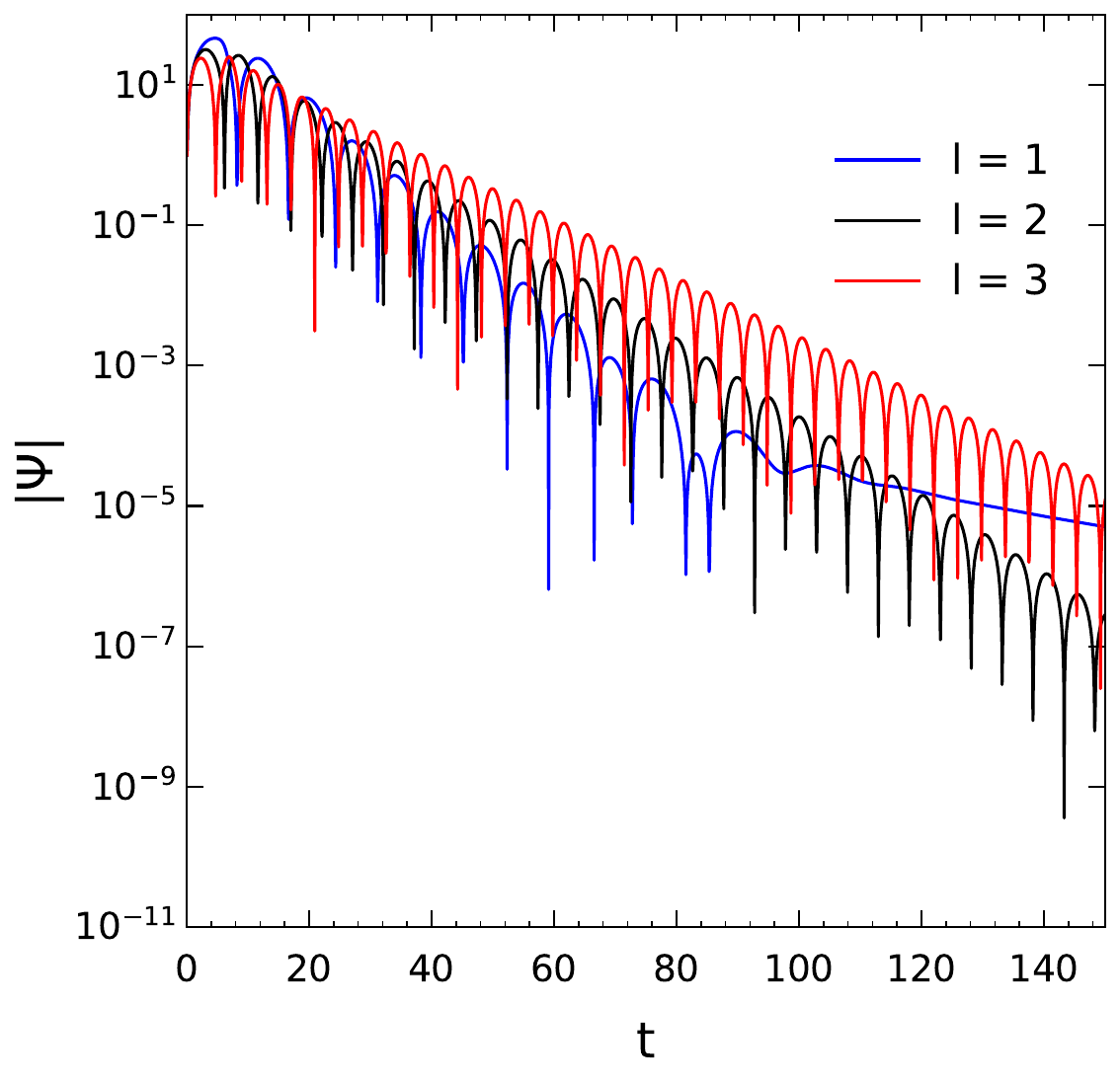}\hspace{0.3cm}
     \includegraphics[scale=0.29]{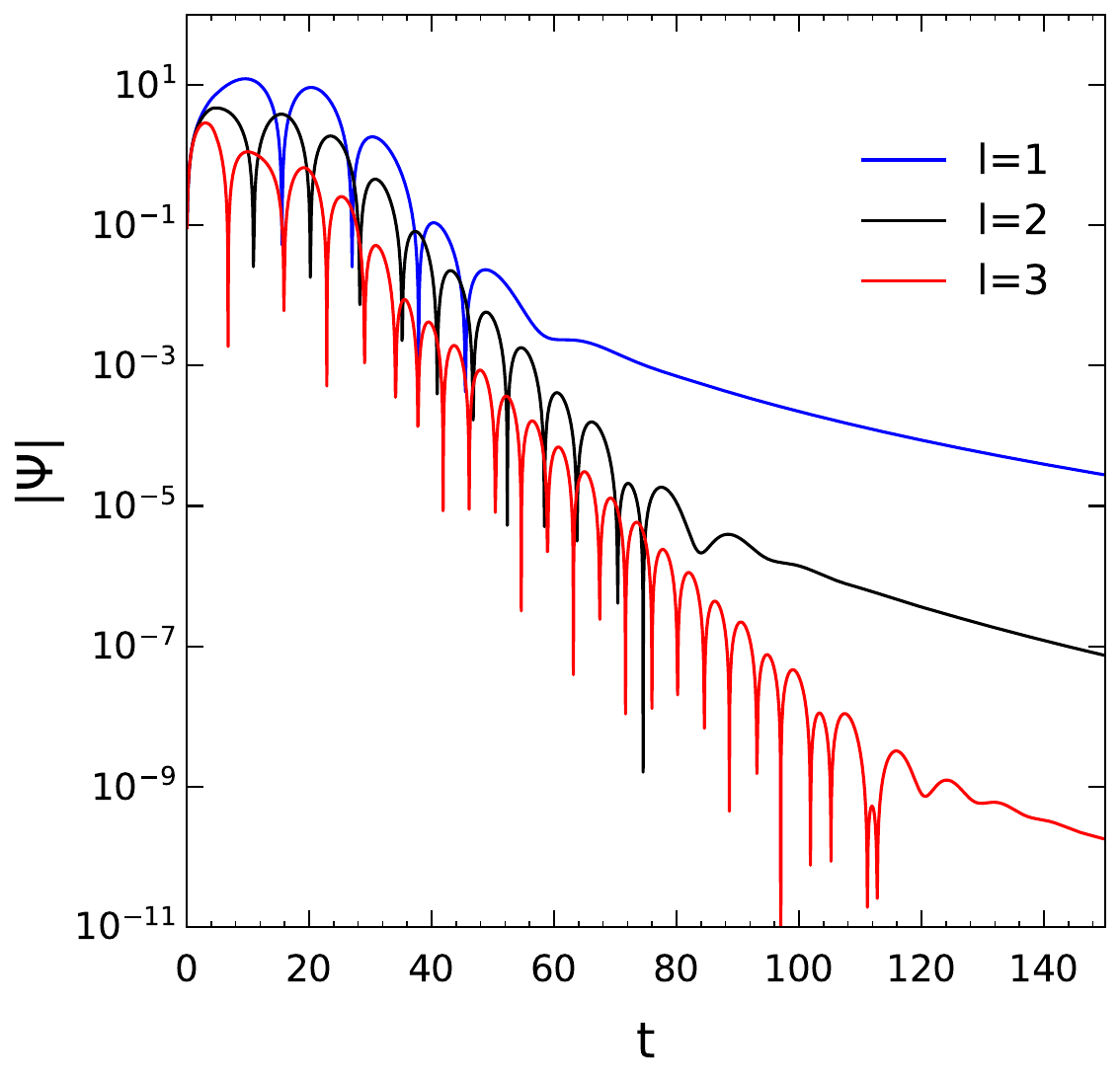}}
    \vspace{-0.2cm}
    \caption{Time domain profiles for the scalar field perturbation 
around the BH spacetime defined by the metric \eqref{eq5}, for different 
combinations of $f_{3,1}$ and $f_{3,2}$ (for details see the related 
explanation in the text).}
    \label{fig6}
\end{figure}
\begin{figure}[!h]
    \centerline{
    \includegraphics[scale = 0.29]{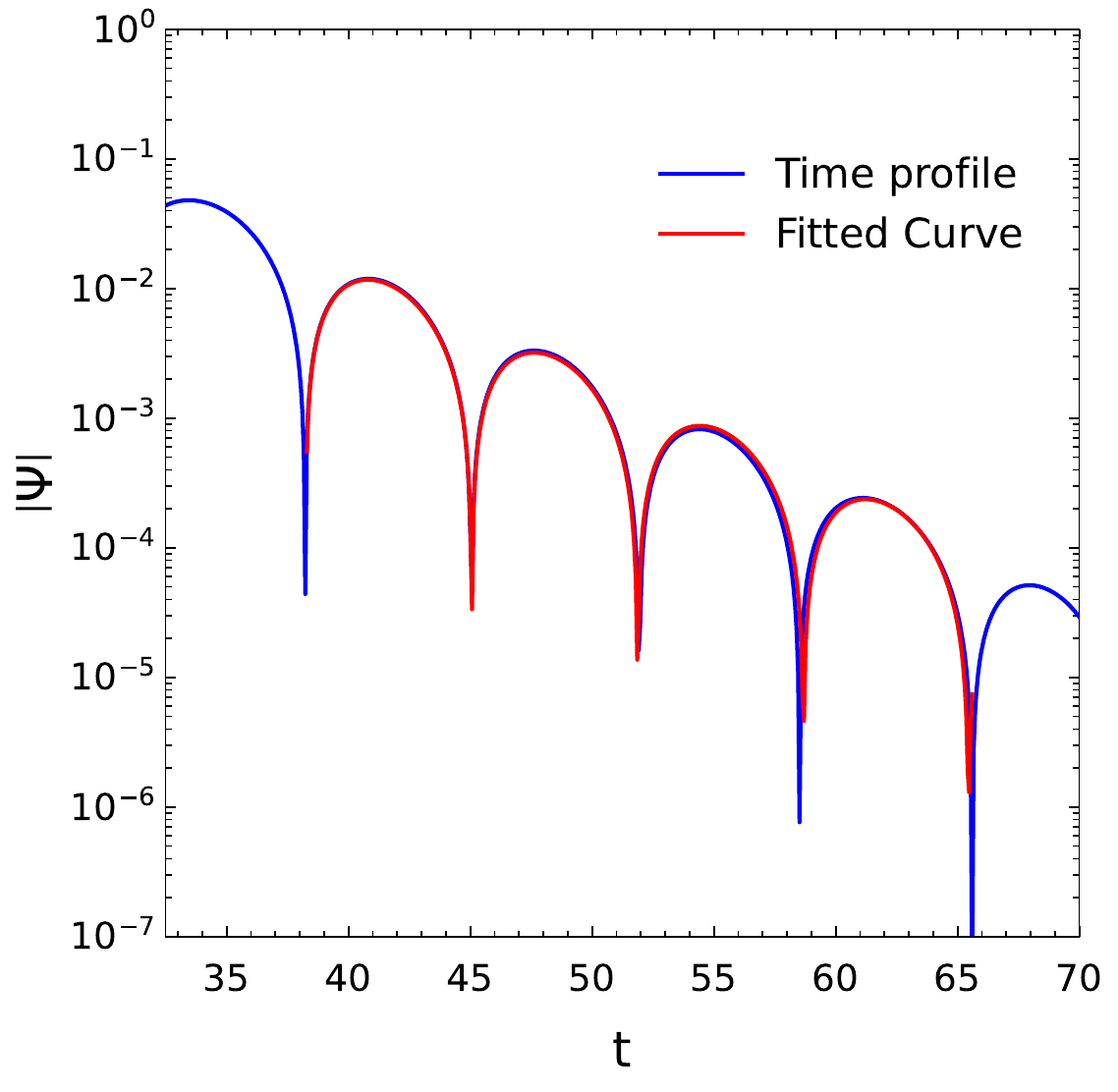}\hspace{0.3cm}
    \includegraphics[scale = 0.29]{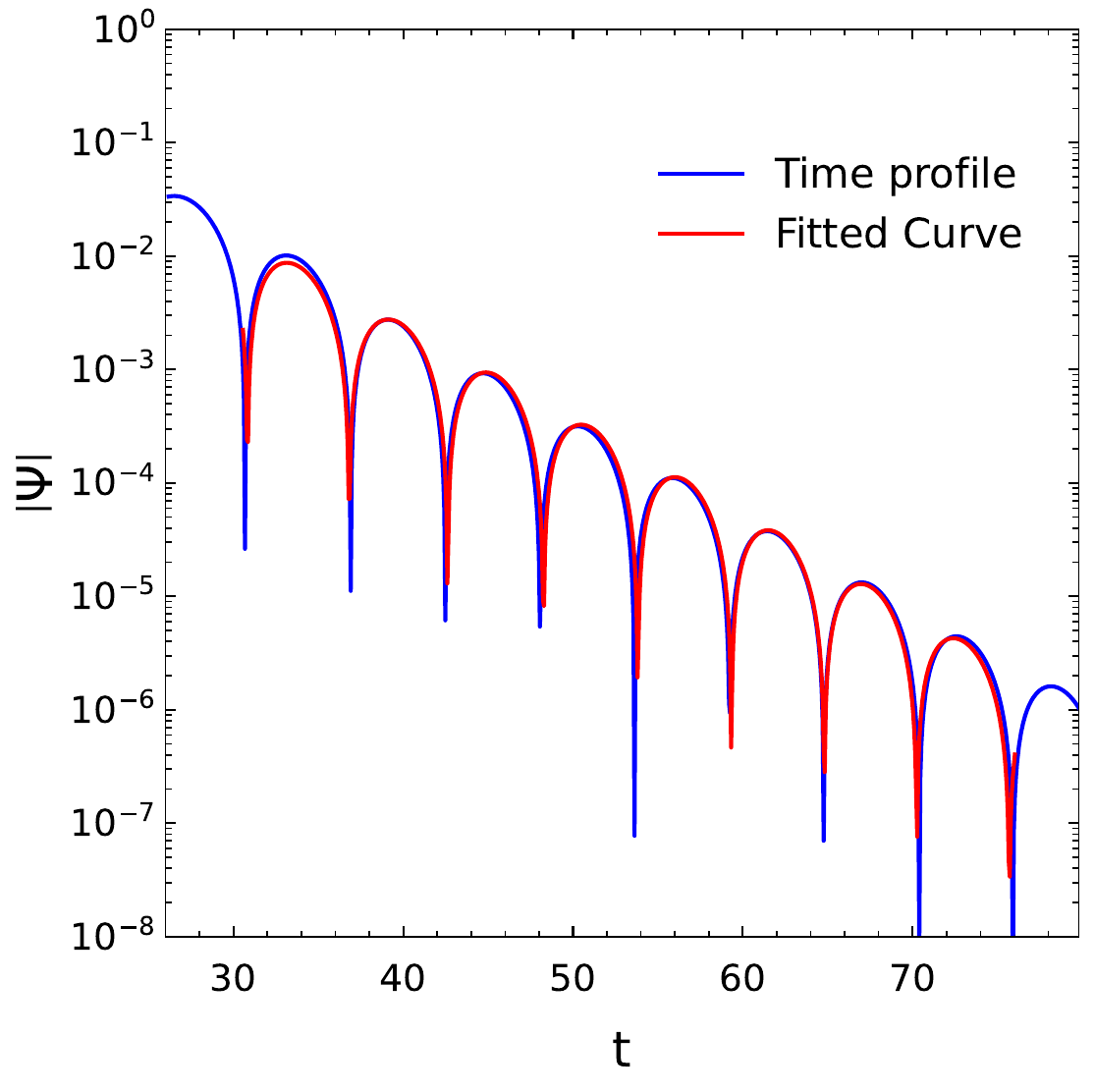}\hspace{0.3cm}
    \includegraphics[scale=0.288]{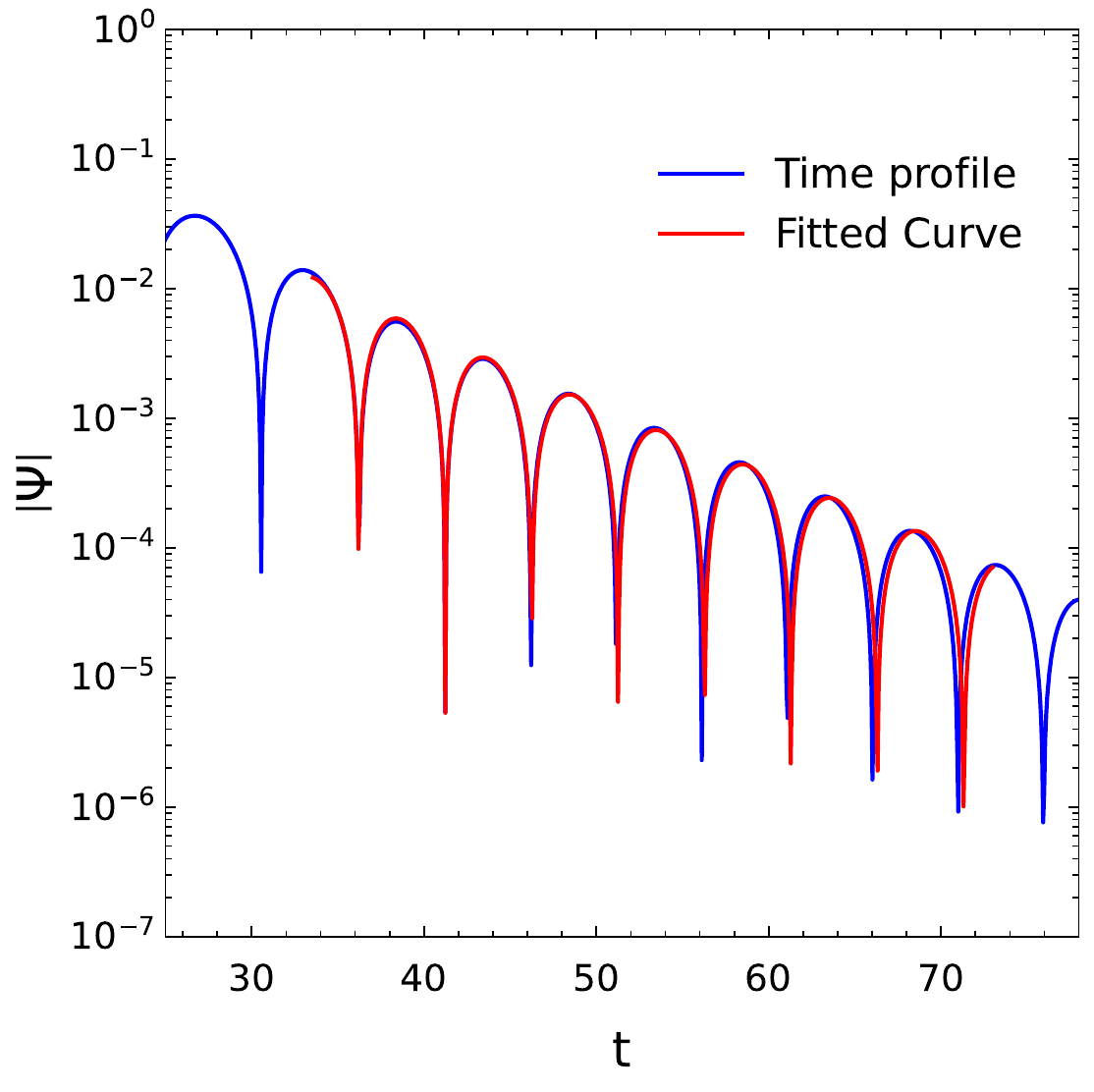}}\vspace{0.5cm}
    \centerline{
     \includegraphics[scale=0.29]{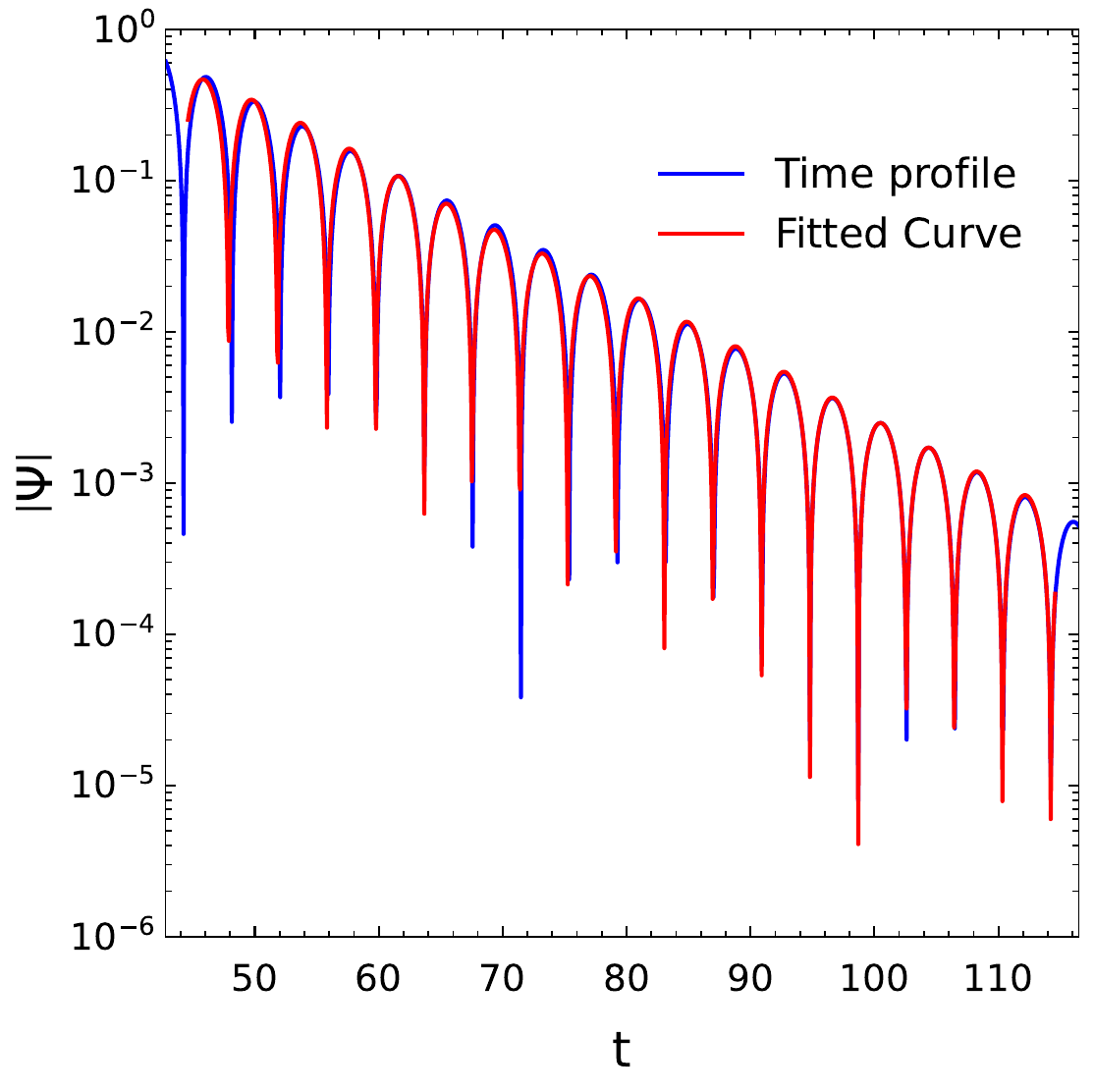}\hspace{0.3cm}
     \includegraphics[scale=0.29]{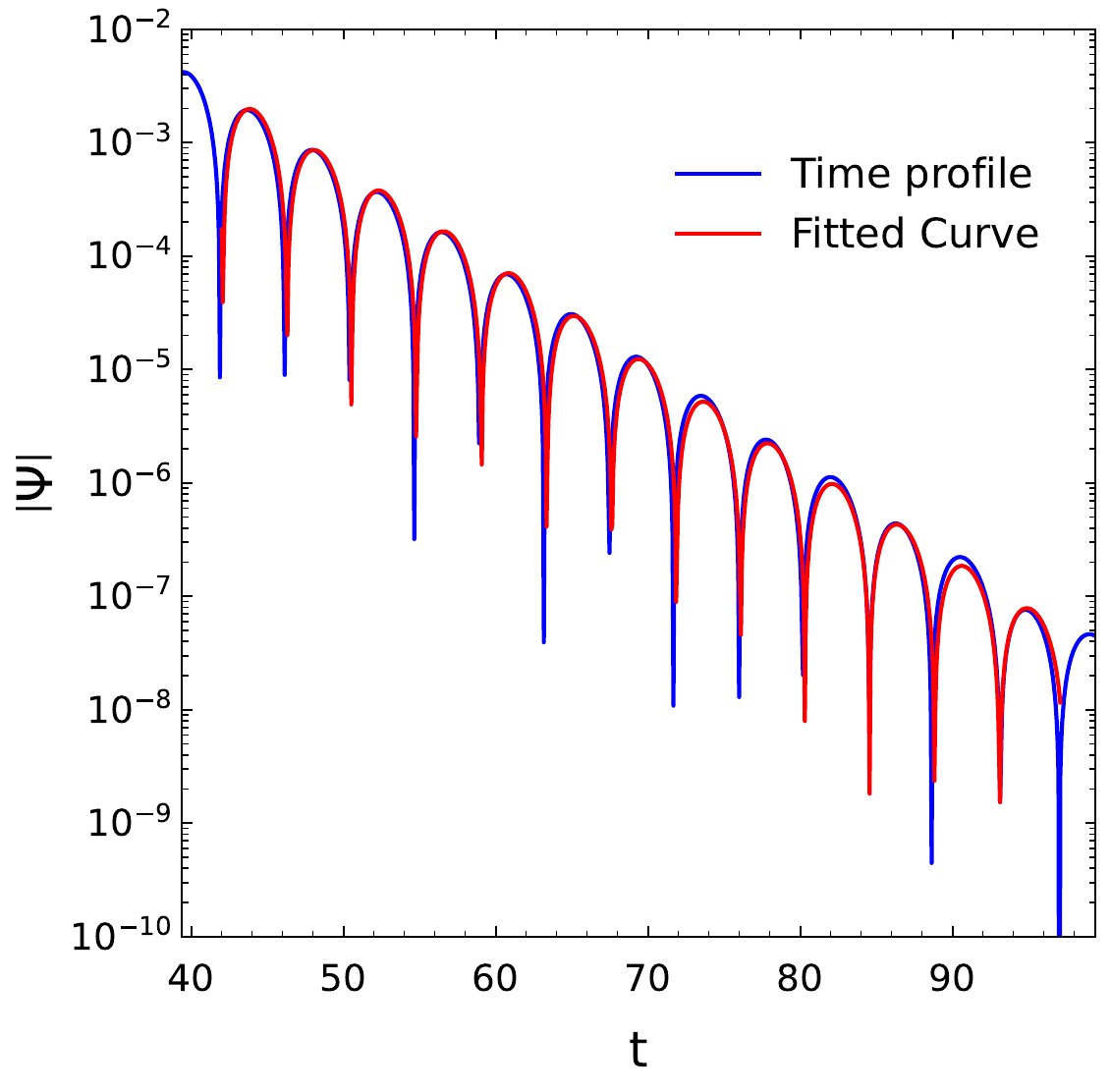}}
    \vspace{-0.2cm}
    \caption{Estimation of QNMs by fitting the time domain profile using 
the Levenberg-Marquardt algorithm \citep{Levenberg1,Levenberg2,Levenberg3}
(for details see the related explanation in the text).}
    \label{fig7}
\end{figure}

In Fig.~\ref{fig6} we plot the time domain profiles obtained from 
the 1st-order and 2nd-order metric corrections for different combinations of 
$f_{3,1}$ and $f_{3,2}$. In the upper row, the left plot shows the time profile 
of the scalar field perturbation for the 1st-order correction with 
$f_{3,1} = -0.02$. The middle plot illustrates the time profile of the 
1st-order correction with $f_{3,1} = 0.01$. In the right plot, we show the 
time profile for the 2nd-order correction with $f_{3,1} = -0.01$ and 
$f_{3,2} = 0.005$. In the lower row, the left plot demonstrates the 
time profile for the 2nd-order correction with $f_{3,1} = -0.01$ and 
$f_{3,2} = -0.002$. The right plot shows the time profile for the 2nd-order 
correction with $f_{3,1} = 0.01$ and $f_{3,2} = 0.002$. From Fig.~\ref{fig6} 
it can be observed that the time domain profile varies with the multiple moment 
$l$, showing a good agreement with the results obtained from the Pad\'e 
averaged WKB approximation method. Further, to extract the QNMs frequencies 
from the time domain analysis, we employ the Levenberg-Marquardt algorithm 
\citep{Levenberg1,Levenberg2,Levenberg3} to perform a nonlinear least-squares 
fit of the waveform to a damped sinusoidal function of the form 
$\Psi(t) = A e^{-\omega_I t} \cos(\omega_R t + \phi)$, where $A$ is the 
amplitude of the wave, $\omega_I$ is the imaginary part of QNM, $\omega_R$ is 
the real part of QNM and $\phi$ is the  initial phase offset of the 
oscillation, over a carefully chosen fitting window that excludes early-time 
transients and late-time noise, as these regions are not dominated by the 
QNM ringing and can introduce significant bias into the extracted frequency. 
The extracted QNMs frequencies are consistent with the results from the 
Pad\'e-WKB approximation method. The QNMs obtained from the time domain method 
are presented up to six significant figures, which demonstrates the fitting 
output and numerical consistency. Fig.~\ref{fig7} demonstrates the estimation 
of QNMs by fitting the time domain profiles. In the upper row, the left plot 
illustrates the fit to the time domain profile obtained from the 1st-order 
metric correction with $f_{3,1} = -\,0.02$ and $l = 1$, yielding the QNM 
frequency $\omega = 0.465790 - 0.154704i$. In the middle plot, we fit the 
time domain profile for the 1st-order metric correction with 
$f_{3,1} = 0.01$ and $l = 2$, from which we obtain the QNM frequency 
$\omega = 0.672137 - 0.081753i$. In the right plot, we fit the time domain 
profile obtained from the 2nd-order metric correction considering 
$f_{3,1} = -\,0.02$, $f_{3,2} = 0.005$ and $l = 2$, and we obtain the QNM as 
$0.605362 - 0.383951i$. In the lower row, the left plot illustrates the 
fitting of the time domain profile for the 2nd-order metric correction with 
$f_{3,1} = -0.01$, $f_{3,2} = -0.002$ and $l = 3$, yielding the QNM frequency 
$\omega = 0.714722 - 0.086763$. The right plot shows the fitting of the time 
domain profile for the 2nd-order metric correction with $f_{3,1} = 0.01$, 
$f_{3,2} = 0.002$ and $l = 3$, yielding the QNM frequency 
$\omega = 0.613452 - 0.101527$. Additionally, we determine 
the difference in the magnitudes of the QNMs obtained from the 6th-order 
Pad\'e averaged WKB approximation method and those obtained using the time 
domain analysis method as follows:
\begin{equation}
\Delta_\text{QNM} = \frac{\left| \text{QNM}_{\text{WKB}} - \text{QNM}_{\text{Time domain}} \right|}{2}. 
\label{eq24}
\end{equation}
\begin{table}[!h]
\caption{Comparison of QNMs obtained from the time domain profiles and 
Pad\'e averaged WKB method for the 1st-order metric correction.}
\vspace{5pt}
    \centering
    \begin{tabular}{|@{\hskip 5pt}c@{\hskip 5pt}| @{\hskip 5pt}c@{\hskip 5pt}|@{\hskip 5pt}c@{\hskip 5pt}|@{\hskip 5pt}c@{\hskip 5pt}|@{\hskip 5pt}c@{\hskip 5pt}|@{\hskip 5pt}c@{\hskip 5pt}|}
        \hline
        \textit{l} & $f_{3,1}$ &Time domain & $R^2$ & Pad\'e averaged WKB & $\Delta_\text{QNM}$ \\[2pt]
        \hline
        $l = 1$ & $-0.02$ &$0.465790 - 0.154704i$ & $0.975360$ & $0.465088 - 0.158462i$ & $0.000191$ \\
        $l = 2$ & $-0.02$ &$0.651392 - 0.155052i$ & $0.977538$ & $0.657593 - 0.158505i$ & $0.003548$\\
        $l = 3$ & $-0.02$ & $0.871569 - 0.165721i$ & $0.968405$ & $0.870989 - 0.162352i$ & $0.00170$ \\
        $l = 1$ & $0.01$ &$0.566243 - 0.007954i$ & $0.967225$ & $0.566109 - 0.072774i$ & $0.032424$ \\
        $l = 2$ & $0.01$ &$0.672138 - 0.081753i$ & $0.9855785$ & $0.674438 - 0.082673i$ & $0.001238$\\
        $l = 3$ & $0.01$ & $0.814257 - 0.087894i$ & $0.989740$ & $0.812411 - 0.089384i$ & $0.001186$ \\
        \hline
    \end{tabular}
    \label{tab:6}
\end{table}

\begin{table}[!h]
\caption{Comparison of QNMs obtained from the time domain profiles and 
Pad\'e averaged WKB method for the 2nd-order metric correction.}\vspace{5pt}
    \centering
    \begin{tabular}{|@{\hskip 5pt}c@{\hskip 5pt}|@{\hskip 5pt}c@{\hskip 5pt}|@{\hskip 5pt}c@{\hskip 5pt}|@{\hskip 5pt}c@{\hskip 5pt}|@{\hskip 5pt}c@{\hskip 5pt}|@{\hskip 5pt}c@{\hskip 5pt}|@{\hskip 5pt}c@{\hskip 5pt}|}
        \hline
        \textit{l} & $f_{3,1}$& $f_{3,2}$& Time domain & $R^2$ & Pad\'e averaged WKB & $\Delta_\text{QNM}$ \\[2pt]
        \hline
        $l = 1$ & $-0.01$ & $0.005$ & $0.603155 - 0.382175i$ & $0.95890551$ & $0.607196 - 0.384308i$ & $0.002286$ \\
        $l = 2$ & $-0.01$ & $0.005$ & $0.571028 - 0.332176i$ & $0.97922510$ & $0.570424 - 0.331178i$ & $0.000584$\\
        $l = 3$ & $-0.01$ & $0.005$ & $0.719532 - 0.069925i$ & $0.96897990$ & $0.718777 - 0.071480i$ & $0.000865$\\
        $l = 1$ & $0.01$ & $0.002$ & $0.316542 - 0.071287i$ & $0.95185421$ & $0.318914 - 0.075432i$ & $0.002388$ \\
        $l = 3$ & $-0.02$ & $-0.007$ & $0.717865 - 0.096987i$ & $0.97454830$ & $0.719945 - 0.097322i$ & $0.001050$\\
        \hline
    \end{tabular}
    \label{tab:7}
\end{table}
Table~\ref{tab:6} shows the QNMs obtained from the time domain profiles along 
with QNMs obtained from the Pad\'e averaged WKB approximation method for 
$l = 1$, $2$ and $ 3$ considering the 1st-order metric correction, for both 
positive and negative values of $f_{3,1}$. Similarly, 
Table~\ref{tab:7} shows the QNMs obtained from the time domain profiles along 
with QNMs obtained from the Pad\'e averaged WKB approximation method for 
$l = 1$, $2$ and $3$ considering the 2nd-order metric correction for different 
combinations of $f_{3,1}$ and $f_{3,2}$. In both the tables we show the $R^2$ 
value of the fitting. From the time domain analysis, it is observed that for 
both first-order and second-order metric corrections, the oscillation 
frequency and decay rate of the QNMs increase with the multipole number $l$. 
Further, it can be also observed that the results obtained from the time 
domain analysis are in good agreement with the results obtained from the 
Pad\'e averaged WKB approximation method. 

\section{Shadow of the black hole} \label{5}
The study of BH shadow provides an opportunity to test different theories of 
gravity as well as helps in the study of gravity in the BH's extreme 
gravity regimes. The shape of the shadow formed by a BH is determined by the 
parameters of the underlying theory and is defined by the photon sphere. The 
photon sphere is a region around the BH where gravity is so strong that 
photons, or light particles, travel in unstable circular orbits around the BH. 
In this section, our aim is to study the behavior of the shadow when 
higher-order corrections of the IDG are included and for different values of 
the free model parameters. 

For a system with spherically symmetric and static spacetime metric the 
Lagrangian $\mathcal{L}(x, \dot{x})$ of that system can be expressed 
as \cite{C1}
\begin{equation}
\mathcal{L}(x, \dot{x}) \equiv \frac{1}{2}g_{\mu\nu}\dot{x}^{\mu}\dot{x}^{\nu} =\frac{1}{2}\Big[-f(r)\,\dot{t}^2+\frac{1}{g(r)}\,\dot{r}^2+r^2(\dot{\theta}^2+Sin^2\theta\dot{\phi}^2)\Big],
\end{equation} 
where the dots over the variables represent differentiation with respect to 
the proper time $\tau$. For our considered BH system the metric functions 
$f(r)$ and $g(r)$ are given respectively by Eqs.~\eqref{eq6} and \eqref{eq7}. 
Now, the null geodesic equation for photons is \citep{C1} 
\begin{equation}
- f(r)\, \dot{t}^2 + \frac{1}{g(r)}\, \dot{r}^2 + r^2 \dot{\phi}^2 = 0. 
\label{eq27}
\end{equation} 
By utilizing the Euler-Lagrange equation $$\frac{d}{d\tau} \left( \frac{\partial \mathcal{L}}{\partial \dot{x}^\mu} \right) - \frac{\partial \mathcal{L}}{\partial x^\mu} = 0$$ we can obtain two conserved quantities of the system, 
viz.~energy $\mathcal{E}$ and angular momentum $L$ which are given by 
$\mathcal{E} = f(r)\, \dot{t}$ and $L = r^2\, \dot{\phi}$. Using these two 
conserved quantities in Eq.~\eqref{eq27} the reduced potential for the 
photons' orbital motion can be obtained as \citep{C1,Ali1} 
\begin{equation}
U_{r} = \frac{{L^2}{g(r)}}{r^2}.
\label{eq29}
\end{equation}
Since angular momentum $L$ is a conserved quantity, the behaviour of the 
reduced potential will not depend on it, so we will consider $L = 1$ in our 
further calculations.
\begin{figure}[!h]
    \centerline{
    \includegraphics[scale = 0.28]{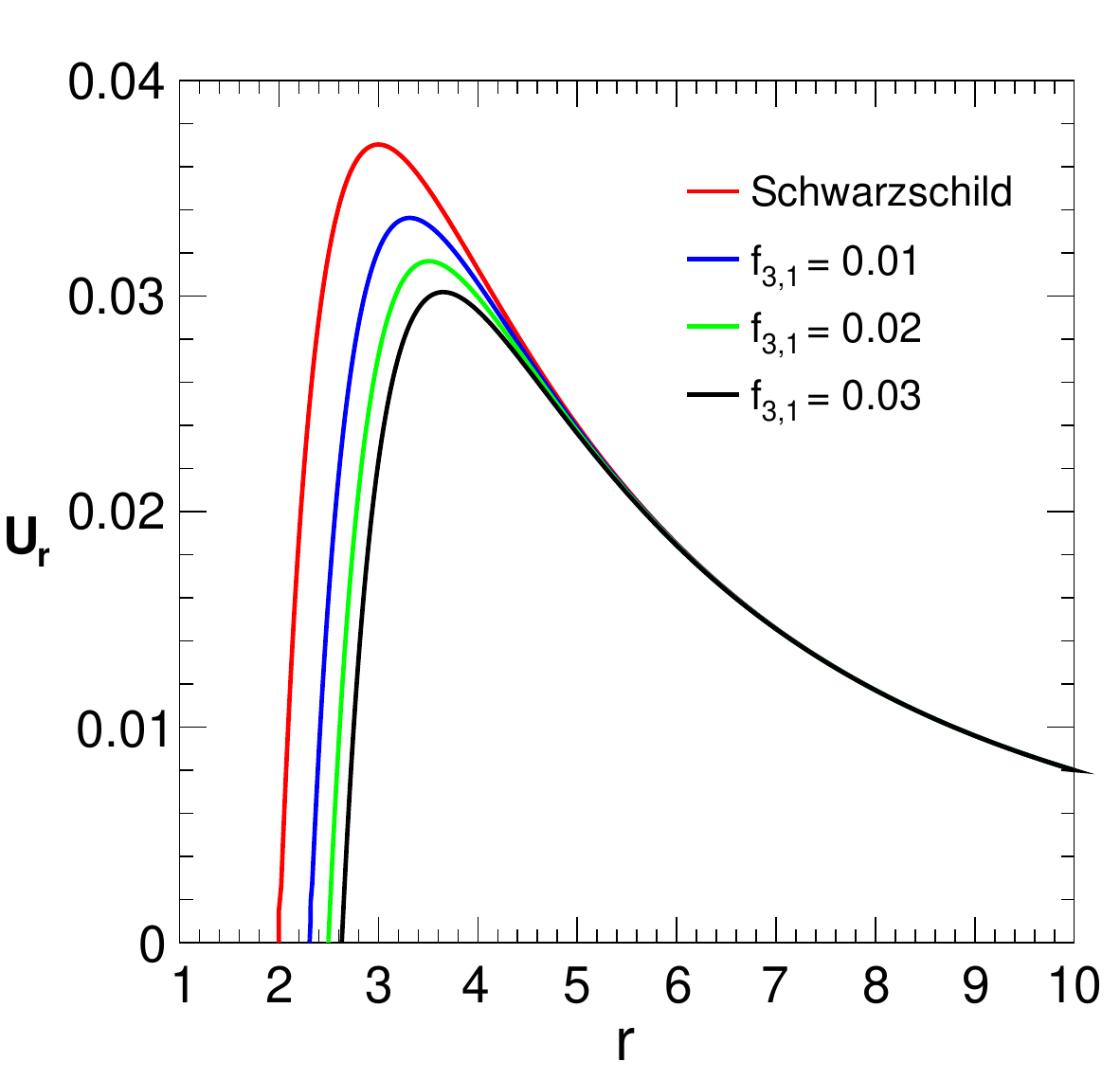}\hspace{0.3cm}
    \includegraphics[scale = 0.28]{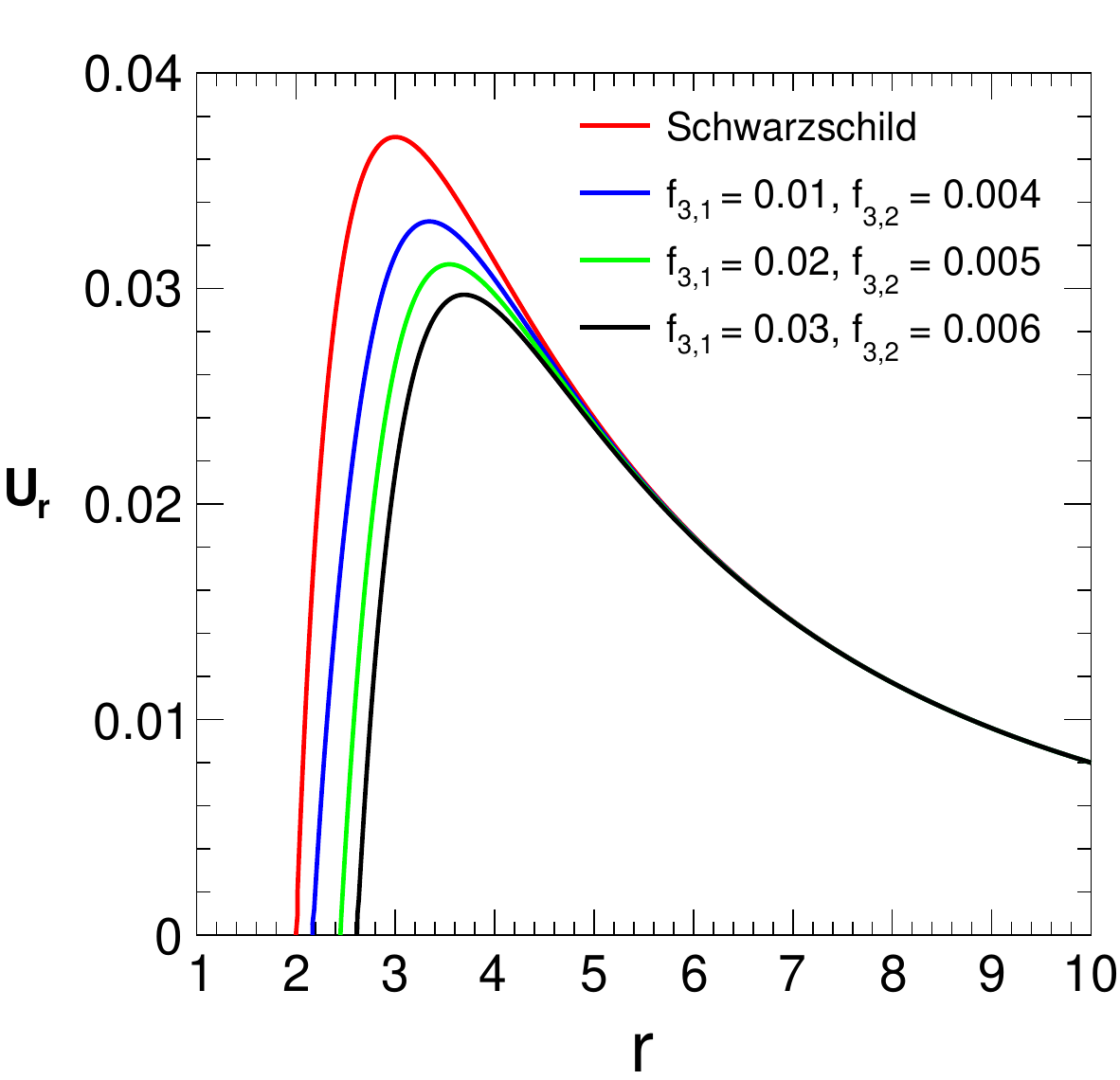}\hspace{0.3cm}
    \includegraphics[scale = 0.28]{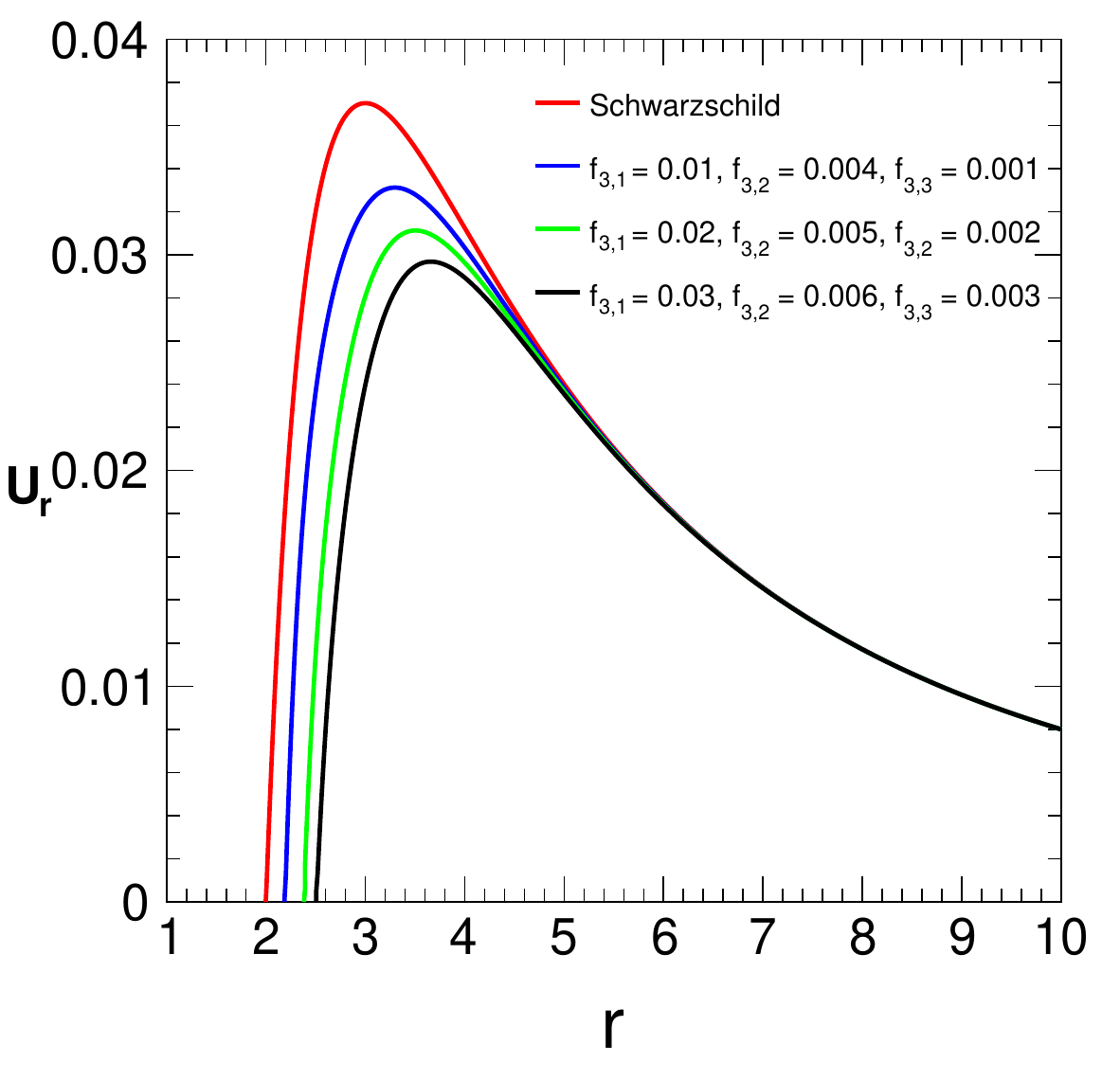}}
    \centerline{
    \includegraphics[scale = 0.28]{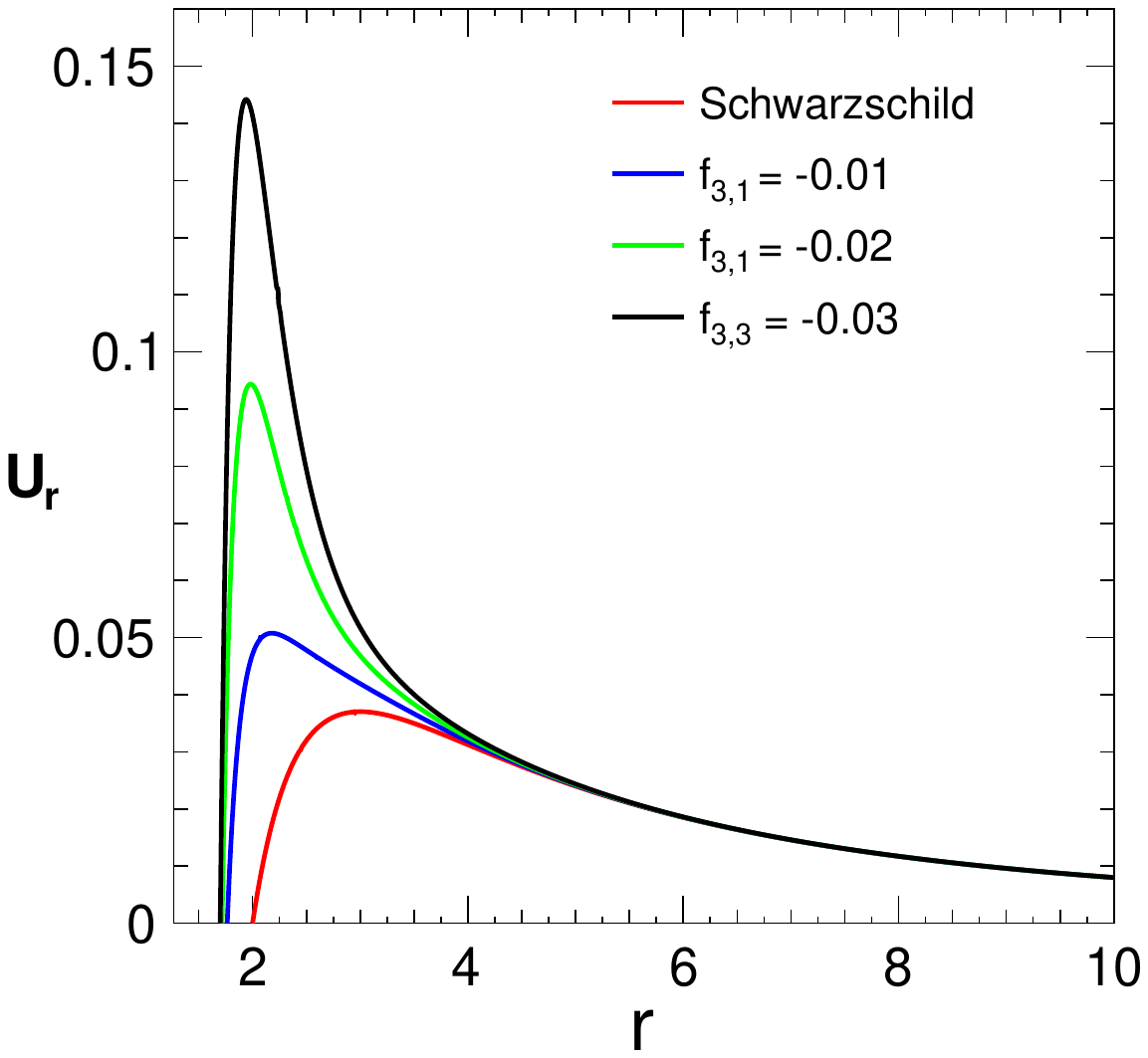 }\hspace{0.3cm}
    \includegraphics[scale = 0.28]{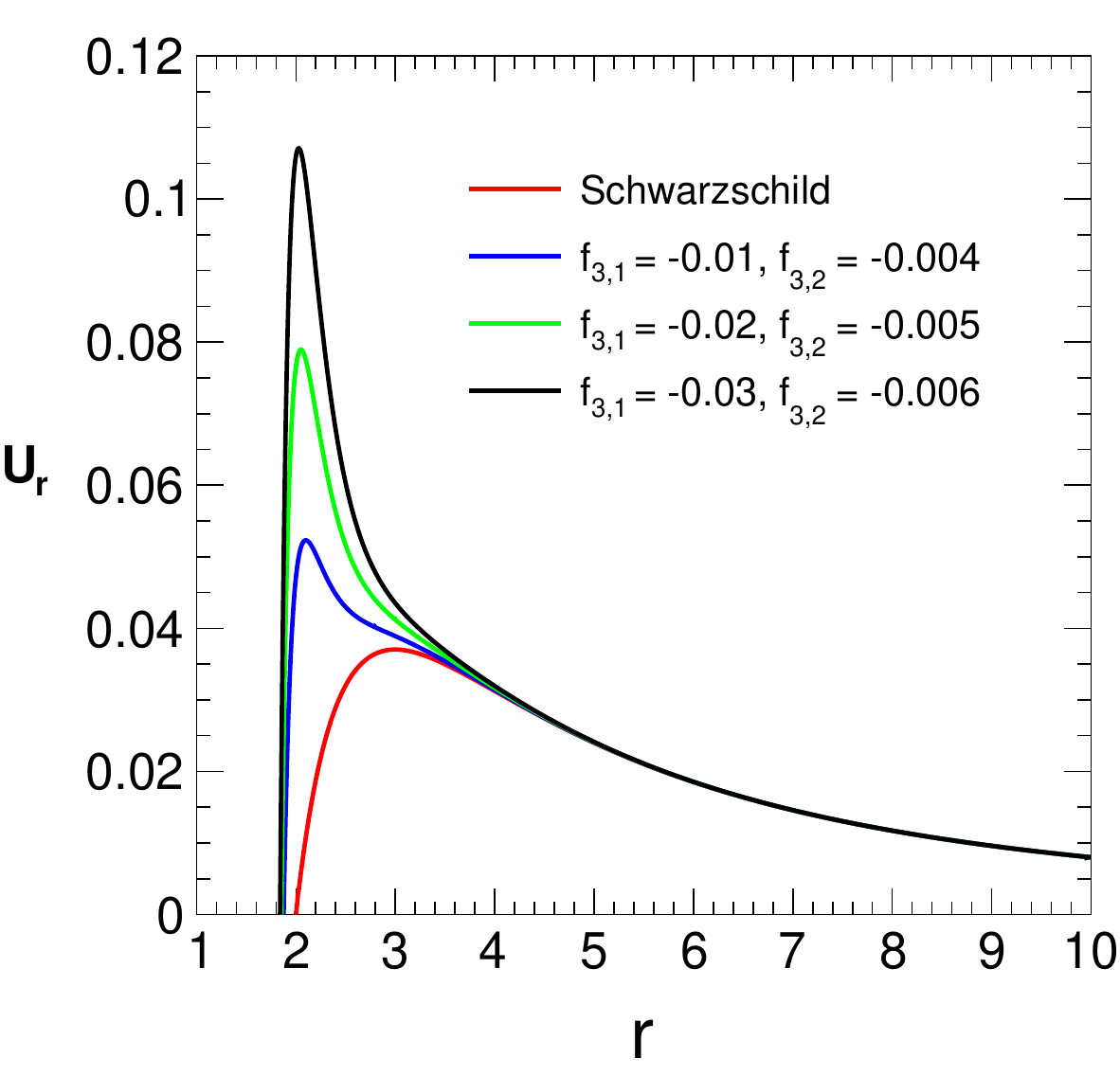}\hspace{0.3cm}
    \includegraphics[scale = 0.28]{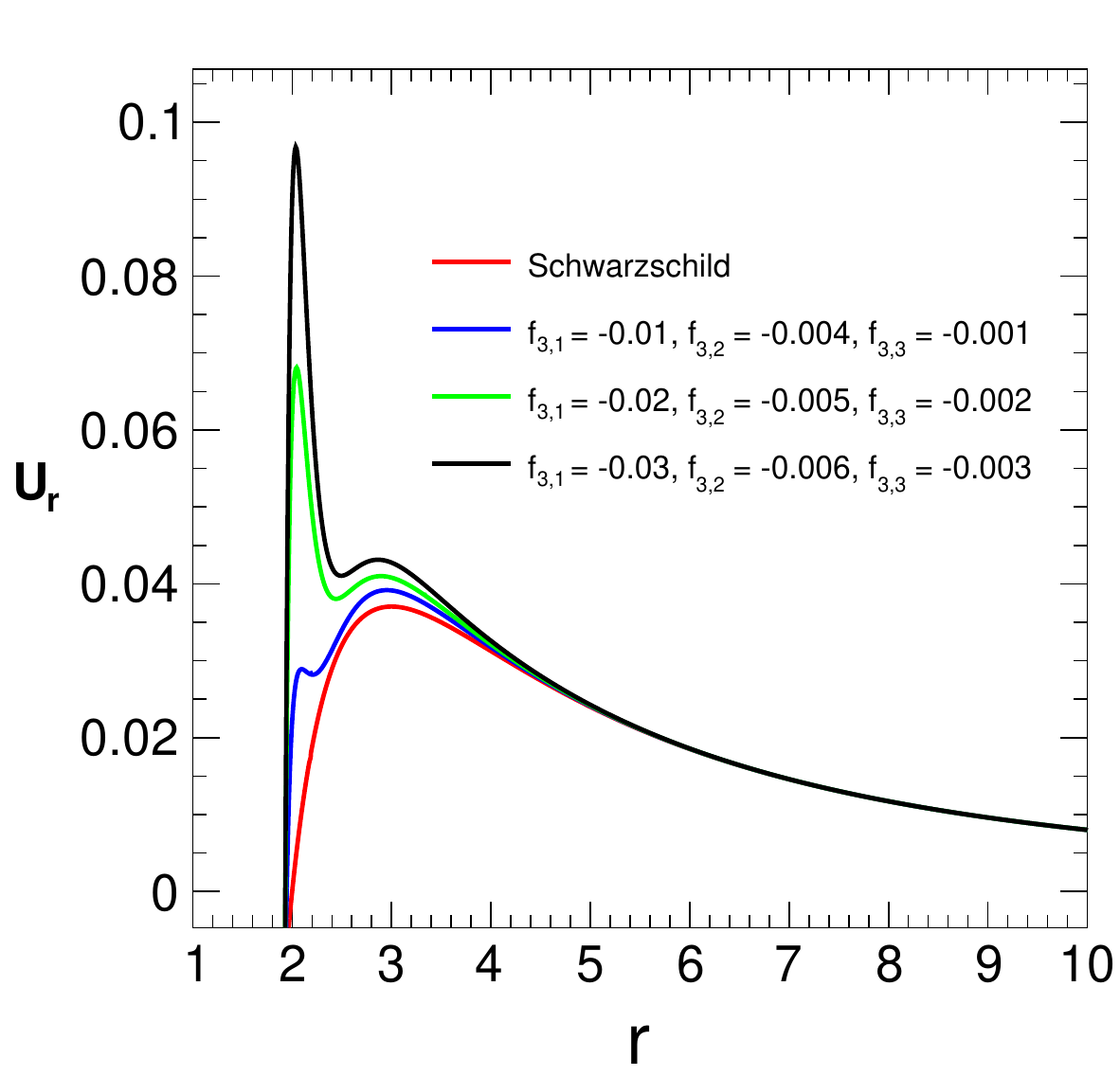}}
     \centerline{
    \includegraphics[scale = 0.28]{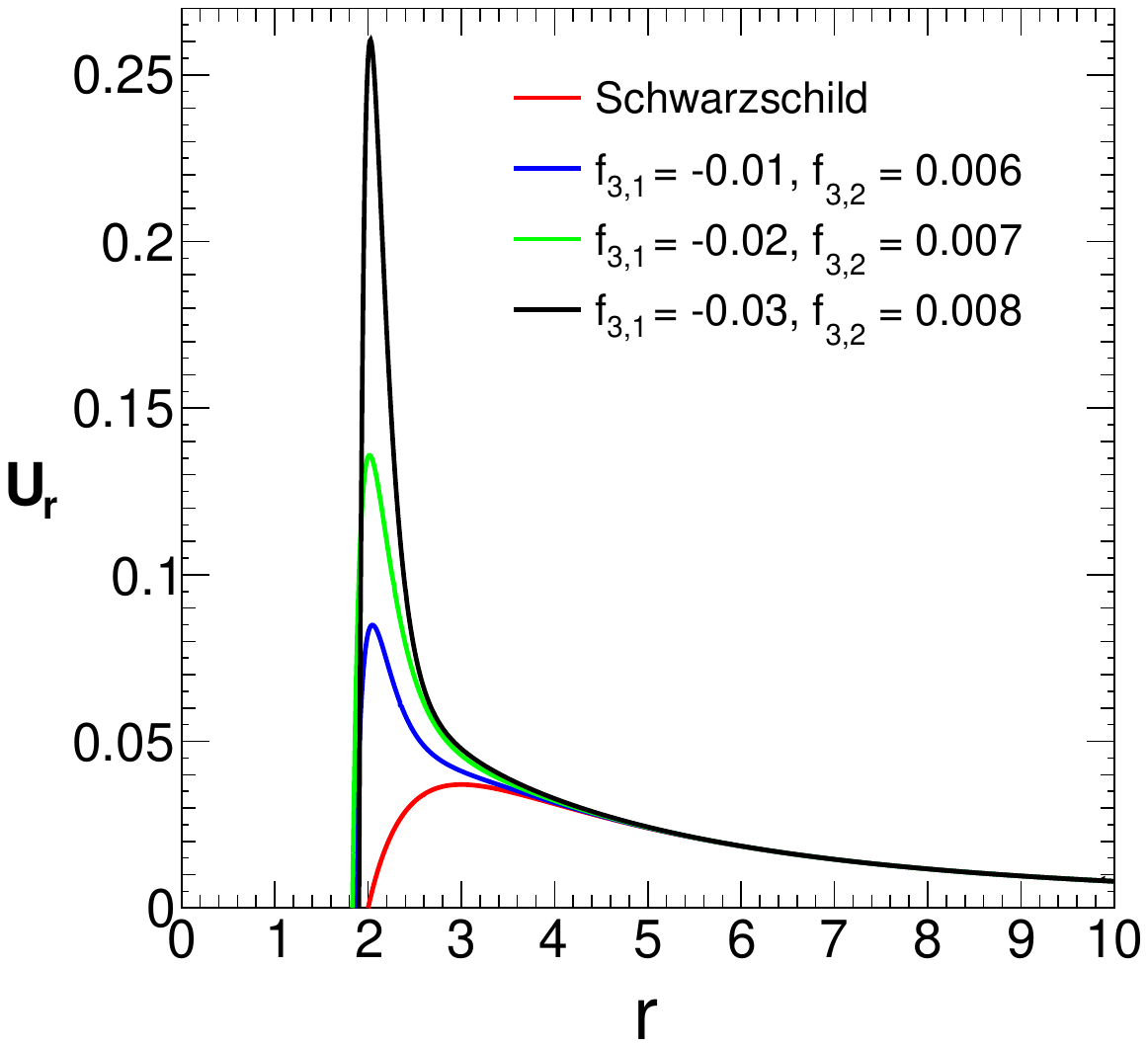 }\hspace{0.3cm}
    \includegraphics[scale = 0.28]{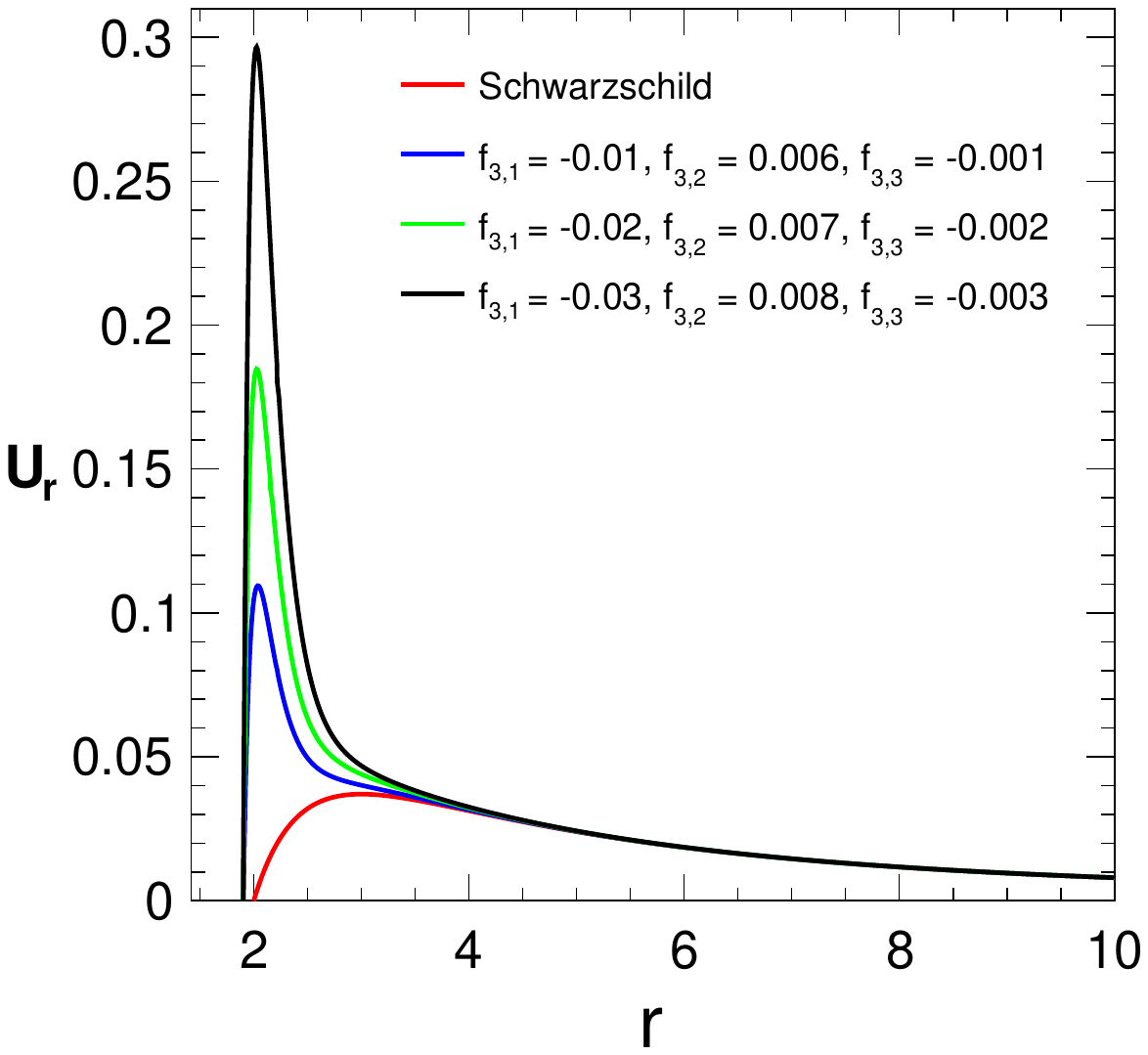}}
    \vspace{-0.2cm}  
    \caption{Variation of reduced potential $U_{r}$ given by 
Eq.~\eqref{eq29} with radial distance $r$ for different metric corrections 
and for different sign combinations of the free parameters.}
    \label{fig8}
\end{figure}

Fig.~\ref{fig8} shows the variation of the reduced potential $U_{r}$ 
with respect to $r$. The upper row displays this variation for positive values 
of the correction parameters up to the 3rd-order. Specifically, the left panel 
corresponds to the 1st-order correction, the middle panel to the 2nd-order 
correction, and the right panel to the 3rd-order correction. It is evident 
from the figure that the peak of $U_r$ decreases as the positive values of 
the correction parameters increase. However, the overall effect of 
higher-order corrections remains very small when the parameters $f_{3,1}$, 
$f_{3,2}$ and $f_{3,3}$ are all positive. The middle row shows the variation 
of the reduced potential $U_r$ with respect to $r$ for negative values of the 
correction parameters. It can be seen from the plots that, compared to the 
Schwarzschild case, the peak of the potential increases for negative parameter 
values and further rises as the magnitude of these parameters increases. 
However, it should be noted that the peak tends to decrease when higher-order 
corrections are included. The bottom row illustrates the variation of the 
reduced potential $U_r$ with respect to $r$ for both positive and negative 
sign combinations of the correction parameters for the 2nd-order and 3rd-order 
cases. The left plot corresponds to the 2nd-order correction. It shows that 
as the positive values of $f_{3,2}$ increase, while $f_{3,1}$ is kept negative 
with increasing magnitude, the peak of the potential rises. The right plot 
corresponds to the 3rd-order correction. It can be observed that when 
$f_{3,1}$ is negative, $f_{3,2}$ is positive, and $f_{3,3}$ takes negative 
values with increasing magnitude, the peak of the potential also increases. 
Moreover, for the last two cases, the peak of the potential becomes higher as 
higher-order corrections are included. Further, all potential values merged 
together after a certain value of $r$ depending on the order of corrections.

\begin{figure}[!h]
    \centerline{
    \includegraphics[scale = 0.3]{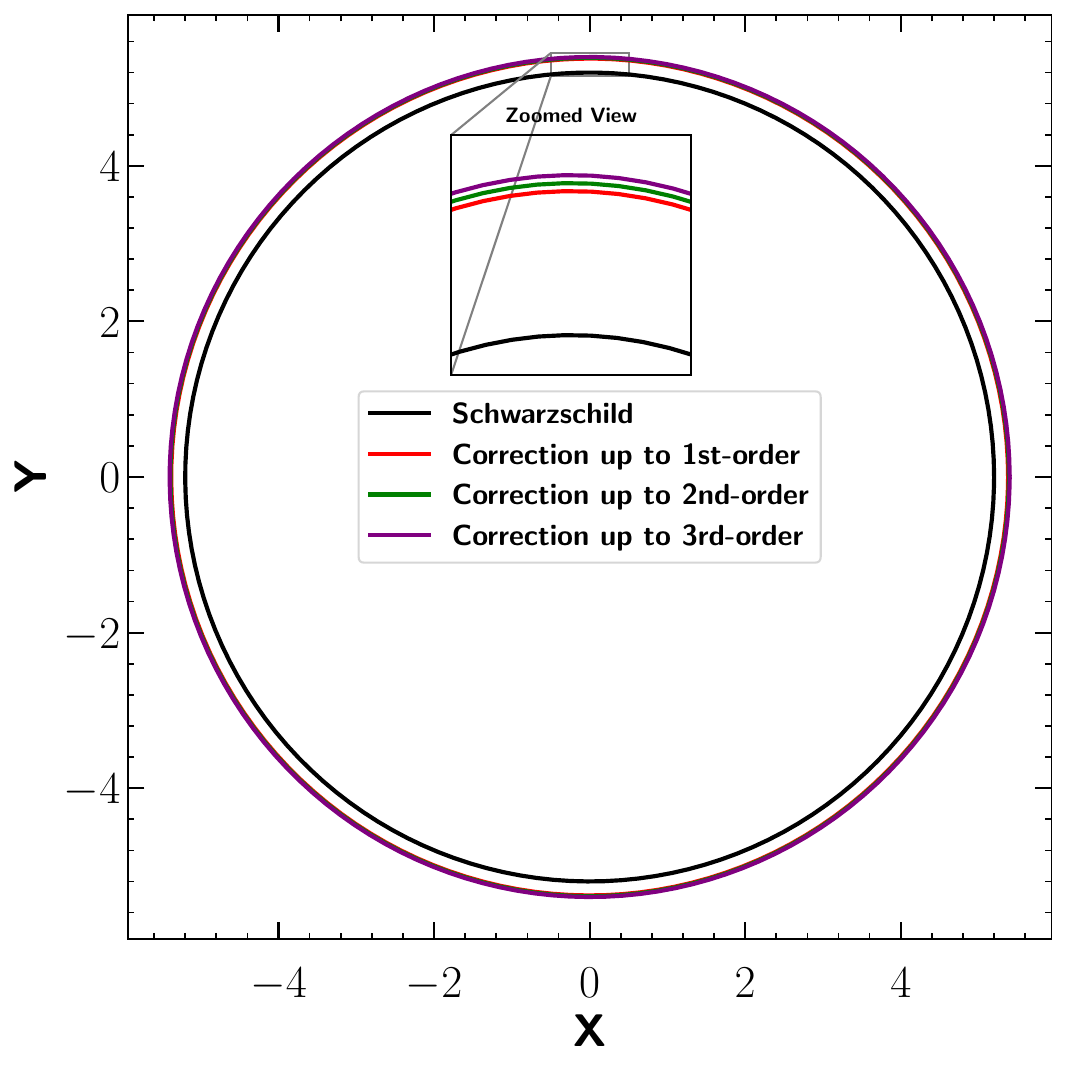}\hspace{0.6cm}
    \includegraphics[scale=0.3]{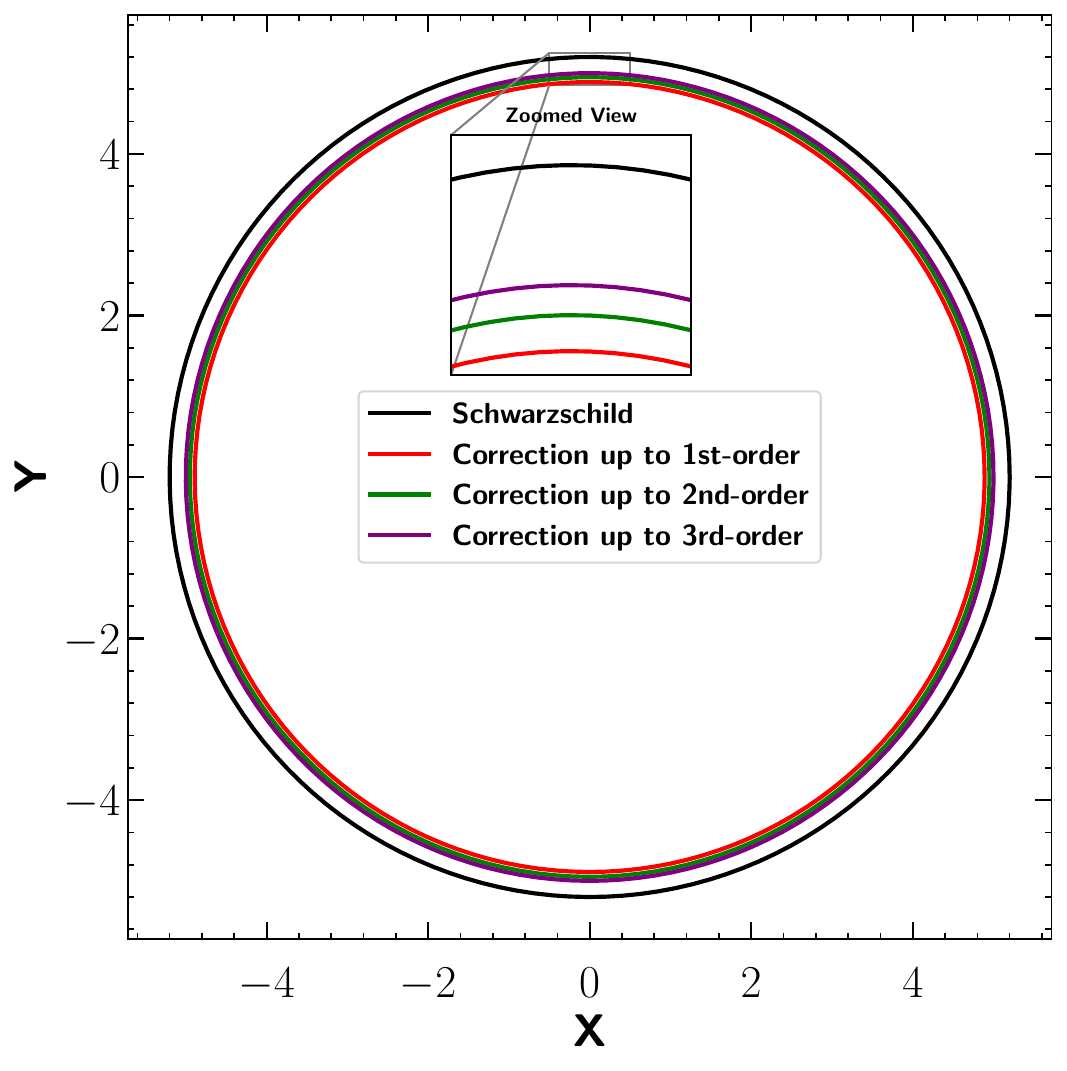}\hspace{0.6cm}
    \includegraphics[scale=0.3]{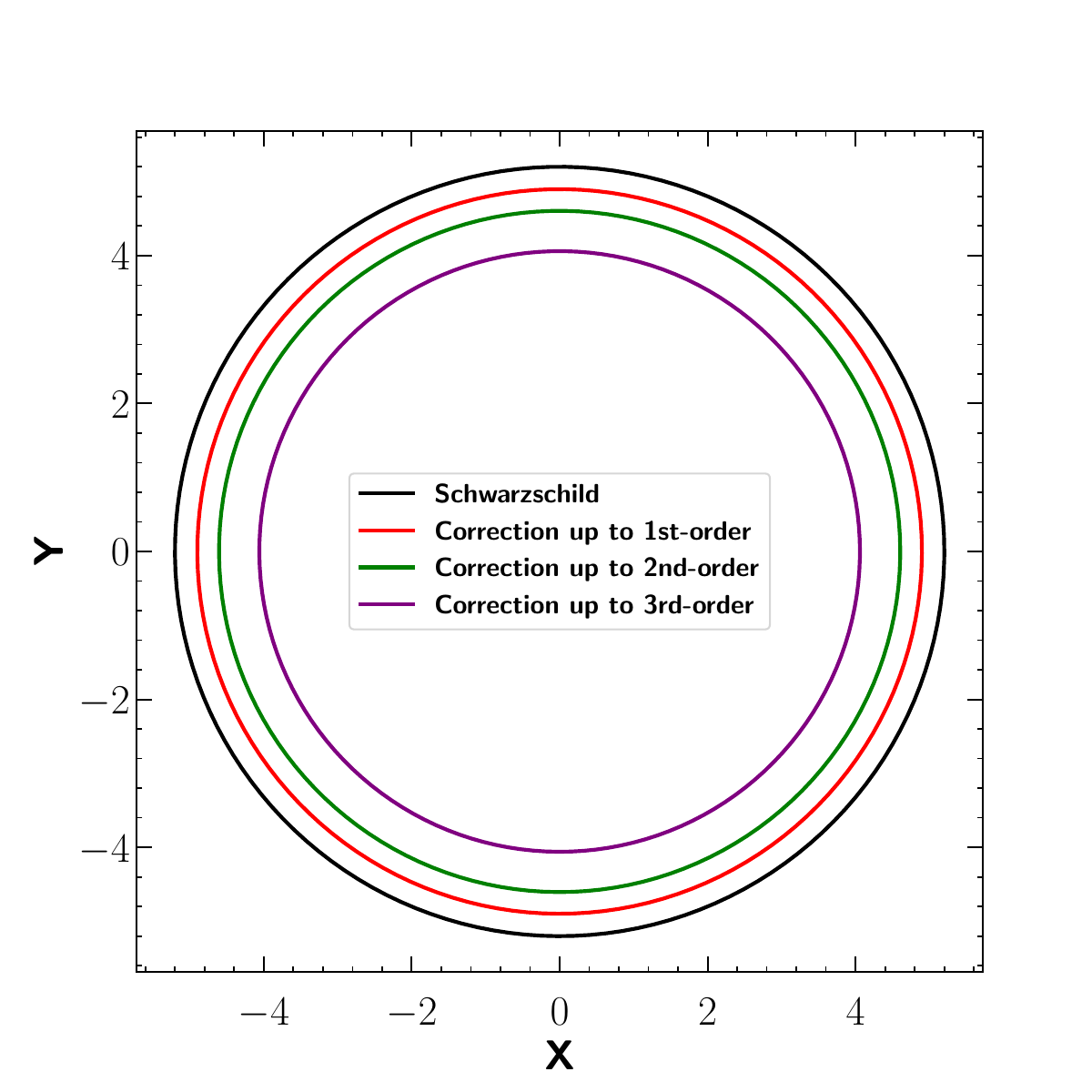}}
    \vspace{-0.2cm}
    \caption{Stereographic mapping of the variation of shadow radius in 
celestial coordinates for the addition of different orders of metric 
corrections with different sign combinations. The left plot is for all 
positive correction parameters, the middle plot is for all negative correction 
parameters, and the right plot is for mixed-sign combinations of the 
correction parameters (for details see the related explanation in the text).}
    \label{fig9}
\end{figure}

Now we move to study the shadow radius. To study the shadow radius one needs 
to first determine the radius of the photon sphere. The photon sphere radius 
can be obtained by solving the following equation \citep{C1,Luminet1,Synge1}: 
\begin{equation}
\left. \frac{f'(r)}{f(r)} \right|_{r\,=\,r_{\text{ph}}} = \left. \frac{h'(r)}{h(r)} \right|_{r\,=\,r_{\text{ph}}},
\end{equation}
where $r_{ph}$ represents the radius of the photon sphere and $h(r) = r^2$. 
This equation leads to the shadow radius $R_s$ for a static observer at a 
large distance as
\begin{equation}
R_s = \frac{r_{ph}}{\sqrt{f(r_{ph})}}. \label{eq30}
\end{equation} 
The stereographic projection of the BH's shadow from the plane of the black 
hole to the observer's observation plane using coordinates $(X, Y)$ can be 
used to determine the shadow's apparent shape. The coordinates $(X,Y)$ is 
defined as \cite{Kumar1} 
\begin{align}
X & = \lim_{r_0 \to \infty} \left( -\,r_0^2 \sin\theta_0\, \frac{d\phi}{dr} \Big|_{r_0} \right),\\[8pt]
Y & = \lim_{r_0 \to \infty} \left( r_0^2\, \frac{d\theta}{dr} \Big|_{(r_0,\theta_0)} \right),
\end{align} 
where $r_0$ is the position of the observer and $\theta_0$ is the angular 
position of the observer with respect to the BH plane. Fig.~\ref{fig9} 
shows the stereographic projection of the variation in shadow radius under 
different orders of metric corrections for various sign combinations of the 
correction parameters. The left panel displays the behavior of the shadow 
radius when all correction parameters are positive. It can be observed that 
in this case, the shadow radius is larger than that of the Schwarzschild black 
hole and increases slightly with the inclusion of higher-order corrections. 
However, the variation remains small, which is also reflected in the behavior 
of the reduced potential under higher-order corrections. The middle panel 
illustrates the case where all correction parameters are negative. Here, the 
shadow radius is smaller than the Schwarzschild case, but it increases 
slightly as higher-order corrections are added. The variation remains modest 
in this scenario as well. A zoomed view is provided for clarity in both the 
left and middle panels. The right panel shows the shadow radius for mixed-sign 
combinations of the correction parameters. For the 1st-order correction, we 
take $f_{3,1} < 0$ as a reference. For the 2nd-order correction, we consider 
$f_{3,1} < 0$ and $f_{3,2} > 0$, and for the 3rd-order correction, we choose 
$f_{3,1} < 0$, $f_{3,2} > 0$ and $f_{3,3} < 0$. In this case, we observe a 
significant deviation from the Schwarzschild case, where the shadow radius 
decreases with the inclusion of higher-order corrections.

\begin{figure}[!h]
    \centerline{
    \includegraphics[scale = 0.8]{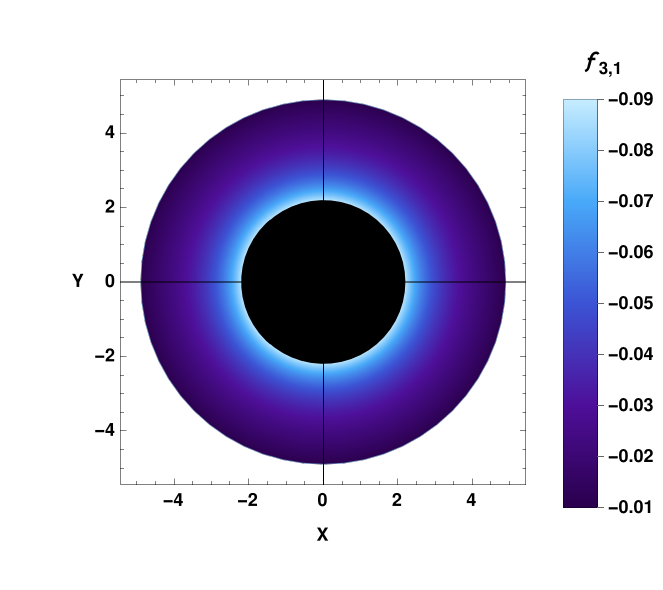}\hspace{0.8cm}
     \raisebox{0.5cm}{\includegraphics[scale = 0.939]{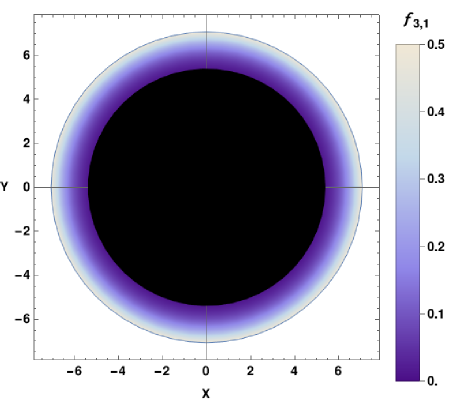}}}
     \vspace{-0.7cm}
    \caption{Stereographic mapping of the variation of shadow radius in 
celestial coordinates for the 1st-order metric correction with different 
signs of $f_{3,1}$. The left plot is for the negative values of $f_{3,1}$, 
while the right plot is for its positive values.}
    \label{fig10}
\end{figure}

\begin{figure}[!h]
    \centerline{
    \includegraphics[scale = 0.688]{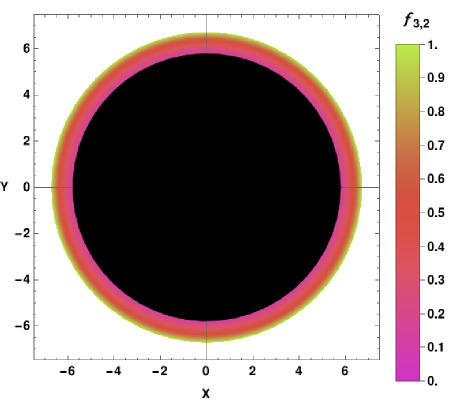}\hspace{0.5cm}
    \includegraphics[scale=0.7]{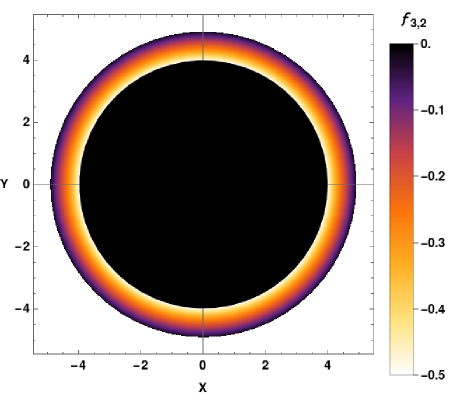}\hspace{-0.2cm}
    \raisebox{-0.5cm}{\includegraphics[scale=0.6]{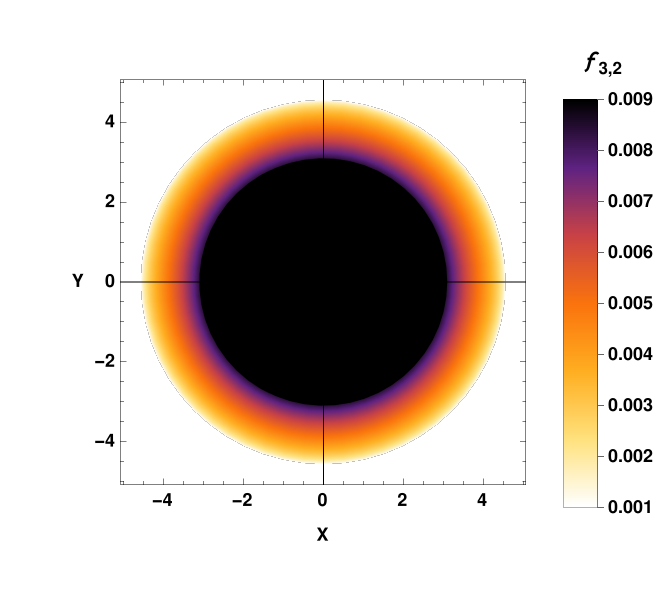}}}
    \vspace{-0.6cm}
    \caption{Stereographic mapping of the variation of shadow radius as a 
function of $f_{3,2}$ in celestial coordinates for the 2nd-order metric 
correction with different sign combinations of $f_{3,1}$ and $f_{3,2}$. In 
the left plot $f_{3,1} = 0.01$, in the middle and right plots $f_{3,1} 
= -\,0.01$ are used.}
    \label{fig10A}
\end{figure}

\begin{figure}[!h]
    \centerline{
    \includegraphics[scale = 0.9]{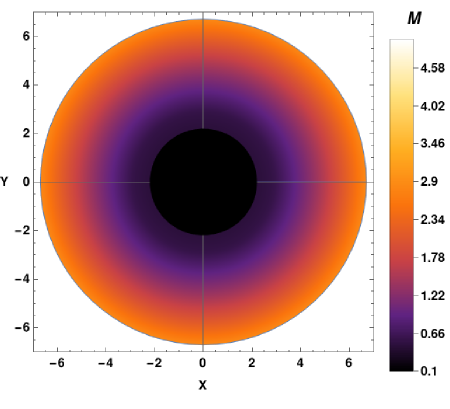}\hspace{0.8cm}
    \includegraphics[scale = 0.91]{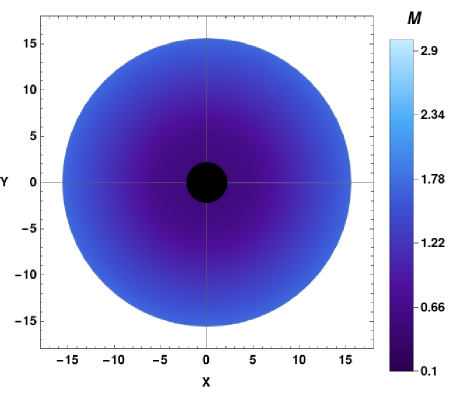}}
    \vspace{-0.3cm}
    \caption{Stereoscopic mapping of the variation of shadow radius 
with respect to mass of the BH in celestial coordinates. The left plot is for
the 1st-order metric correction with $f_{3,1} = -0.01$ and the right one is 
for the 2nd-order metric correction with $f_{3,1} = -0.015$ and 
$f_{3,2} = 0.005$.}
    \label{fig10B}
\end{figure}

Fig.~\ref{fig10} shows the stereographic projection of the shadow 
radius as a function of the free parameter $f_{3,1}$ for the 1st-order metric 
correction. The left panel presents the variation of the shadow radius with 
negative values of $f_{3,1}$, where it is clearly seen that the shadow radius 
decreases as the magnitude of the negative $f_{3,1}$ increases. The right 
panel displays the variation with positive values of $f_{3,1}$, showing that 
the shadow radius increases slightly with increasing $f_{3,1}$. It should be 
noted that the rate of increase is very small for positive values of $f_{3,1}$.
Fig.~\ref{fig10A} shows the stereographic projection of the shadow 
radius as a function of the free parameter $f_{3,2}$ for the 2nd-order metric 
correction with different fixed values of $f_{3,1}$. The left panel displays 
the variation of the shadow radius with increasing positive values of 
$f_{3,2}$ for $f_{3,1} = 0.01$. It can be observed from the plot that the 
shadow radius increases with $f_{3,2}$, although the rate of increase is 
small, similar to the trend seen in the right panel of Fig.~\ref{fig10}. The 
middle panel illustrates the variation of the shadow radius with negative 
values of $f_{3,2}$ for $f_{3,1} = -0.01$. In this case, the shadow radius 
decreases with the magnitude of $f_{3,2}$. The right panel presents the 
variation of the shadow radius with positive values of $f_{3,2}$, again for 
$f_{3,1} = -0.01$. It can be observed that the shadow radius decreases with 
increasing $f_{3,2}$, and in this case, the decrease is more rapid compared 
to the scenario where both $f_{3,1}$ and $f_{3,2}$ are negative.
Furthermore, since both metric functions $f(r)$ and $g(r)$ depend 
explicitly on the BH mass $M$, it is essential to treat $M$ as a variable. 
Fig.~\ref{fig10B} illustrates the variation of the shadow radius as a function 
of $M$. The left panel displays the shadow radius for the 1st-order metric 
correction with $f_{3,1} = -0.01$, while the right panel corresponds to the 
2nd-order correction, taking $f_{3,1} = -0.015$ and $f_{3,2} = 0.005$. It is 
observed from the figure that the shadow radius increases with increasing mass 
for both metric corrections. However, it should be noted that the rate of 
increase of the shadow radius with respect to mass is greater in the case of 
the 2nd-order correction than in the 1st-order correction.
   
To constrain the coefficients $f_{3,1}$ and $f_{3,2}$, we utilize the 
technique described in Ref.~\cite{constrain1}. It is to be noted that due to
the complexity of the process we could not include the coefficient 
$f_{3,3}$ for the constraining. Here, we outline some key steps 
of this approach. The main idea is to compare the observed angular radius of 
the Sgr A* BH, which was recently measured by the EHT collaboration with 
the theoretically computed shadow radius from Eq.~\eqref{eq30}. This 
comparison helps us impose constraints on the model parameters. It is 
important to have a prior value for the mass-to-distance ratio of the Sgr A*. 
Another crucial aspect of this method is the calibration factor, which 
connects the observed shadow radius to the theoretical prediction. Similar 
methods have been used in previous studies 
\cite{constrain2,constrain3,constrain4} and we will follow the same 
strategy in this regard. The EHT group introduced a new parameter $\delta$, 
which represents the fractional deviation between the observed shadow radius 
$R_s$ and the shadow radius of a Schwarzschild BH $R_{sch}$. Thus the parameter 
$\delta$ can be defined as follows \cite{constrain1}: 
\begin{equation}
\delta = \frac{R_s}{R_{sch}} - 1 = \frac{R_s}{3{\sqrt{3}}M} - 1. 
\label{eq33}
\end{equation}
The estimated values of the parameter $\delta$ from the measurements of Keck 
and VLTI experiments are \cite{constrain1} $\delta = -\,0.04^{+0.09}_{-0.10}$ 
(Keck) and $\delta = -\,0.08^{+0.09}_{-0.09}$ (VLTI) respectively. For 
simplicity, we consider the mean of these two observational values as in the 
case of Ref.~\cite{constrain1}. The mean value of $\delta$ is 
\begin{equation}
\delta = -\,0.060 \pm 0.065.
\end{equation} 
As mentioned in Ref.~\cite{constrain1}, considering a Gaussian distribution of 
$\delta$ with $\sigma=0.065$ as the standard deviation, calculated the 
$1\sigma$ and $2\sigma$ intervals for the $\delta$ parameter as  
\begin{align}
-\,0.125& \lesssim \delta \lesssim 0.005 \quad (1\sigma), \label{eq35}\\[5pt]
-\,0.190& \lesssim \delta \lesssim 0.070 \quad (2\sigma). \label{eq36}
\end{align}
These bounds for $\delta$ in Eq.~\eqref{eq35} and Eq.~\eqref{eq36} when used 
in Eq.~\eqref{eq33}, give the bounds on the Schwarzschild radius $R_{sch}$ as
\begin{align}
4.55 & \lesssim \frac{r_{sh}}{M} \lesssim 5.22 \quad (1\sigma),\\[5pt]
4.21 & \lesssim \frac{r_{sh}}{M} \lesssim 5.56 \quad (2\sigma).
\end{align}

\begin{figure}[!h]
    \centerline{\hspace{-0.3cm}
    \includegraphics[scale = 0.5]{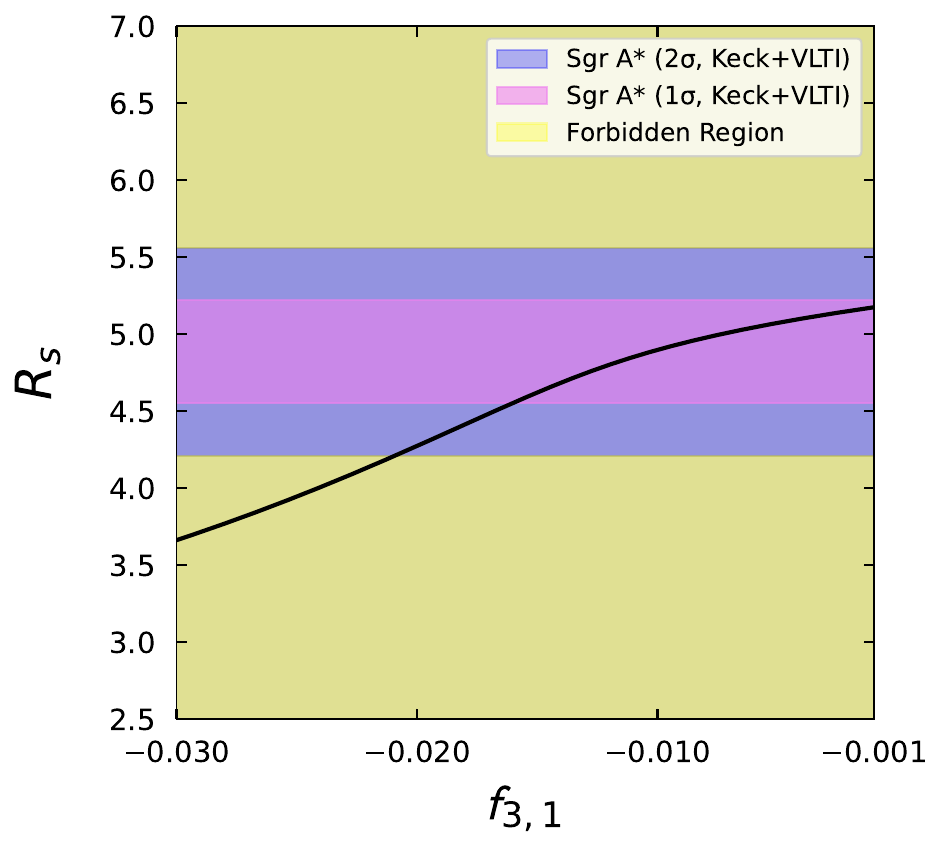}\hspace{0.3cm}
    \includegraphics[scale = 0.5]{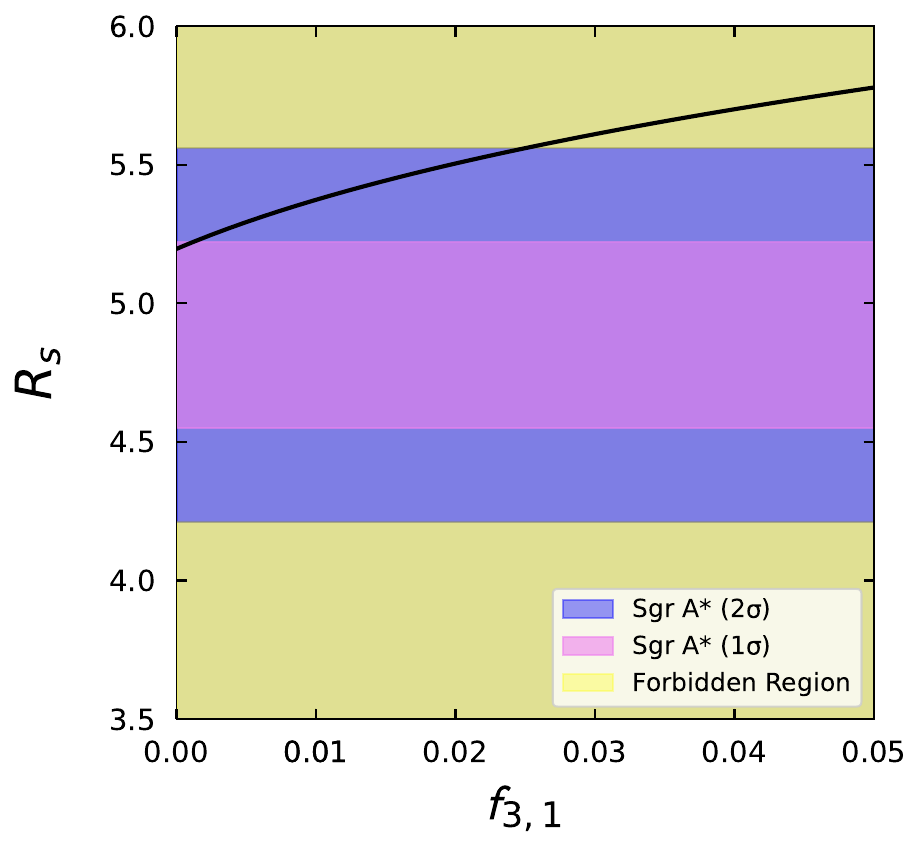}}
   
    \centerline{
     \includegraphics[scale=0.5]{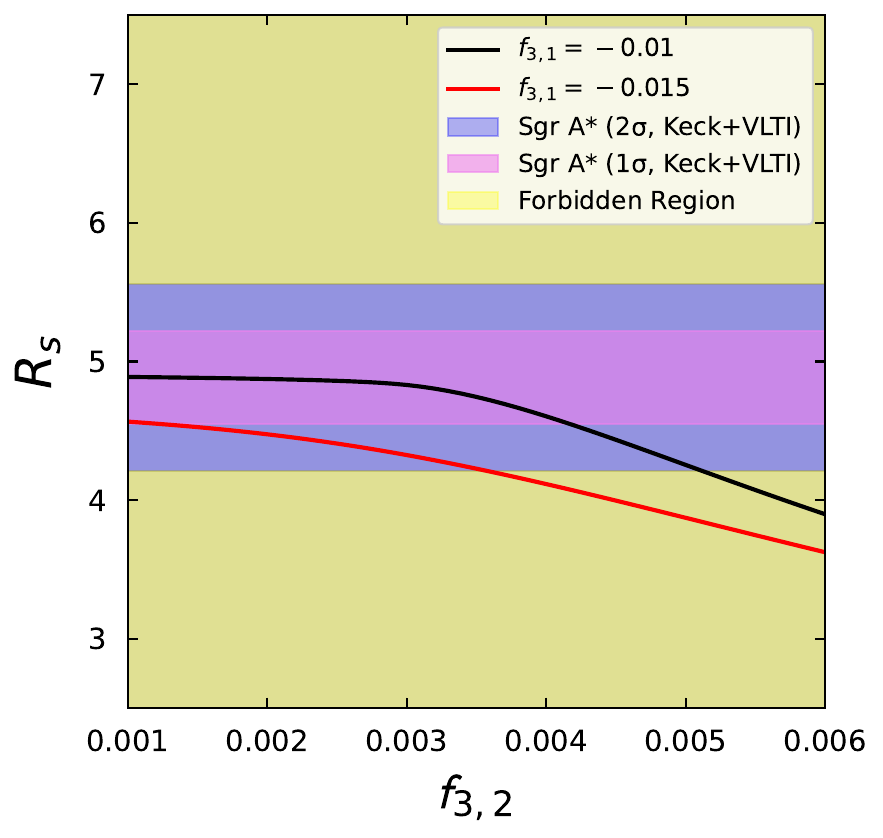}\hspace{0.8cm}
    \includegraphics[scale=0.5]{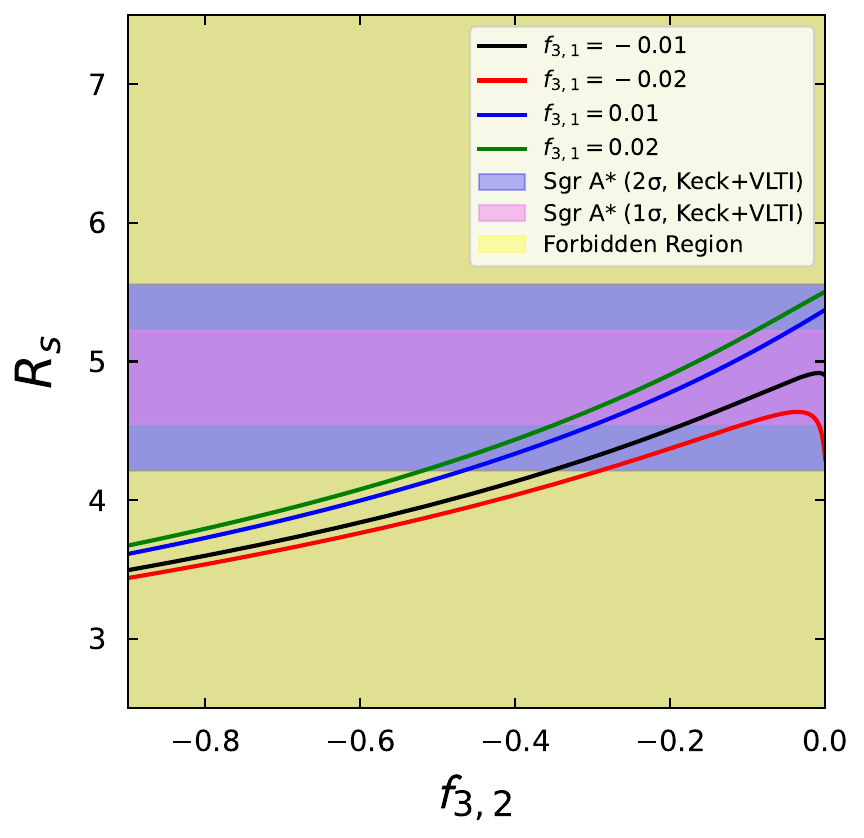}}
    \vspace{-0.2cm}
    \caption{Shadow radius versus the coefficients $f_{3,1}$ and $f_{3,2}$ 
in the background of Keck and VLTI experiments' constrains \cite{constrain1} 
obtained from the observations of Sgr $A^*$.}
    \label{fig11}
\end{figure}

Fig.~\ref{fig11} shows the shadow radius as a function of the 
coefficients $f_{3,1}$ and $f_{3,2}$ for different sign combinations, 
overlaid with observational bounds from the Keck and VLTI experiments. Here, 
the yellow regions show the forbidden zones, the blue regions are the allowed 
regions given by $2\sigma$ bound, and the violet region is the allowed region 
given by $1\sigma$ bound. In the upper row, the left panel presents the 
constraints on negative values of $f_{3,1}$ for the 1st-order correction. As 
evident from the figure, larger negative values of $f_{3,1}$ push the shadow 
radius rapidly into the lower forbidden region, whereas less negative values 
shift it into the allowed parameter space. The upper-right panel shows the 
variation of the shadow radius with positive values of $f_{3,1}$. It is clear 
from the plot that the shadow radius moves quickly into the upper forbidden 
region as $f_{3,1}$ becomes more positive. In the bottom row, we present the 
constraints on $f_{3,2}$ for the 2nd-order correction, with different fixed 
values of $f_{3,1}$. The left panel illustrates the variation of $R_s$ with 
positive values of $f_{3,2}$. As $f_{3,2}$ increases, the shadow radius 
shifts into the lower forbidden region, while for smaller positive values of 
$f_{3,2}$ shadow radius remains within the $2\sigma$ or $1\sigma$ allowed 
regions, depending on the value of $f_{3,1}$. More negative values of 
$f_{3,1}$ tend to push the shadow radius into the lower forbidden zone at 
smaller values of $f_{3,2}$ compared to less negative $f_{3,1}$ values. The 
right panel shows the variation of the shadow radius with negative values of 
$f_{3,2}$ for other sign combinations of $f_{3,1}$. This plot reveals that the 
shadow radius enters the lower forbidden region more rapidly when $f_{3,1}$ is 
negative compared to the case when it is positive. Furthermore, as the 
positive value of $f_{3,1}$ increases, the shadow radius tends to enter the 
allowed region more quickly.

This approach to constraining theoretical parameters has been utilized in the 
literature \cite{constrain2,constrain3} and by the EHT collaboration 
\cite{constrain1}, offering a reliable method for parameter estimation. 
However, when dealing with multiple model parameters, additional complementary 
techniques are required to isolate individual parameters and achieve more 
stringent constraints. We leave this aspect for future exploration.

In addition to the analysis of shadow constraints, we also 
investigated the possible existence of naked singularities (NSs) \cite{NS} in 
the parameter space of the modified metric function $f(r)$. An NS arises when 
the metric fails to admit a real, positive root of $f(r) = 0$, indicating the 
absence of an event horizon. To explore this, we systematically scanned a 
wide range of values for the correction parameters $f_{3,1}$ and $f_{3,2}$ and 
solved $f(r) = 0$ numerically. Our results show that for all physically 
meaningful and extended parameter values, the function $f(r)$ admits at least 
one real positive root. This confirms the absence of NS in the region 
considered. This is also confirmed by the horizon structures of the BH as 
shown in Fig.~\ref{figA}. Consequently, all parameter combinations shown in 
Fig.~\ref{fig11} correspond to regular BH solutions, and the shadow radius 
analysis is limited to the BH branch.
\section{Summery and Conclusion} \label{6}
The primary goal of this study is to investigate how different corrections to 
the Schwarzschild metric alter the BH's observables. We also studied the 
effects of the variation of free parameters on BH's observables. In 
this direction, we studied the effects on BH's QNMs and the BH's shadow 
that are introduced by the metric corrections up to 2nd-order terms. The study 
up to 2nd-order metric correction can provide us valuable information about 
the low energy QG effects.  

In Section~\ref{3} we studied the QNMs of the BH. Figs.~\ref{fig1}, \ref{fig2} 
and \ref{fig3} shows the behavior of the BH's effective  potential with 
respect to $r$ for different corrections to the metric and for different 
values of the free parameters. It is seen that the peak of the effective 
potential decreases compared to Schwarzschild case for the positive 
values of $f_{3,1}$, $f_{3,2}$ and $f_{3,3}$ and hight of the peak remains 
same after the inclusion of higher-order corrections. For negative values of 
these parameters, the peak of the effective potential increases compared to 
the Schwarzschild case as the parameters become more negative the peak also 
increases with the inclusion of the higher order corrections. Further, for 
negative values of $f_{3,1}$, positive values of $f_{3,2}$ and negative value 
of $f_{3,3}$, it was found that the peak of the effective potential increases 
compared to Schwarzschild case as the values of the free parameters increase 
in this case also the peak of the effective potential increases with the 
inclusion of higher-order corrections. Moreover all potential values merged 
together after a certain value of $r$ depending on the order of corrections. 
Thus we can say that the sign and magnitude of these parameters has an 
important role in the behavior of the effective potential $V(r)$. In this 
section we also calculated the QNM's of the BH using 6th order Padé-averaged 
WKB approximation method. Figs.~\ref{fig4}--\ref{fig5B} shows the behavior of 
the real part and the  imaginary part of the QNMs with respect to the 
parameters up to 2nd-order metric corrections. The calculated values of QNMs 
were presented in Tables.~\ref{table1}--\ref{table5} for different sign 
combinations of the parameters. It is evident from Tables 
\ref{table1}--\ref{table5} that different combinations of the parameters 
result in distinct patterns of QNMs, leading to significant deviations from 
the GR case. 

In section~\ref{4} we studied the evolution of a scalar field around our 
considered BH. It is clearly visible from the time profile that the 
oscillation and decay of QNMs are in agreement with the results that obtained 
from the 6th order Padé-averaged WKB approximation method. We also calculated 
the QNMs from the time profile of the evolution using Levenberg Marquardt 
algorithm. It can be seen that the QNMs frequencies are almost equal for both 
the methods used. Further the discrepancy between the results of these two 
methods is quantified using the parameter $\Delta_{QNM}$, defined in 
Eq.~\ref{eq24}.

In Section~\ref{5}, we studied the shadow of the BH. 
Figure~\ref{fig8} presents the behavior of the reduced potential $U_{r}$ as a 
function of $r$ for different values and sign combinations of the free 
parameters, corresponding to various orders of corrections. It can be observed 
that the peak of the reduced potential increases or decreases depending on 
both the sign of the parameters and the correction order. Furthermore, we 
provided a stereographic mapping of the shadow radius. As shown in 
Fig.~\ref{fig9}, the shadow radius is significantly influenced by the order 
of the correction. We also examined the variation of the shadow radius with 
respect to $f_{3,1}$ and $f_{3,2}$ for different sign combinations. It was 
found that, for corrections up to 1st-order, the shadow radius decreases with 
negative values of $f_{3,1}$ and increases with positive values. For 2nd-order 
corrections, the variation of the shadow radius depends on the combined signs 
of $f_{3,1}$ and $f_{3,2}$. Moreover, we studied the dependence of the shadow 
radius on the BH mass for both 1st-order and 2nd-order corrections. As shown 
in Fig.~\ref{fig10B}, the shadow radius increases with the mass of the BH in 
both cases. Finally, we constrained the parameters $f_{3,1}$ and $f_{3,2}$ 
using data from the Keck and VLTI observations. In addition, we investigated 
the possibility of NSs by analyzing whether the metric function $f(r)$ admits 
a real, positive root for various parameter combinations. Our analysis shows 
that for all physically meaningful values of $f_{3,1}$ and $f_{3,2}$ 
considered, the function $f(r)$ possesses at least one real, positive root, 
confirming the existence of an event horizon. Therefore, no NSs were found in 
the parameter space studied.

Our study mainly focus on how the corrections to the metric made change to the BH's observables viz. QNMs and shadow and how the observables behaves with respect to the parameters of the theory. To study the effects arises from the metric corrections we consider up to 2nd-order corrections for simplicity, the more higher-order corrections will create more mathematical complexity. However the corrections to Schwarzschild metric up to 2nd-order correction introduced by IDG has introduced a significant deviation to the BH's QNMs and shadow radius from the results that obtained from GR. This could be due to the tiny QG effects included by the corrections. We hope this study will contribute towards a better understanding on the effects on BH's observables arises from the higher order correction to the metric in IDG. This study sheds light on our understanding of low-energy QG effects and offers valuable insights for future research.

\section*{Acknowledgements} UDG is thankful to the Inter-University Centre 
for Astronomy and Astrophysics (IUCAA), Pune, India for the Visiting 
Associateship of the institute.


\end{document}